\documentclass[12pt, leqno]{article}
\usepackage{amsmath,caption,setspace,multirow}
\usepackage[top=1.1in, bottom=1.1in, left=1.1in, right=1.1in]{geometry}

\usepackage[round]{natbib}
\usepackage{color,soul}

\DeclareCaptionStyle{italic}[justification=centering]{labelfont={bf},textfont={it},labelsep=colon}
\usepackage{graphicx,psfrag,epsf,textcomp,epstopdf}
\usepackage{enumerate, dsfont}
\usepackage{natbib}
\usepackage{url,xcolor}
\usepackage{booktabs, subfig, bm, paralist,mathpazo,tikz,todonotes,longtable,microtype,algorithm}

\usepackage[pdftex,colorlinks=true]{hyperref}
\definecolor{darkblue}{rgb}{0,0,.6}
\hypersetup{citecolor=darkblue,linkcolor=darkblue,urlcolor=darkblue}
\definecolor{DarkRed}{rgb}{.7,0,.4}

\newcommand\Tau{\mathcal{T}}
\usepackage{comment}

\newcommand{\blind}{0}

\newcommand{\X}{\mathcal{X}}
\newcommand{\Y}{\mathcal{Y}}

\graphicspath{{plots/}}

\newsavebox\CBox

 \newtheorem{@definition}{\sc Definition}[section]

  \renewcommand\X{\mathcal{X}}

\begin{document}

\def\spacingset#1{\renewcommand{\baselinestretch}{#1}\small\normalsize} \spacingset{1}

\if0\blind
{
  \title{\bf A comparison of parameter estimation in function-on-function regression}
    \author{
Ufuk Beyaztas \\
Department of Statistics \\
Bartin University \\
\\
Han Lin Shang \\
{Research School of Finance, Actuarial Studies, and Statistics} \\
{Australian National University}
 }
  \maketitle
} \fi

\if1\blind
{
  \bigskip
  \bigskip
  \bigskip
  \begin{center}
    {\LARGE\bf A comparison of parameter estimation in function-on-function regression}
\end{center}
  \medskip
} \fi

\maketitle

\begin{abstract}
Recent technological developments have enabled us to collect complex and high-dimensional data in many scientific fields, such as population health, meteorology, econometrics, geology, and psychology. It is common to encounter such datasets collected repeatedly over a continuum. Functional data, whose sample elements are functions in the graphical forms of curves, images, and shapes, characterize these data types. Functional data analysis techniques reduce the complex structure of these data and focus on the dependences within and (possibly) between the curves. A common research question is to investigate the relationships in regression models that involve at least one functional variable. However, the performance of functional regression models depends on several factors, such as the smoothing technique, the number of basis functions, and the estimation method. This paper provides a selective comparison for function-on-function regression models where both the response and predictor(s) are functions, to determine the optimal choice of basis function from a set of model evaluation criteria. We also propose a bootstrap method to construct a confidence interval for the response function. The numerical comparisons are implemented through Monte Carlo simulations and two real data examples.
\end{abstract}


\newpage
\spacingset{1.56}

\section{Introduction\label{sec:intro}}

Multivariate statistical techniques are best suited to analyze the data obtained from a discrete data matrix. On the other hand, recent technological advances have led to collecting functional data that are measured repeatedly over discrete time points, and frequently occur in many research fields. Existing multivariate methods may not be capable of analyzing such data due to common technical, issues such as multicollinearity, high dimensionality, and the possible high correlation among observations. Thus, the need for functional data analysis (FDA) techniques is increasing. FDA has several substantial advantages over conventional methods; for example, 
\begin{inparaenum}
\item[1)] it reduces the dimensionality of the data,
\item[2)] it bypasses the problems of missing data and the high correlation between sequential observations,
\item[3)] it minimizes data noise, and 
\item[4)] it provides additional information about the data, such as smoothness and derivatives. See \cite{ramsay2002, ramsay2006}, \cite{ferraty2006}, \cite{horvath2012} and \cite{cuevas2014} for more information about FDA and its applications.
\end{inparaenum}

Functional linear models, among many others are used to explain the relationship between a response and its predictor(s), and they have received considerable attention in the literature. Several functional regression analysis techniques have been proposed, depending on whether the response/predictors are scalar or function: 
\begin{inparaenum}
\item[1)] function-on-scalar; 
\item[2)] scalar-on-function; and 
\item[3)] function-on-function. 
\end{inparaenum}
For the first two cases, well-known examples include \cite{ramsay1991}, \cite{cardot1999, cardot2003}, \cite{james2002}, \cite{hu2004}, \cite{hall2007}, \cite{reiss2007}, \cite{ferraty2009}, \cite{cook2010}, \cite{malloy2010}, \cite{chen2011}, \cite{goldsmith2011}, \cite{dou2012}, and \cite{mclean2012}. For the last case, the focus of this paper, consult \cite{ramsay1991}, \cite{fan1999}, \cite{senturk2005, senturk2008}, \cite{yao2005}, \cite{harezlak2007}, \cite{matsui2009}, \cite{valderrama2010}, \cite{he2010}, \cite{jiang2011}, \cite{ivanescu2015}, \cite{chiou2016}, and \cite{zhang2018}, and references therein.

Early studies on functional linear regression models were conducted by \cite{ramsay1991}, who constructed a functional regression model for a functional response and functional predictors. \cite{ramsay2006} proposed the least squares (LS) method to estimate this regression model, while \cite{yamanishi2003} suggested a weighted LS method. \cite{matsui2009} pointed out that the LS method produces unstable/unfavorable estimates; thus, they applied the maximum penalized likelihood (MPL) method to obtain stable estimates in functional linear models. In this study, we restrict our attention to the LS and MPL methods, since they are commonly used methods for these analyses. In addition to the estimation methods, the accuracy of functional linear models also depends on the chosen smoothing technique, smoothing parameter, the choice of the number of basis functions, and the model evaluation criteria.

The first step in FDA is to smooth the functional data by a suitable basis function system. In most studies, \textit{B-spline basis} and \textit{Fourier basis} functions have been used to express discretely observed data as smooth functions. The \textit{Gaussian basis} function, which is part of the general class of radial basis functions, is also a suitable instrument to obtain smooth curves from discrete data. For more information about these smoothing techniques, refer to \cite{ramsay2006} and \cite{ando2008}. Throughout this study, all three bases mentioned above are considered to smooth functional data. Four model evaluation criteria-\textit{Generalized Bayesian Information Criterion} (GBIC), \textit{Generalized Information Criterion} (GIC), \textit{Modified Akaike Information Criterion} (MAIC), and \textit{Generalized Cross-Validation} (GCV)-are used to choose the appropriate smoothing parameter and the number of basis functions. Apart from comparing these smoothing techniques, we propose a case-resampling-based bootstrap method to evaluate estimation accuracy, and focus on constructing a confidence interval for the response function.

The remaining of this paper is organized as follows. Section~\ref{sec:meth} provides an overview of the functional data, the functional linear model and its estimation strategies, as well as the model evaluation criteria. Section~\ref{sec:num} compares the performance of the estimation methods and smoothing techniques under several model evaluation criteria via a Monte Carlo experiment. Section~\ref{sec:real} reports the results obtained by implementing the smoothing tools using two data analyses. Section~\ref{sec:conc} concludes the paper, and offers some ideas on how the methodology presented could be further extended.

\section{Methodology\label{sec:meth}}

\subsection{Notations and nomenclature}

Let $t = \left\lbrace t_1, \cdots, t_J \right\rbrace$ denote the discrete time points where the sample elements (or random functions) of a functional dataset $\left\lbrace \X_i(t): i = 1, \cdots, N, ~ t \in \Tau \right\rbrace$ are recorded (where $\Tau$ is a closed and bounded interval). Denote the probability space $\left( \Omega, \mathcal{F}, P\right)$, where $\Omega$, $\mathcal{F}$, and $P$ are the sample space, sigma algebra, and the probability measure, respectively. Also, denote $\left( \mathcal{H}, \langle\cdot,\cdot\rangle \right)$ as the separable Hilbert space with norm $\parallel \cdot \parallel$ generated by the inner product $\langle\cdot,\cdot\rangle$. Then, the functional random variable $\X = \left\lbrace \X(t): t \in \Tau \right\rbrace$ is defined as $\X: \left( \Omega, \mathcal{F}, P\right) \rightarrow \mathcal{H}$. Most of the FDA processes are canalized within the $\mathcal{L}_2 = \mathcal{L}_2(\Tau)$ separable Hilbert space, which is the space of square integrable and real-valued functions defined on $\Tau$, $f: \Tau \rightarrow \mathbb{R}$ satisfying $\int_{\Tau} f^2(t) dt < \infty$. The inner product on $\mathcal{H}$ is defined by 
\begin{equation*}
\langle f,g \rangle = \int_{\Tau} f(t) g(t) dt, ~ \qquad \forall f, g \in \mathcal{L}_2
\end{equation*}
We assume that the functional random variable is an element of $\mathcal{L}_2$. We further assume that $\X(t) \in \mathcal{L}_2$ is a second-order stochastic process, so that it has a finite second-order moment: $\text{E} \left( \vert \X \vert \right) = \int_{\Omega} \vert \X \vert^2 dP < \infty$.

\subsection{Basis function expansion\label{sec:smooth}}

An element of functional data $\X_i(t)$ can be approximated by a linear combination of basis functions $\phi_k(t)$ and associated coefficients $c_{ik}$ for a sufficiently large number of $K$; that is:
\begin{align*}
\X_i(t) &= \sum_{k=1}^{\infty} c_{ik} \phi_k(t), \\
\widehat{\X}_i(t) &= \sum_{k=1}^K c_{ik} \phi_k(t),\qquad k=1,\dots,K,
\end{align*}
where $\widehat{\X}_i(t)$ is the approximation of $\X_i(t)$ and converges to $\X_i(t)$ as $K \rightarrow \infty$. The beauty of the basis function expansion is that the key features, as well as the non-linearity of the data, are captured by the basis functions and the model remains linear in the transformations. An important task in the basis function expansion is to choose an appropriate basis function. Here, we consider three basis functions: Fourier, B-spline, and Gaussian.

The Fourier basis is the most appropriate for approximating the periodic functions defined on $\Tau$. Let $\omega = \frac{2 \pi}{\Tau}$ denote the frequency. The Fourier basis functions for an even integer $K$ are defined as follows:
\begin{equation*}
\phi_0(t) = \frac{1}{\sqrt{\Tau}} \qquad \phi_{2r-1} = \frac{\sin r \omega t}{\sqrt{\Tau / 2}} \qquad \phi_{2r} = \frac{\cos r \omega t}{\sqrt{\Tau / 2}},
\end{equation*}
for $r = 1, \cdots, K/2$. If the observations are equally spaced at discrete time points, then the Fourier basis functions are orthonormal basis functions. The Fourier basis is useful when the functions are stable (i.e., when no strong local features are present in the functions). Further, it computes the coefficients quickly and efficiently since it is commonly implemented with the fast Fourier transformation algorithm. However, it is not useful for functions with strong local features.

The B-spline basis is one of the most commonly used basis function expansions in FDA for non-periodic data. Conceptually, B-splines are the linear combinations of the piecewise polynomial functions neighboring smoothly at a set of breakpoints (called knots). Let $\tau = \left\lbrace \tau_1, \cdots, \tau_{v+1} \right\rbrace$ denote an increasing sequence of breakpoints, $\tau_0 < \cdots < \tau_{v}$, which divide the interval $\Tau$ into $v$ subintervals (knots). In this case, the $k$\textsuperscript{th} B-spline basis function is defined as $\phi_k(t) = B_{k,v}(t)$. To construct B-spline basis functions, one needs extra knots outside the boundary of the knot sequence $\left[ \tau_0, \tau_{v} \right]$. Define $\cdots < \tau_{-1} < \tau_0 < \tau_1 < \cdots < \tau_{v} < \tau_{v+1} < \cdots$ by the augmented knot sequence. Then, the constant B-splines are defined as follows:
\begin{equation*}
B_{k,1}(t) = \begin{cases} 
      1 & \tau_k \leq t < \tau_{k+1} \\
      0 & \text{otherwise} ~~~~.
   \end{cases}
\end{equation*}
Using constant B-splines, the high order B-splines are constructed via the following recursion:
\begin{equation*}
B_{k,v}(t) = \frac{t - \tau_k}{\tau_{k+v-1}-\tau_k} B_{k, v-1}(t) + \frac{\tau_{k+v} - t}{\tau_{k+v} - \tau_{k+1}} B_{k+1, v-1}(t).
\end{equation*}
The advantage of the B-spline basis is its flexibility and it is computationally fast for computing derivatives as it is locally nonzero and has compact support.

The Gaussian basis is another frequently used basis function in FDA, and is part of the general class of radial basis functions proposed by \cite{ando2008}. The Gaussian basis functions are defined as follows:
\begin{equation*}
\phi_k(t) = \exp \left\lbrace - \frac{\left( t- \tau_{k+2} \right)^2}{2 \sigma^2} \right\rbrace ,
\end{equation*}
where $\tau_k$s are evenly spaced knots, which determine the centers of the basis functions, satisfying $\tau_1 < \cdots < \tau_4 = \min(t) < \tau_5 < \cdots < \tau_{K+2} = \max(t) < \cdots < \tau_{K+4}$, and $\sigma = \frac{\tau_{k+2} - \tau_k}{3}$ is the width parameter. In a Gaussian basis function, parameters are identified based on the structure of the data. In addition, it produces a sparse design matrix, which enables faster computation. Note that B-splines and Gaussian bases can also be used to fit periodic data if the smoothing parameter and the number of basis functions are appropriately determined.

The accuracy of converting discrete observations into a functional form depends on the choice of the appropriate number of basis functions $K$. On the one hand, discrete data are well fitted by smooth functions when $K$ is large, but the noise present in the data may not be eliminated. On the other hand, critical features of the data cannot be captured by the smooth functions when $K$ is small. As a tradeoff, the penalized LS method aims to minimize the residual sum of squares (RSS) between the original data and smooth data with an optimally chosen $K$. It is given in Equation~\eqref{mpl1}:
\begin{equation}\label{mpl1}
\text{RSS} = \sum_{j=1}^J \left( \X_{i}(t_j) - \sum_{k=1}^K c_{ik} \phi_k(t_j) \right)^2 + \lambda \int_{\Tau} \left[ d(t) \right]^2 dt, ~~ i = 1, \cdots, N,
\end{equation}
where $d(t)$, the $n$\textsuperscript{th} derivative of $\X(t)$; $\int_{\Tau} \left[ d(t) \right]^2 dt = \int_{\Tau} \left[ D^{(n)} \X(t) \right]^2 dt$,  measures the roughness of the expansion, and $\lambda$ is the smoothing parameter that controls the degree of roughness. More precisely, the roughness term is defined as follows:
\begin{align}
\int_{\Tau} \left[ d(t) \right]^2 dt &= \int_{\Tau} \left[ D^{(n)} \X(t) \right]^2 dt \nonumber \\
&= \int_{\Tau} \left( c^\top D^{(n)} \phi(t) D^{(n)} \phi^\top(t) c \right) dt \nonumber \\
&= \mathbf{c}^\top \mathbf{R}_n \mathbf{c}, \label{mpl2}
\end{align}
where $\mathbf{R}_n = \int_{\Tau} D^{(n)} \phi(t) D^{(n)} \phi^\top(t) dt$. From equations~\eqref{mpl1} and~\eqref{mpl2}, the penalized least squares estimate of $\mathbf{c}$ is obtained as 
\begin{equation*}
\mathbf{\widehat{c}} = \left( \mathbf{\Phi}^\top \mathbf{\Phi} + \widehat{\lambda} \mathbf{R}_n \right)^{-1} \mathbf{\Phi}^\top \mathbf{\X}, 
\end{equation*}
where $\mathbf{\Phi} = \left\lbrace \phi_1(t), \cdots, \phi_K(t) \right\rbrace$. Analogously, the projection matrix $S_{\lambda}$ is obtained as $S_{\lambda} = \mathbf{\Phi} \left( \mathbf{\Phi}^\top \mathbf{\Phi} + \widehat{\lambda} \mathbf{R}_n \right)^{-1} \mathbf{\Phi}^\top$. The degrees of freedom of roughness is given by $\text{df}(\lambda) = \text{trace} \left( S_{\lambda} \right)$. The function fits every point of the discrete data when $\lambda \rightarrow 0$, and tends to take the same form as the standard regression curve as $\lambda \rightarrow \infty$. A number of techniques, including GBIC \citep{konishi2004}, GIC \citep{konishi2008}, MAIC \citep{fujikoshi1997}, and GCV \citep{craven1979} have been proposed to choose the optimal smoothing parameter, $\lambda$, and the number of basis functions $K$. A brief description of these  techniques is given in Section~\ref{sec:2.4}. Please see \cite{ramsay2006} and \cite{matsui2009} for more information and a derivation of the information criteria considered in this study.

\subsection{Function-on-function regression model}

Let $\Y_i(t)$ for $i=1,\dots,N$ denote a set of functional responses, and $\X_{im}(s)$ for $m=1,\dots,M$ denote a set of functional predictors, where $s \in S$ and $t \in \Tau$ are closed and bounded intervals on the real line. We consider the regression of $\Y_i(t)$ on $\X_{im}(s)$ to explain the functional relationship between the functional response and $m$ functional predictors, which can be formulated by the following multiple functional linear model \citep{ramsay2006}:
\begin{equation}
\Y_i(t) = \beta_0(t) + \sum_{m=1}^M \int_{S} \X_{im}(s) \beta_m(s,t) ds + \epsilon_i(t), \label{reg1}
\end{equation}
where $\beta_0(t)$, $\beta_m(s,t)$, and $\epsilon_i(t)$ denote the intercept function, bivariate coefficient function linking the response with the $m$\textsuperscript{th} predictor, and the random error function having a Gaussian process ($\epsilon_i(t) \sim \text{GP}(\mathbf{0}, \mathbf{\Sigma}_{\epsilon})$), respectively. In~\eqref{reg1}, the linear relationship is characterized by the surface $\beta_m(s,t)$. In practice, the functional response and functional predictors are centered, and thus, without loss of generality, the intercept function $\beta_0(t)$ is eliminated from~\eqref{reg1}. Let $\Y^*_i(t) = \Y_i(t) - \overline{\Y}(t)$ and $\X^*_{im}(s) = \X_{im}(s) - \overline{\X}_m(s)$, where $\overline{\Y}(t) = N^{-1} \sum_{i=1}^N \Y_i(t)$ and $\overline{\X}_m(s) = N^{-1} \sum_{i=1}^N \X_{im}(s)$  denote the centered counterparts of $\Y_i(t)$ and $\X_{im}(s)$, respectively. Then, the regression model given in Equation~\eqref{reg1} is re-expressed as in~\eqref{regc}.
\begin{equation}
\Y^*_i(t) = \sum_{m=1}^M \int_{S} \X^*_{im}(s) \beta_m(s,t) ds + \epsilon_i(t). \label{regc}
\end{equation}

The usual approach before fitting the functional regression model is to represent the functional response and predictors, as well as the bivariate coefficient function, as basis function expansions. Following Section~\ref{sec:smooth}, the centered functional response and functional predictors can be written as follows:
\begin{align*}\label{eq:smv}
\Y^*_i(t) &= \sum_{k=1}^{K_{\Y}} c_{ik} \phi_k(t) = \mathbf{c}^\top_i \mathbf{\Phi}(t), \qquad \forall t \in \Tau \nonumber \\
\X^*_{im}(s) &= \sum_{j=1}^{K_{m,\X}} d_{imj} \psi_{mj}(s) = \mathbf{d}^\top_{im} \mathbf{\Psi}(s), \qquad \forall s \in S,
\end{align*}
where $\mathbf{\Phi}(t) = \lbrace \phi_1(t), \cdots, \phi_{K_{\Y}}(t) \rbrace^\top$ and $\mathbf{\Psi}(s) = \lbrace \psi_{m1}(s), \cdots, \psi_{mK_{m,\X}}(s) \rbrace^\top$ denote the vectors of the basis functions, $\mathbf{c}_{i} = \lbrace c_{i1}, \cdots, c_{iK_{\Y}} \rbrace^\top$ and $\mathbf{d}_{im} = \lbrace d_{im1}, \cdots, d_{imK_{m,\X}} \rbrace^\top$ are the coefficient vectors, and $K_{\Y}$ and $K_{m,\X}$ are the number of basis functions used for approximating the functional response and functional predictors, respectively. Similarly, the bivariate coefficient function is defined as:
\begin{equation*}\label{smc}
\beta_m(s,t) = \sum_{j,k} \psi_{mj}(s) b_{mjk} \phi_k(t) = \mathbf{\Psi}^\top_m(s) \mathbf{B}_m \mathbf{\Phi}(t),
\end{equation*}
where $\mathbf{B}_m = (b_{mjk})_{j,k}$ denote a $K_{m,\X} \times K_{\Y}$ dimensional coefficient matrix. Accordingly, the regression model defined in~\eqref{regc} can be expressed in the discrete form, as follows:
\begin{align}\label{regs}
\mathbf{c}^\top_i \mathbf{\Phi}(t) &= \sum_{m=1}^M \mathbf{d}^\top_{im} \bm{\zeta}_{\psi_m} \mathbf{B}_m \mathbf{\Phi}(t) + \epsilon_i(t) \nonumber \\
&= z^\top_i \mathbf{B} \mathbf{\Phi}(t) + \epsilon_i(t),
\end{align}
where $\bm{\zeta}_{\psi_m} = \int_{S} \psi_m(s) \psi^\top_m(s) ds$ is a matrix with dimension $K_{m,\X} \times K_{m,\X}$, $z_i = \left( \mathbf{d}^\top_{i1} \mathbf{\zeta}_{\psi_1}, \cdots, \mathbf{d}^\top_{iM} \mathbf{\zeta}_{\psi_M} \right)^\top$ is a vector with length $\sum_{m=1}^M K_{m,\X}$, and $\mathbf{B} = \left( \mathbf{B}_1, \cdots, \mathbf{B}_M \right)^\top$ is a $\sum_{m=1}^M K_{m,\X} \times K_{\Y}$ dimensional coefficient matrix. Note that the matrix $\bm{\zeta}_{\psi_m}$ is equal to $\mathbb{I}_{K_{m,\X}}$ when the basis functions are orthogonal. On the other hand, for non-orthogonal basis functions, such as a Gaussian basis, $(j-k)$\textsuperscript{th} elements are calculated as follows \citep{matsui2009}:
\begin{equation*}
\bm{\zeta}^{j,k}_{\psi_m} = \sqrt{\pi \sigma^2} \exp \left\lbrace - \frac{\tau_{j+2} - \tau_{k+2}}{4 \sigma^2} \right\rbrace. 
\end{equation*}

The LS method, proposed by \cite{ramsay2006} to estimate the coefficient matrix $\mathbf{B}$ works as follows. Let $\mathbf{C} = \left( \mathbf{c}_1, \cdots, \mathbf{c}_N \right)^\top$ and $\mathbf{Z} = \left( \mathbf{z}_1, \cdots, \mathbf{z}_N \right)^\top$. The first step is to minimize the integrated sum of squares,
\begin{align*}
& \sum_{i=1}^N \int_{\Tau} \left[ \Y^*_i(t) - \sum_{m=1}^M \int_{S} \X^*_{im}(s) \beta_m(s,t) ds \right]^2 dt \nonumber \\
&= \int_{\Tau} \text{trace} \left\lbrace \left[ \mathbf{C} \mathbf{\Phi}(t) - \mathbf{Z} \mathbf{B} \mathbf{\Phi}(t) \right] \left[ \mathbf{C} \mathbf{\Phi}(t) - \mathbf{Z} \mathbf{B} \mathbf{\Phi}(t) \right]^\top \right\rbrace dt \nonumber \\
&= \int_{\Tau} \text{trace} \left\lbrace \left( \mathbf{C} - \mathbf{Z} \mathbf{B} \right) \mathbf{\Phi}(t) \mathbf{\Phi}^\top(t) \left( \mathbf{C} - \mathbf{Z} \mathbf{B} \right)^\top \right\rbrace dt \nonumber \\
&= \text{trace} \left\lbrace \left( \mathbf{C} - \mathbf{Z} \mathbf{B} \right) \bm{\zeta}_{\phi} \left( \mathbf{C} - \mathbf{Z} \mathbf{B} \right)^\top \right\rbrace.
\end{align*}
Then, the LS estimator of $\mathbf{B}$ is obtained as:
\begin{equation}
\text{vec} \left( \widehat{\mathbf{B}} \right)  = \left( \bm{\zeta}_{\phi} \otimes \mathbf{Z}^\top \mathbf{Z} \right)^{-1} \text{vec} \left( \mathbf{Z}^\top \mathbf{C} \bm{\zeta}_{\phi} \right), \label{LSe}
\end{equation} 
where $\text{vec}$ and $\otimes$ denote the column-stacking operator and Kronecker product, respectively.

Hereafter, we describe the MPL method proposed by \cite{matsui2009} in detail. Suppose the error function $\pmb{\epsilon}^*_i(t)$ has the form $\pmb{\epsilon}^*_i(t) = \mathbf{e}^\top_i \mathbf{\Phi}(t)$, where $\mathbf{e}_i = \left( e_{i1}, \cdots, e_{iK} \right)^\top$ is a vector consisting of independent and identically distributed Gaussian random variables with mean $\mathbf{0}$ and variance-covariance matrix $\mathbf{\Sigma}$. Then, the regression equation given in~\eqref{regs} has the following form:
\begin{equation}
\mathbf{c}^\top_i \mathbf{\Phi}(t) = z^\top_i \mathbf{B} \mathbf{\Phi}(t) + \mathbf{e}^\top_i \mathbf{\Phi}(t). \label{rmax}
\end{equation}
Multiplying both sides of equation~\eqref{rmax} from the right by $\mathbf{\Phi}^\top(t)$ and integrating with respect to the function support $\Tau$ yields:
\begin{align}
\mathbf{c}^\top_i \mathbf{\Phi}(t) \mathbf{\Phi}^\top(t) &= z^\top_i \mathbf{B} \mathbf{\Phi}(t) \mathbf{\Phi}^\top(t) + \mathbf{e}^\top_i \mathbf{\Phi}(t) \mathbf{\Phi}^\top(t) \nonumber \\
\int_{\Tau} \mathbf{c}^\top_i \mathbf{\Phi}(t) \mathbf{\Phi}^\top(t) dt &= \int_{\Tau} z^\top_i \mathbf{B} \mathbf{\Phi}(t) \mathbf{\Phi}^\top(t) dt + \int_{\Tau} \mathbf{e}^\top_i \mathbf{\Phi}(t) \mathbf{\Phi}^\top(t) dt \nonumber \\
\mathbf{c}^\top_i \bm{\zeta}_{\phi} &= z^\top_i \mathbf{B} \bm{\zeta}_{\phi} + \mathbf{e}^\top_i \bm{\zeta}_{\phi} \nonumber \\
\mathbf{c}_i &= \mathbf{B}^\top z_i + \mathbf{e}_i, \label{reglin}
\end{align}
since $\bm{\zeta}_{\phi}$ is nonsingular. For \eqref{reglin}, the probability density function is given by
\begin{equation*}
f \left( \mathbf{\Y}_i | \mathbf{\X}_i; \bm{\theta} \right) = \frac{1}{\left( 2 \pi \right)^{K/2} \vert \mathbf{\Sigma} \vert^{1/2}} \exp \left\lbrace  - \frac{1}{2} \left( \mathbf{c}_i - \mathbf{B}^\top z_i \right)^\top \mathbf{\Sigma}^{-1/2} \left( \mathbf{c}_i - \mathbf{B}^\top z_i \right) \right\rbrace, 
\end{equation*}
where $\bm{\theta} = \left( \mathbf{B}, \mathbf{\Sigma} \right)$ is the parameter vector. Denote the penalized log-likelihood function by $\ell_{\lambda}(\bm{\theta})$,
\begin{equation*}
\ell_{\lambda}(\bm{\theta}) = \sum_{i=1}^N \ln f \left( \mathbf{\Y}_i | \mathbf{\X}_i; \bm{\theta} \right) - \frac{N}{2} \text{trace} \left\lbrace \mathbf{B}^\top \left( \mathbf{\Lambda}_M \odot \mathbf{\Omega} \right) \right\rbrace,
\end{equation*}
where $\mathbf{\Lambda}_M = \pmb{\lambda}_M \pmb{\lambda}^\top_M$ with $\pmb{\lambda}_M = \left( \sqrt{\lambda_1} \mathds{1}^\top_{K_{1,x}}, \cdots, \sqrt{\lambda_M} \mathds{1}^\top_{K_{M,\X}} \right) $ is a $\left( \sum_{m=1}^M K_{m,\X} \right) \times \left( \sum_{m=1}^M K_{m,\X} \right)$ dimensional  matrix of penalty parameters, $\mathbf{\Omega}$ is a positive semi-definite matrix, and $\odot$ denotes the Hadamard product. The penalized maximum likelihood estimator of $\theta$, $\widehat{\theta} = \left( \widehat{\mathbf{B}}, \widehat{\mathbf{\Sigma}} \right)$ is obtained by equating the derivatives of the penalized log-likelihood function with respect to $ \bm{\theta} = \left( \mathbf{B}, \mathbf{\Sigma} \right) $ to $\mathbf{0}$,
\begin{align}
\text{vec}(\widehat{\mathbf{B}}) &= \left( \widehat{\mathbf{\Sigma}}^{-1} \otimes \mathbf{Z}^\top \mathbf{Z} + N \mathbf{I}_{K_{\Y}} \otimes \left( \mathbf{\Lambda}_M \odot \mathbf{\Omega} \right) \right)^{-1} \left( \widehat{\mathbf{\Sigma}}^{-1} \otimes \mathbf{Z}^\top \right) \text{vec}(\mathbf{C}) \label{PMLe} \\
\widehat{\mathbf{\Sigma}} &= \frac{1}{N} \left( \mathbf{C} - \mathbf{Z} \widehat{\mathbf{B}} \right)^\top \left( \mathbf{C} - \mathbf{Z} \widehat{\mathbf{B}} \right) \nonumber.
\end{align}
Finally, the penalized maximum likelihood estimator of $\mathbf{C}$ is obtained as:
\begin{align*}
\text{vec} \left( \widehat{\mathbf{C}} \right) &= \text{vec} \left( \mathbf{Z} \widehat{\mathbf{B}} \right)  \nonumber \\
&= \left( \mathbf{I}_{K_{\Y}} \otimes Z\right) \left( \widehat{\mathbf{\Sigma}}^{-1} \otimes \mathbf{Z}^\top \mathbf{Z} + N \mathbf{I}_{K_{\Y}} \otimes \left( \mathbf{\Lambda}_M \odot \mathbf{\Omega} \right) \right)^{-1} \times \left( \widehat{\mathbf{\Sigma}}^{-1} \otimes \mathbf{Z}^\top \right) \text{vec}(\mathbf{C}).
\end{align*}

As stated above, the LS method estimates the coefficient matrix $\widehat{\mathbf{B}}$ by minimizing the integrated sum of squares of the differences between the observed and fitted functions. It works well in certain circumstances; however, it produces unstable estimates when the data have degenerate structures \citep[see also][]{matsui2009}. Also, because of the ill-posed nature of the function-on-function regression model, the LS method encounters a singular matrix problem when a large number of functional predictors are included in the model. Compared with the LS, the MPL produces stable estimates for the coefficient matrix $\widehat{\mathbf{B}}$. However, it is computationally more intensive than the LS method.

The performance of this method is based on the penalty parameters, and thus an information criterion is needed to select the best overall model. The information criterion techniques mentioned in Section~\ref{sec:2.4} are used to evaluate the model selection in both LS and MPL methods.

\subsection{Model selection criteria}\label{sec:2.4}

This section briefly describes the information criteria considered to determine the optimal number of basis functions and roughness parameter as well as to select the best approximating model.

The GBIC is proposed by \cite{konishi2004} by extending the usual Bayesian information criterion to select the optimal penalty parameter as well as to evaluate estimated models as follows:
\begin{align*}
\text{GBIC} &= -2 \sum_{i=1}^N \ln f \left( \Y_i \vert \X_i; \hat{\pmb{\theta}} \right) + N \text{trace} \left\lbrace \mathbf{B}^\top \left( \mathbf{\Lambda}_M \odot \mathbf{\Omega} \right) \mathbf{B} \right\rbrace \\
& + \left( r + Kq \right) \ln N - \left( r + Kq\right) \ln \left( 2 \pi \right) \\
& - K \ln \vert \mathbf{\Lambda}_M \odot \pmb{\Omega} \vert_{+} + \ln \vert R_{\lambda} \left(\hat{\pmb{\theta}} \right) \vert,
\end{align*}
where $q = p - \text{rank} \left(\pmb{\Omega}\right)$, $p = \sum_m K_{m, \X}$, $r = K_{\Y} \left(K_{\Y} + 1 \right) / 2$ and 
\begin{equation*}
R_{\lambda} \left(\hat{\pmb{\theta}} \right) =  \frac{1}{N} \sum_{i=1}^N \frac{\partial^2}{\partial \pmb{\theta} \partial \pmb{\theta}^\top} \left\lbrace  \ln f \left( \Y_i \vert \X_i; \pmb{\theta} \right) - \frac{1}{2} \text{trace} \left\lbrace \mathbf{B}^\top \left( \mathbf{\Lambda}_M \odot \mathbf{\Omega} \right) \mathbf{B} \right\rbrace \right\rbrace.
\end{equation*}

The GIC of \cite{konishi2008} evaluates the estimated models in a following way:
\begin{equation*}
\text{GIC} = -2 \sum_{i=1}^N \ln f \left( \Y_i \vert \X_i; \hat{\pmb{\theta}} \right) + 2 \text{trace} \left\lbrace R_{\lambda} \left( \hat{\pmb{\theta}} \right) ^{-1} Q_{\lambda} \left( \hat{\pmb{\theta}} \right) \right\rbrace,
\end{equation*}
where
\begin{equation*}
Q_{\lambda} \left( \hat{\pmb{\theta}} \right) = \sum_{i=1}^N \frac{\partial}{\partial \pmb{\theta}} \left\lbrace \ln f \left( \Y_i \vert \X_i; \pmb{\theta} \right) - \frac{1}{2} \text{trace} \left\lbrace \mathbf{B}^\top \left( \mathbf{\Lambda}_M \odot\mathbf{\Omega} \right) \mathbf{B} \right\rbrace \right\rbrace \frac{\partial}{\partial \pmb{\theta}^\top} \ln f \left( \Y_i \vert \X_i; \pmb{\theta} \right).
\end{equation*}

The MAIC of \cite{fujikoshi1997} is used to select the best estimated model as follows:
\begin{equation*}
\text{MAIC} = -2 \sum_{i=1}^N  \ln f \left( \Y_i \vert \X_i; \pmb{\theta} \right) + 2 \text{trace} \left( S_{\lambda} \right),
\end{equation*}
where $S_{\lambda} = \left( \widehat{\mathbf{\Sigma}}^{-1} \otimes \mathbf{Z}^\top \mathbf{Z} + N \mathbf{I}_{K_{\Y}} \otimes \left( \mathbf{\Lambda}_M \odot \mathbf{\Omega} \right) \right)^{-1} \left( \widehat{\mathbf{\Sigma}}^{-1} \otimes \mathbf{Z}^\top \right)$.
The only problem related to the use of MAIC may be the theoretical justificiation of the bias-correction terms in the MAIC since the usual Akaike information criterion includes only models estimated by the ML \citep{matsui2009}.

The GCV proposed by \cite{craven1979} is as follows:
\begin{equation*}
\text{GCV} = \frac{\text{trace} \left\lbrace \left( \mathbf{C} - \mathbf{Z} \mathbf{B} \right)^\top \left( \mathbf{C} - \mathbf{Z} \mathbf{B} \right) \right\rbrace }{N K \left( 1 - \text{trace} \left(S_{\lambda} \right) / \left( N K \right) \right)^2}
\end{equation*}
Compared with other information criteria, the GCV is computationally expensive.

\subsection{Bootstrapping}

In functional linear models, estimating the variability associated with the predicted response functions and constructing their confidence intervals are of great interest. However, calculating the asymptotic variance is even more difficult than in standard regression settings. The nonparametric bootstrap method is a frequently used technique to overcome this difficulty. For example, in a nonparametric functional regression, \cite{fv2011} used a bootstrap technique based on the residuals to construct the confidence interval of the regression function. To construct the bootstrap confidence interval for the response function, we assume that the regression model has the probability structure given by~\eqref{rmax}. Denote the $N$ sets of coefficient matrices as $\mathbf{G} = \left( \mathbf{C}^\top, \mathbf{Z}^\top \right)$ for the response and predictor functions. Herein, we use a case-resampling method in which there are two sources of errors that must be taken into account when constructing the confidence interval: smoothing errors $\bm{\epsilon}_i^s(t) = \mathbf{\Y}_i(t) - \sum_{k=1}^{K_{\Y}} b_{ik} \kappa_k(t)$, where $\sum_{k=1}^{K_{\Y}} b_{ik} \kappa_k(t)$ denotes the approximated response function by a pre-determined basis and the number of basis functions $K_{\Y}$, and predicted model errors $\bm{\epsilon}_i^p(t) = \mathbf{\Y}_i(t) - \widehat{\mathbf{\Y}}_i(t)$. Let $\left\lbrace \bm{\epsilon}^s(t) \right\rbrace  = \left\lbrace  (\bm{\epsilon}_1^s(t))^\top, \cdots, (\bm{\epsilon}_N^s(t))^\top \right\rbrace$ and $\left\lbrace \bm{\epsilon}^p(t) \right\rbrace  = \left\lbrace  (\bm{\epsilon}_1^p(t))^\top, \cdots, (\bm{\epsilon}_N^p(t))^\top \right\rbrace$ denote the error matrices. Then, the following algorithm is used to calculate the confidence interval for the response function.
\begin{itemize}
\item[Step 1.] Obtain a bootstrap resample $\mathbf{G}^* = \left( \mathbf{C}^{* \top}, \mathbf{Z}^{* \top} \right) $ by sampling with replacement from $\mathbf{G}$.
\item[Step 2.] Calculate the bootstrap LS and MPL estimates, $\widehat{\mathbf{B}}^*$, using ~	\eqref{LSe} and~\eqref{PMLe}, respectively. 
\item[Step 3.] Draw bootstrap samples $\bm{\epsilon}^{s *}(t)$ and $\bm{\epsilon}^{p *}(t)$ from $\bm{\epsilon}^{s}(t)$ and $\bm{\epsilon}^{p}(t)$, respectively.
\item[Step 4.] Obtain the fitted response functions as $\widehat{\mathbf{\Y}}^*(t) = \mathbf{Z}^* \widehat{\mathbf{B}}^* \mathbf{\Phi}(t) +  \bm{\epsilon}^{s *}(t) + \bm{\epsilon}^{p *}(t)$.
\item[Step 5.] Repeat steps 1-4 $B$ times to obtain bootstrap replicates of the fitted response functions $\lbrace \widehat{\mathbf{\Y}}^{*,1}(t), \cdots,$ $\widehat{\mathbf{\Y}}^{*,B}(t) \rbrace$. 
\end{itemize}
Let $Q_i^{\alpha}(t)$ denote the $\alpha$\textsuperscript{th} quantile of the generated $B$ sets of bootstrap replicates of the fitted response function $\widehat{\mathbf{\Y}}_i^*(t)$. Then we obtain the $100(1-\alpha)\%$ bootstrap confidence interval for $\Y_i(t)$ as $\left[Q_i^{\alpha/2}(t), Q_i^{1-\alpha/2}(t) \right]$.

\section{Numerical results\label{sec:num}}

Through Monte Carlo simulations, we present the finite sample performance of three basis functions (Fourier, B-spline, and Gaussian), two estimation methods (LS and MPL), and four roughness parameter selection and model evaluation criteria (GBIC, GIC, MAIC, and GCV). Throughout the simulations, we consider the simple functional regression model $\left\lbrace \left(\Y_i(t), \X_{i}(s)\right); s \in S, t \in \Tau \right\rbrace$:
\begin{equation*}
\Y_i(t) = \beta_0(t) + \int_{\Tau} \X_i(s) \beta_1(s,t) ds + \epsilon_i(t), \qquad i = 1, \cdots, N,
\end{equation*}
where $N = 25, 50$ and $100$ sets of functional variables are considered. For the data-generating process, we consider three cases: Case-I, where both the functional response and functional predictor have a non-periodic structure; Case-II, where the functional response has a periodic structure and the functional predictor has a non-periodic structure; and Case-III, where both functional variables have a periodic structure. 

For all cases, the number of basis functions is fixed at $K = 10$ to compare the performances of the smoothing methods under the same conditions. The generated data are converted into a functional form using the smoothing methods noted above. The LS and MPL methods are applied to estimate the model from the data, and four model selection criteria are used to evaluate these estimated models. The number of Monte Carlo simulations is set to $\text{MC} = 500$. For each simulation replication, the average mean squared error (AMSE) is defined as:
\begin{equation*}
\text{AMSE} = \sum_{i=1}^N \left( \Y_i(t) - \widehat{\Y}_i(t) \right)^2/N. 
\end{equation*}
To construct confidence intervals for the generated response functions, $B = 500$ bootstrap simulations are performed and $\alpha$ is set to 0.05 to obtain 95\% pointwise confidence intervals. To compare the smoothing techniques for each response function, we calculate the bootstrap coverage probability ($\text{CP}_i$), length of confidence interval ($\text{width}_i$), and the interval score ($\text{score}_i$) as follows:
\begin{align*}
\text{CP} &= \frac{1}{\text{MC}} \sum_{i = 1}^{\text{MC}} \mathds{1} \left\lbrace Q_i^{\alpha/2}(t) \leq \eta_i(t) \leq Q_i^{1-\alpha/2}(t) \right\rbrace \\
\text{width}_i &= \sum_{j=1}^J \left( Q_{ij}^{1-\alpha/2}(t) - Q_{ij}^{\alpha/2}(t) \right)  \\
\text{score}_i &= \frac{1}{J} \sum_{j=1}^J \left\lbrace \left(Q_{ij}^{1-\alpha/2}(t) - Q_{ij}^{\alpha/2}(t) \right) \right. \\
&+ \frac{2}{\alpha} \left(Q_{ij}^{\alpha/2}(t) - \eta_i(t) \right) \mathds{1} \left\lbrace \eta_i(t) < Q_{ij}^{\alpha/2}(t)\right\rbrace  \\
&+ \left. \frac{2}{\alpha} \left(\eta_i(t) - Q_{ij}^{1-\alpha/2}(t) \right) \mathds{1} \left\lbrace \eta_i(t) > Q_{ij}^{1-\alpha/2}(t)\right\rbrace \right\rbrace ,
\end{align*}
where $\mathds{1} \lbrace \cdot \rbrace$ denotes the indicator function.

The figures plotted in this section have been relegated to the Appendix (Section~\ref{appendix}) to make this section more readable.

\subsection{Case-I}

The data points for the response and predictor variables are generated as:
\begin{itemize}
\item For $j = 1, \cdots, 50$, the predictor is generated using the following process:
\begin{equation*}
\X_{ij} = \nu_i(s_j) + \epsilon_{ij},
\end{equation*}
where $\epsilon_{ij} \sim \text{N}(0,1)$ and $s_j \sim \text{Uniform}(-1,1)$. The term $\nu_i(s)$ is generated as: 
\begin{equation*}
\nu_i(s) = 2 \exp(a_{1,i} s) \sin (\pi s^2)/a_{1,i} + a_{2,i} \cos (\pi s), 
\end{equation*}
where $a_{1,i} \sim \text{N}(2, 0.02^2)$ and $a_{2,i} \sim \text{N}(-3, 0.03^2)$.
\item For the response variable, the design points are generated as:
\begin{equation*}
\Y_{ij} = \upsilon_i(t_j) + \epsilon_{ij}, 
\end{equation*}
where $\epsilon_{ij} \sim \text{N}(0,1)$ and $t_j \sim \text{Uniform}(-1,1)$. $\upsilon_i(t)$ is generated as $\upsilon_i(t) = \eta_i(t) + \varepsilon_i$ where $\eta_i(t) = 2 a_{1,i}^2 \sin(\pi t^2) + 2 a_{2,i} \cos(\pi t^2)$ with the same $a_1$ and $a_2$ as in the predictor variable,  and $\bm{\varepsilon}$ follows a multivariate normal distribution with mean $\mathbf{0}$ and variance-covariance matrix $\mathbf{\Sigma} = [(0.5^{\vert k - l \vert}) \rho ]_{k,l}$. Four different $\rho$ values are considered: $\rho = [0.5, 1, 2, 4]$. Herein, the parameter $\rho$ can be considered to noise-to-signal ratio; the smoothing methods are expected to perform less effectively at a higher noise-to-signal ratio.
\end{itemize}

For this case, graphical displays of the generated data and related smooth functions obtained via the three basis functions, as well as an example of the constructed bootstrap confidence intervals are presented in Appendix~\ref{app:sim_Case_I}. 

The obtained results are reported in Table~\ref{tab:resultsim}. Note that the values given in brackets are the estimated standard errors for the calculated performance metrics.
\begin{center}
\tabcolsep 0.065in
\begin{small}
\begin{longtable}{@{}llllrrrrr@{}} 
\caption{Simulation results: Average performance metrics (Case-I).}\\
\toprule
$\rho$ & {Basis} & {Method} & {Metric} & {GCV} & {GIC} & {MAIC} & {GBIC} \\
\midrule
0.5 & Gaussian & LS & AMSE & 12.3265 (2.6624) & 11.4823 (1.6106) & 11.9778 (2.2182) & 18.6575 (8.2373) \\
& & & $\text{CP}$ & 0.9179 (0.0178) & 0.9120 (0.0189) & 0.9140 (0.0178) & 0.9278 (0.0159)  \\
& & & $\text{width}$ & 17.4999 (0.0422) & 17.4918 (0.0403) & 17.4946 (0.0429) & 17.6609 (0.0481)  \\
& & & $\text{score}$ & 5.9121 (0.2317) & 5.9034 (0.2213) & 5.9078 (0.2238) & 5.9802 (0.2606)  \\
\\
& & MPL & AMSE & 12.1907 (2.7728) & 11.2973 (1.7616) & 11.8369 (2.3623) & 18.6797 (8.3906) \\
& & & $\text{CP}$ & 0.9116 (0.0170) & 0.9089 (0.0215) & 0.9110 (0.0178) & 0.9248 (0.0160) \\
& & & $\text{width}$ & 17.4775 (0.0421) & 17.4671 (0.0400) & 17.4691 (0.0433) & 17.6394 (0.0484)  \\ 
& & & $\text{score}$ & 5.8762 (0.2285) & 5.8738 (0.2195) & 5.8730 (0.2207) & 5.9539 (0.2561)  \\ \cmidrule{3-8}

& B-spline & LS & AMSE & 11.7764 (3.0730) & 10.7497 (1.5801) & 11.3308 (2.5519) & 11.1945 (2.4353) \\
& & & $\text{CP}$ & 0.9147 (0.0162) & 0.9130 (0.0180) & 0.9124 (0.0177) & 0.9110 (0.0195)  \\
& & & $\text{width}$ & 17.4630 (0.0424) & 17.4438 (0.0405) & 17.4558 (0.0423) & 17.4520 (0.0407)  \\
& & & $\text{score}$ & 5.8813 (0.2301) & 5.8638 (0.2252) & 5.8610 (0.2216) & 5.8621 (0.2228)  \\
\\
& & MPL & AMSE & 11.7758 (3.0730) & 10.7494 (1.5800) & 11.3303 (2.5519) & 11.1945 (2.4353) \\
& & & $\text{CP}$ & 0.9142 (0.0176) & 0.9115 (0.0193) & 0.9135 (0.0183) & 0.9107 (0.0191) \\
& & & $\text{width}$ & 17.4628 (0.0409) & 17.4449 (0.0413) & 17.4556 (0.0415) & 17.4509 (0.0408)  \\
& & & $\text{score}$ & 5.8791 (0.2284) & 5.8651 (0.2232) & 5.8657 (0.2227) & 5.8600 (0.2225)  \\ \cmidrule{3-8}
 
& Fourier & LS& AMSE & 17.2901 (1.2974) & 17.2901 (1.2967) & 17.2901 (1.2974) & 17.4412 (1.4333) \\
& & & $\text{CP}$ & 0.9161 (0.0183) & 0.9174 (0.0189) & 0.9169 (0.0198) & 0.9199 (0.0191)  \\
& & & $\text{width}$ & 17.6876 (0.0442) & 17.6889 (0.0433) & 17.6882 (0.0445) & 17.6867 (0.0443)  \\
& & & $\text{score}$ & 6.0575 (0.2078) & 6.0570 (0.2060) & 6.0537 (0.2076) & 6.0677 (0.2078)  \\
\\
& & MPL & AMSE & 17.2901 (1.2974) & 17.2901 (1.2967) & 17.2901 (1.2974) & 17.4407 (1.4333) \\
& & & $\text{CP}$ & 0.9192 (0.0194) & 0.9195 (0.0170) & 0.9161 (0.0188) & 0.9176 (0.0177) \\
& & & $\text{width}$ & 17.6853 (0.0436) & 17.6895 (0.0445) & 17.6891 (0.0430) & 17.6873 (0.0439)  \\
& & & $\text{score}$ & 6.0549 (0.2066) & 6.0585 (0.2060) & 6.0543 (0.2056) & 6.0677 (0.2078)  \\
\midrule
1 & Gaussian & LS & AMSE & 15.0139 (9.5717) & 13.4454 (2.2136) & 13.9274 (3.2014) & 20.0004 (10.1121) \\
& & & $\text{CP}$ & 0.9373 (0.0157) & 0.9331 (0.0166) & 0.9359 (0.0150) & 0.9429 (0.0149)  \\
& & & $\text{width}$ & 18.3528 (0.0435) & 18.3269 (0.0412) & 18.3329 (0.0430) & 18.4607 (0.0465)  \\
& & & $\text{score}$ & 6.6440 (0.2830) & 6.6277 (0.2496) & 6.6277 (0.2494) & 6.7719 (0.3179)  \\
\\
& & MPL & AMSE & 15.4397 (9.8959) & 13.8643 (2.3770) & 14.3338 (3.4484) & 20.6246 (10.3833) \\
& & & $\text{CP}$ & 0.9356 (0.0166) & 0.9319 (0.0176) & 0.9324 (0.0157) & 0.9416 (0.0139) \\
& & & $\text{width}$ & 18.3174 (0.0467) & 18.2907 (0.0433) & 18.2990 (0.0418) & 18.4271 (0.0469)  \\
& & & $\text{score}$ & 6.5990 (0.2850) & 6.5854 (0.2476) & 6.5848 (0.2504) & 6.7305 (0.3184)  \\ \cmidrule{3-8}

& B-spline & LS & AMSE & 13.2939 (3.7424) & 12.5298 (2.8052) & 12.7804 (3.1652) & 12.4539 (2.6479) \\
& & & $\text{CP}$ & 0.9330 (0.0154) & 0.9306 (0.0146) & 0.9338 (0.0147) & 0.9336 (0.0187)  \\
& & & $\text{width}$ & 18.2887 (0.0377) & 18.2758 (0.0389) & 18.2785 (0.0400) & 18.2711 (0.0396)  \\
& & & $\text{score}$ & 6.6062 (0.2560) & 6.5823 (0.2472) & 6.5908 (0.2525) & 6.5806 (0.2496)  \\
\\
& & MPL & AMSE & 13.2930 (3.7425	) & 12.5290 (2.8053) & 12.7796 (3.1653) & 12.4539 (2.6479) \\
& & & $\text{CP}$ & 0.9330 (0.0160) & 0.9338 (0.0142) & 0.9313 (0.0163) & 0.9332 (0.0160) \\
& & & $\text{width}$ & 18.2878 (0.0408) & 18.2752 (0.0401) & 18.2804 (0.0384) & 18.2687 (0.0429)  \\
& & & $\text{score}$ & 6.6076 (0.2561) & 6.5810 (0.2515) & 6.5878 (0.2531) & 6.5787 (0.2485)  \\ \cmidrule{3-8}
 & Fourier & LS & AMSE & 22.5813 (1.7323) & 22.5603 (1.7293) & 22.5803 (1.7325) & 23.2732 (3.0764) \\
& & & $\text{CP}$ & 0.9387 (0.0142) & 0.9382 (0.0136) & 0.9373 (0.0146) & 0.9391 (0.0140)  \\
& & & $\text{width}$ & 18.6029 (0.0422) & 18.6064 (0.0415) & 18.6028 (0.0410) & 18.6109 (0.0430)  \\
& & & $\text{score}$ & 6.8164 (0.2380) & 6.8172 (0.2337) & 6.8169 (0.2376) & 6.8345 (0.2391)  \\
\\
& & MPL & AMSE & 22.5813 (1.7323) & 22.5603 (1.7293) & 22.5803 (1.7325) & 23.2724 (3.0764) \\
& & & $\text{CP}$ & 0.9371 (0.0139) & 0.9383 (0.0139) & 0.9402 (0.0159) & 0.9387 (0.0151) \\
& & & $\text{width}$ & 18.6032 (0.0431) & 18.6062 (0.0419) & 18.6016 (0.0410) & 18.6122 (0.0433)  \\
& & & $\text{score}$ & 6.8148 (0.2337) & 6.8167 (0.2310) & 6.8159 (0.2365) & 6.8335 (0.2377)  \\
\midrule
2 & Gaussian & LS & AMSE & 16.8402 (4.8670) & 16.0597 (4.1472) & 16.4771 (4.6607) & 24.8920 (10.3739) \\
& & & $\text{CP}$ & 0.9605 (0.0125) & 0.9593 (0.0137) & 0.9601 (0.0129) & 0.9676 (0.0108)  \\
& & & $\text{width}$ & 19.5171 (0.0452) & 19.5113 (0.0475) & 19.6679 (0.0466) & 19.5130 (0.0482)  \\ 
& & & $\text{score}$ & 7.5505 (0.2306) & 7.5378 (0.2074) & 7.5370 (0.2154) & 7.7720 (0.3487)  \\
\\
& & MPL & AMSE & 16.2248 (5.0129) & 15.3785 (4.2993) & 15.8208 (4.8083) & 24.5539 (10.6644) \\
& & & $\text{CP}$ & 0.9591 (0.0109) & 0.9593 (0.0145) & 0.9659 (0.0119) & 0.9600 (0.0147) \\
& & & $\text{width}$ & 19.4854 (0.0444) & 19.4771 (0.0463) & 19.4806 (0.0462) & 19.6341 (0.0495)  \\
& & & $\text{score}$ & 7.5057 (0.2321) & 7.4926 (0.2071) & 7.4918 (0.2165) & 7.7280 (0.3487)  \\ \cmidrule{3-8}

& B-spline & LS & AMSE & 15.5255 (5.4462) & 14.1798 (3.4623) & 14.5592 (4.0295) & 14.6967 (4.5171) \\
& & & $\text{CP}$ & 0.9590 (0.0113) & 0.9590 (0.0137) & 0.9585 (0.0141) & 0.9559 (0.0129)  \\
& & & $\text{width}$ & 19.4576 (0.0478) & 19.4379 (0.0494) & 19.4435 (0.0456) & 19.4397 (0.0476)  \\
& & & $\text{score}$ & 7.4914 (0.2154) & 7.4910 (0.2069) & 7.4905 (0.2163) & 7.4886 (0.2184)  \\
\\
& & MPL & AMSE & 15.5231 (5.4465	) & 14.1781 (3.4624) & 14.5573 (4.0295) & 14.6966 (4.5170) \\
& & & $\text{CP}$ & 0.9598 (0.0133) & 0.9592 (0.0133) & 0.9596 (0.0121) & 0.9584 (0.0144) \\
& & & $\text{width}$ & 19.4564 (0.0470) & 19.4378 (0.0481) & 19.4417 (0.0463) & 19.4419 (0.0475)  \\
& & & $\text{score}$ & 7.4888 (0.2178) & 7.4901 (0.2049) & 7.4888 (0.2148) & 7.4886 (0.2209)  \\ \cmidrule{3-8}
 
& Fourier & LS & AMSE & 22.6742 (1.9884) & 22.6093 (1.7653) & 22.5457 (1.5752) & 24.2863 (5.1498) \\
& & & $\text{CP}$ & 0.9589 (0.0132) & 0.9592 (0.0140) & 0.9596 (0.0132) & 0.9590 (0.0142)  \\
& & & $\text{width}$ & 19.7074 (0.0445) & 19.7103 (0.0435) & 19.7095 (0.0444) & 19.7295 (0.0461)  \\ 
& & & $\text{score}$ & 7.7047 (0.2003) & 7.7038 (0.1995) & 7.7028 (0.2025) & 7.7172 (0.2113)  \\
\\
& & MPL & AMSE & 22.6737 (1.9884) & 22.6092 (1.7653) & 22.5456 (1.5752) & 24.2845 (5.1499) \\
& & & $\text{CP}$ & 0.9579 (0.0136) & 0.9586 (0.0146) & 0.9605 (0.0145) & 0.9605 (0.0147) \\
& & & $\text{width}$ & 19.7104 (0.0438) & 19.7119 (0.0435) & 19.7096 (0.0455) & 19.7277 (0.0440)  \\
& & & $\text{score}$ & 7.7031 (0.1989) & 7.7072 (0.1991) & 7.7006 (0.2008) & 7.7149 (0.2084)  \\
 \midrule
4 & Gaussian & LS & AMSE & 26.9477 (14.6018) & 23.3340 (7.2316) & 24.8102 (8.7881) & 42.3500 (12.3765) \\
& & & $\text{CP}$ & 0.9790 (0.0146) & 0.9792 (0.0143) & 0.9823 (0.0136) & 0.9857 (0.0115)  \\
& & & $\text{width}$ & 21.7015 (0.0995) & 21.6777 (0.0996) & 21.6754 (0.1019) & 21.8960 (0.1033)  \\
& & & $\text{score}$ & 9.2127 (0.2526) & 9.1998 (0.2468) & 9.1962 (0.2405) & 9.4311 (0.2753)  \\
\\
& & MPL & AMSE & 25.7649 (16.2824) & 21.5729 (7.8746) & 23.3009 (9.5613) & 42.5390 (13.5095) \\
& & & $\text{CP}$ & 0.9805 (0.0163) & 0.9767 (0.0145) & 0.9780 (0.0161) & 0.9865 (0.0106) \\
& & & $\text{width}$ & 21.6612 (0.1088) & 21.6329 (0.1037) & 21.6335 (0.1071) & 21.8466 (0.1048)  \\
& & & $\text{score}$ & 9.1597 (0.2510) & 9.1414 (0.2444) & 9.1412 (0.2395) & 9.3671 (0.2714)  \\ \cmidrule{3-8}

 & B-spline & LS & AMSE & 20.8527 (7.2445) & 20.1549 (5.4145) & 20.8963 (7.2521) & 24.4487 (11.6453) \\
& & & $\text{CP}$ & 0.9755 (0.0170) & 0.9765 (0.0159) & 0.9784 (0.0175) & 0.9796 (0.0163)  \\
& & & $\text{width}$ & 21.5770 (0.1026) & 21.5840 (0.1010) & 21.5806 (0.1025) & 21.6134 (0.1015)  \\
& & & $\text{score}$ & 9.1347 (0.2493) & 9.1089 (0.2213) & 9.1116 (0.2271) & 9.1513 (0.2474)  \\
\\
& & MPL & AMSE & 20.8461 (7.2449) & 20.1516 (5.4149) & 20.8915 (7.2526) & 24.4486 (11.6451) \\
& & & $\text{CP}$ & 0.9771 (0.0156) & 0.9769 (0.0145) & 0.9771 (0.0145) & 0.9788 (0.0159) \\
& & & $\text{width}$ & 21.5777 (0.1027) & 21.5843 (0.1037) & 21.5813 (0.1016) & 21.6148 (0.0979)  \\
& & & $\text{score}$ & 9.1327 (0.2465) & 9.1087 (0.2218) & 9.1081 (0.2247) & 9.1487 (0.2540)  \\ \cmidrule{3-8}
 
& Fourier & LS & AMSE & 30.1646 (3.4746) & 29.8787 (2.2884) & 29.8964 (2.3014) & 33.9865 (9.9590) \\
& & & $\text{CP}$ & 0.9751 (0.0170) & 0.9763 (0.0149) & 0.9761 (0.0132) & 0.9751 (0.0180)  \\
& & & $\text{width}$ & 21.8285 (0.0986) & 21.8360 (0.0997) & 21.8332 (0.1045) & 21.8668 (0.1040)  \\
& & & $\text{score}$ & 9.3395 (0.2197) & 9.3445 (0.2180) & 9.3392 (0.2183) & 9.4050 (0.2525)  \\
\\
& & MPL & AMSE & 30.1641 (3.4743) & 29.8784 (2.2884) & 29.8963 (2.3013) & 33.9815 (9.9589) \\
& & & $\text{CP}$ & 0.9738 (0.0142) & 0.9725 (0.0169) & 0.9738 (0.0161) & 0.9759 (0.0172) \\
& & & $\text{width}$ & 21.8305 (0.0998) & 21.8345 (0.0993) & 21.8356 (0.1030) & 21.8687 (0.1073)  \\
& & & $\text{score}$ & 9.3409 (0.2167) & 9.3395 (0.2179) & 9.3401 (0.2164) & 9.4047 (0.2484)  \\
\bottomrule
\label{tab:resultsim}
\end{longtable}
\end{small}
\end{center}

Our records show that:
\begin{itemize}
\item The LS and MPL methods tend to produce similar AMSE values for all basis functions and model evaluation criteria.
\item The largest and smallest AMSE values, respectively, are produced when the Fourier and B-spline basis functions are used to smooth the generated data.
\item The largest and smallest AMSE values, respectively, are generally produced when GBIC and GIC are used to control the roughness parameter and evaluate the estimated model. 
\item Compared with GCV, GIC, and MAIC, the AMSE values of the LS and MPL are more affected by the variance parameter $\rho$ when GBIC is used to control the roughness parameter and evaluate the model. For example, when $\rho$ is increased from 0.5 to 4 and the Gaussian basis is used to smooth the data, the AMSE values of the estimation methods increase by about 14, 11, 12, and 23 units when GCV, GIC, MAIC, and GBIC are used as information criterion, respectively. From the same perspective, the results show that, between Gaussian and Fourier bases, the AMSE values of the estimation methods are less affected by $\rho$ when the B-spline basis is used to smooth the data. For instance, when $\rho$ increases from 0.5 to 4, the AMSE values of the LS and MPL increase by about 14, 9, and 12 units, respectively, when Gaussian, B-spline and Fourier bases are used to smooth the data and GCV is used as the information criterion.
\item The proposed bootstrap method produces similar coverage probabilities, lengths of confidence intervals and interval scores for all estimation methods and information criteria considered. It produces coverage probabilities close to the customarily nominal confidence level $(1 - \alpha) = 95\%$ when the variance parameter $\rho = 1$ and $\rho = 2$, while coverage probabilities are observed away from this nominal level for other $\rho$ values.
\end{itemize}

\subsection{Case-II and Case-III}

For Case-II, the observations of the functional response and predictor variables are generated using the following processes:
\begin{align*}
\Y_{ij} &= 15 + \cos(\pi j /12) + 2 (a_{1,i} + a_{2,i}) + \epsilon_{ij} \\
\X_{ij} &= \left( 15 + \sin(\pi j/12) + 2(a_{1,i}+a_{2,i}) \right) /(2 \exp(a_{1,i}))+a_{2,i} + \epsilon_{ij},
\end{align*}
where $\epsilon_{ij} \sim N(0, 0.5^2)$, $a_{1,i} \sim N(0, 0.1^1)$, and $a_{2,i} \sim N(0, 0.02^2)$. For Case-III, the observations of the functional response variable are generated as for Case-II. In contrast, for the functional predictor, the following process is employed to generate the observations:
\begin{equation*}
\X_{ij} = 15 + \sin(\pi j/12) + 2(a_{1,i}+a_{2,i}) +a_{2,i} + \epsilon_{ij}.
\end{equation*}

For both Case-II and Case-III, an example of the observed data is presented in Appendix \ref{app:sim_Case_II_III}. We also show the noise and fitted smooth functions for the generated response and predictor variables. The calculated average performance metrics for these two cases are presented in Table~\ref{tab:resultsim2}. The results demonstrate that:
\begin{itemize}
\item Both estimation methods (LS and MPL) produce similar AMSE values for all basis functions and model evaluation criteria, as in Case-I.
\item The smallest AMSE values are, in general, produced when GCV and MAIC are used to control the roughness parameter and evaluate the estimated model. On the other hand, the largest AMSE values are produced when GBIC is used as the information criterion.
\item The largest and smallest AMSE values, respectively, are produced when the Fourier and Gaussian basis functions are used to smooth the generated data.
\item The proposed bootstrap method produces similar coverage probabilities, lengths of confidence intervals and interval scores for all estimation and smoothing methods (except GBIC). When GBIC is used as the information criterion, estimation accuracy deteriorates and the bootstrap method produces coverage probabilities far from the nominal confidence level $(1 - \alpha) = 95\%$ for both cases.
\end{itemize}

\begin{center}
\tabcolsep 0.065in
\begin{small}
\begin{longtable}{@{}llllrrrrr@{}} 
\caption{Simulation results: Average performance metrics (Case-II and Case-III).}\\
\toprule
{Case} & {Basis} & {Method} & {Metric} & {GCV} & {GIC} & {MAIC} & {GBIC} \\
\midrule
1   & Gaussian & LS & AMSE & 0.6258 (0.0861) & 0.7043 (0.0827) & 0.6286 (0.0853) & 23.0925 (8.2623) \\
& & & $\text{CP}$ & 0.9975 (0.0230) &  0.9961 (0.0307) & 0.9961 (0.0292) & 0.9993 (0.0051)  \\
& & & $\text{width}$ & 9.9880 (0.4369) & 10.0556 (0.4321) & 9.9896 (0.4367) & 12.2777 (0.9861)  \\
& & & $\text{score}$ & 1.9995 (0.0425) & 2.0240 (0.0423) & 1.9991 (0.0420) & 3.0324 (0.3896)  \\
\\
& & MPL & AMSE & 0.6252 (0.2897) & 0.7043 (0.0827) & 0.6262 (0.2826) & 23.0775 (8.3374) \\
& & & $\text{CP}$ & 0.9961 (0.0261) & 0.9954 (0.0327) & 0.9967 (0.0240) & 0.9997 (0.0025) \\
& & & $\text{width}$ & 9.9838 (0.4447) & 10.0541 (0.4315) & 9.9855 (0.4451) & 12.1602 (1.1058)  \\ 
& & & $\text{score}$ & 1.9940 (0.0485) & 2.0240 (0.0418) & 1.9965 (0.0481) & 2.9250 (0.4005)  \\ \cmidrule{3-8}

& B-spline & LS & AMSE & 0.6585 (0.1021) & 0.7047 (0.0822) & 0.6642 (0.0892) & 23.3336 (6.9270) \\
& & & $\text{CP}$ & 0.9957 (0.0332) & 0.9954 (0.0343) & 0.9948 (0.0379) & 0.9997 (0.0025)  \\
& & & $\text{width}$ & 10.0245 (0.4315) & 10.0547 (0.4345) & 10.0245 (0.4318) & 12.3287 (0.8601)  \\
& & & $\text{score}$ & 2.009 (0.0410) & 2.0241 (0.0427) & 2.0110 (0.0403) & 3.0484 (0.3235)  \\
\\
& & MPL & AMSE & 0.6581 (0.3454) & 0.7047 (0.0822) & 0.6666 (0.2840) & 23.2187 (6.9810) \\
& & & $\text{CP}$ & 0.9957 (0.0317) & 0.9954 (0.0343) & 0.9944 (0.0361) & 0.9997 (0.0025) \\
& & & $\text{width}$ & 9.9934 (0.4398) & 10.0511 (0.4304) & 10.0000 (0.4397) & 12.0905 (0.9283)  \\
& & & $\text{score}$ & 2.0010 (0.0495) & 2.0241 (0.0411) & 2.0050 (0.0485) & 2.9425 (0.3363)  \\ \cmidrule{3-8}
 
& Fourier & LS& AMSE & 0.8118 (0.0898) & 0.8117 (0.0899) & 0.8118 (0.0898) & 0.8120 (0.0900) \\
& & & $\text{CP}$ & 0.9954 (0.0294) & 0.9944 (0.0371) & 0.9944 (0.0389) & 0.9961 (0.0261)  \\
& & & $\text{width}$ & 10.0547 (0.4235) & 10.0547 (0.4270) & 10.0547 (0.4258) & 10.0547 (0.4231)  \\
& & & $\text{score}$ & 2.0234 (0.0416) & 2.0251 (0.0415) & 2.0234 (0.0422) & 2.0244 (0.0413)  \\
\\
& & MPL & AMSE & 0.8118 (0.0898) & 0.8117 (0.0899) & 0.8118 (0.0898) & 0.8120 (0.0900) \\
& & & $\text{CP}$ & 0.9954 (0.0343) & 0.9961 (0.0292) & 0.9944 (0.0389) & 0.9951 (0.0338) \\
& & & $\text{width}$ & 10.0547 (0.4255) & 10.0547 (0.4272) & 10.0547 (0.4254) & 10.0547 (0.4237)  \\
& & & $\text{score}$ & 2.0242 (0.0422) & 2.0242 (0.0419) & 2.0242 (0.0418) & 2.0253 (0.0414)  \\
\midrule
2 & Gaussian & LS & AMSE & 0.6753 (0.0773) & 0.7410 (0.0853) & 0.6753 (0.0773) & 19.9258 (11.0442) \\
& & & $\text{CP}$ & 0.9954 (0.0400) &  0.9946 (0.0454) & 0.9956 (0.0368) & 0.9996 (0.0040)  \\
& & & $\text{width}$ & 9.9991 (0.4463) & 10.0550 (0.4413) & 9.9985 (0.4480) & 11.9442 (1.2427)  \\
& & & $\text{score}$ & 2.0030 (0.0494) & 2.0253 (0.0469) & 2.0028 (0.0497) & 2.8839 (0.5187)  \\
\\
& & MPL & AMSE & 0.6753 (0.0773) & 0.7410 (0.0853) & 0.6753 (0.0773) & 19.6615 (11.0730) \\
& & & $\text{CP}$ & 0.9968 (0.0284) & 0.9938 (0.0523) & 0.9948 (0.0433) & 1.0000 (0.0000) \\
& & & $\text{width}$ & 9.9986 (0.4456) & 10.0562 (0.4419) & 9.9990 (0.4485) & 11.8229 (1.3469)  \\ 
& & & $\text{score}$ & 2.0027 (0.0486) & 2.0259 (0.0478) & 2.0032 (0.0485) & 2.7796 (0.5384)  \\ \cmidrule{3-8}

& B-spline & LS & AMSE & 0.7084 (0.0822) & 0.7424 (0.0854) & 0.7084 (0.0822) & 21.6790 (9.0010) \\
& & & $\text{CP}$ & 0.9946 (0.0426) & 0.9944 (0.0463) & 0.9948 (0.0460) & 1.0000 (0.0000)  \\
& & & $\text{width}$ & 10.0231 (0.4421) & 10.0559 (0.4433) & 10.0261 (0.4413) & 12.1490 (1.0387)  \\
& & & $\text{score}$ & 2.0124 (0.0481) & 2.0258 (0.0477) & 2.0137 (0.0481) & 2.9734 (0.4175)  \\
\\
& & MPL & AMSE & 0.7084 (0.0822) & 0.7424 (0.0854) & 0.7084 (0.0822) & 21.4825 (9.0272) \\
& & & $\text{CP}$ & 0.9948 (0.0393) & 0.9950 (0.0447) & 0.9944 (0.0435) & 0.9998 (0.0020) \\
& & & $\text{width}$ & 10.0246 (0.4420) & 10.0576 (0.4414) & 10.0245 (0.4427) & 11.9274 (1.1166)  \\
& & & $\text{score}$ & 2.0131 (0.0479) & 2.0264 (0.0469) & 2.0131 (0.0476) & 2.8700 (0.4350)  \\ \cmidrule{3-8}
 
& Fourier & LS& AMSE & 0.8425 (0.0905) & 0.8424 (0.0905) & 0.8425 (0.0905) & 0.8425 (0.0905) \\
& & & $\text{CP}$ & 0.9946 (0.0453) & 0.9964 (0.0364) & 0.9948 (0.0361) & 0.9966 (0.0292)  \\
& & & $\text{width}$ & 10.0557 (0.4351) & 10.0552 (0.4345) & 10.0567 (0.4342) & 10.0569 (0.4346)  \\
& & & $\text{score}$ & 2.0257 (0.0478) & 2.0254 (0.0484) & 2.0260 (0.0477) & 2.0262 (0.0480)  \\
\\
& & MPL & AMSE & 0.8425 (0.0905) & 0.8424 (0.0905) & 0.8425 (0.0905) & 0.8425 (0.0905) \\
& & & $\text{CP}$ & 0.9960 (0.0301) & 0.9954 (0.0410) & 0.9958 (0.0358) & 0.9948 (0.0448) \\
& & & $\text{width}$ & 10.0559 (0.4341) & 10.0561 (0.4348) & 10.0572 (0.4354) & 10.0566 (0.4340)  \\
& & & $\text{score}$ & 2.0258 (0.0474) & 2.0261 (0.0479) & 2.0260 (0.0476) & 2.0253 (0.0479)  \\
\bottomrule
\label{tab:resultsim2}
\end{longtable}
\end{small}
\end{center}

Overall, the results presented in Tables~\ref{tab:resultsim} and \ref{tab:resultsim2} demonstrate that the combination of basis function and information criterion of B-spline-GIC in general performs better in terms of AMSE values when both the response and predictor functions have a non-periodic structure. On the other hand, when the response and/or predictor functions have a periodic structure, the Gaussian basis with GCV (or MAIC) tend to produce lower AMSE values than other basis function-information criterion combinations. Checking the details of the simulation results (not presented here) shows that the GBIC tends to produce positive-larger $\lambda$ values than other information criteria. Thus, both estimation methods have their worst performances when GBIC is used to control the roughness parameter and evaluate the estimated model. 

Moreover, we compare the computing time of the estimation methods considered in this study. The calculations were carried out using R 3.6.0. on an Intel Core i7 6700HQ 2.6 GHz PC. Our records show that the MPL requires more computing time than the LS method. For example, we observe that the LS and MPL methods require the following computing times (in seconds) to estimate the coefficient matrix $\mathbf{B}$: $[\text{LS}, \text{MPL}] = [1.73, 2.94]$ when $K = 10$ and $[\text{LS}, \text{MPL}] = [2.42, 60.40]$ when $K = 40$. The computational time of MPL increases exponentially with increasing $K$. Note that the interpretation of the results for $N = 25$ and $50$ are qualitatively similar to those presented in this paper, and can be obtained on request from the corresponding author.

\section{Data analyses\label{sec:real}}

We compare the performance of the smoothing techniques using two empirical data examples. Similar to Section~\ref{sec:num}, some figures and tables constructed for the empirical data examples are provided in the Appendix to make the data analysis section more readable.

The first dataset is monthly Japanese meteorological data spanning January 1961 to December 2018. This dataset has six variables-wind power, mean temperature, humidity, vapor pressure, $\log$-sunshine duration, and global solar radiation-and was collected from 27 meteorological stations across Japan (see Appendix~\ref{app:jmmd_fig}, the data were obtained from the Japanese Meteorological Agency).

The second dataset is North Dakota weekly weather data, for January 2000-December 2018. It includes three variables-wind power, mean temperature, and global solar radiation-and was collected from 45 meteorological stations (see Appendix~\ref{app:ndwwd_fig}; the data were obtained from the North Dakota Agricultural Weather Network Center). 

\subsection{Japanese monthly meteorological data}

The data are averaged for each meteorological variable over the whole period. The time series plots of the averaged monthly meteorological variables for all 27 stations are presented in Appendix~\ref{app:jmmd_fig}.

We consider predicting monthly global solar radiation using the remaining five meteorological variables. For this purpose, the discretely observed data are first converted to functional forms using the basis functions and information criteria considered in this study. The estimated number of basis functions $\widehat{K}$ and roughness parameter $\log_{10}\widehat{\lambda}$ for each basis function and information criterion are presented in Table~\ref{tab:paramN}.

\begin{table}[!htbp]
\centering
\tabcolsep 0.056in
\caption{Estimated number of basis functions and penalty parameters for the Japanese meteorological data.}
\begin{tabular}{@{}l l c c c c c c c@{}}
\toprule
& &  & \multicolumn{6}{c}{Variables} \\ \cmidrule{4-9}
& &  & Wind & Temperature & Humidity & Vapor & Sunshine  & Solar \\
Basis & & Parameter & & & & pressure & duration & radiation \\  
\midrule
\multirow{8}{*}{Gaussian} & \multirow{2}{*}{GCV} & $\widehat{K}$ & 10 & 10 & 9 & 10 & 7 & 6 \\ 
& & $\widehat{\lambda}$ & -0.7070 & -2.1212 & -2.7272 & -2.1212 & -0.7070 &	-2.1212 \\
\\
& \multirow{2}{*}{GIC} & $\widehat{K}$ & 10 & 10 & 10 & 10 & 10 & 10 \\
& & $\widehat{\lambda}$ & -1.1111 & -2.9292 & -3.7373 & -3.3333 & -1.7171 &	-3.1313 \\ \\
& \multirow{2}{*}{MAIC} & $\widehat{K}$ & 10 & 10 & 10 & 10 & 10 & 10 \\
& & $\widehat{\lambda}$ & -1.5151 & -2.7272 & -3.3333 & -2.9292 & -1.3131 &	-2.7272 \\ \\
& \multirow{2}{*}{GBIC} & $\widehat{K}$ & 7 & 10 & 10 & 10 & 10 & 10 \\
& & $\widehat{\lambda}$ & 10.000 & 10.000 & 10.000 & 10.000 & 10.000 &	10.000 \\
\midrule
\multirow{8}{*}{B-spline} & \multirow{2}{*}{GCV} & $\widehat{K}$ & 10 & 10 & 7 & 10 & 5 & 6 \\
& & $\widehat{\lambda}$ & -0.5050 & -2.3232 & -3.7373 & -2.3232 & -0.7070 &	-2.7272 \\ \\
& \multirow{2}{*}{GIC} & $\widehat{K}$ & 10 & 10 & 10 & 10 & 10 & 10 \\
& & $\widehat{\lambda}$ & -1.5151 & -2.9292 & -3.3333 & -2.9292 & -1.1111 &	-3.1313 \\ \\
& \multirow{2}{*}{MAIC} & $\widehat{K}$ & 10 & 10 & 10 & 10 & 10 & 10 \\
& & $\widehat{\lambda}$ & -1.5151 & -3.1313 & -3.3333 & -2.9292 & -0.9090 &	-3.1313 \\ \\
& \multirow{2}{*}{GBIC} & $\widehat{K}$ & 7 & 10 & 10 & 10 & 10 & 10 \\
& & $\widehat{\lambda}$ & 10.000 & 10.000 & 10.000 & 10.000 & 10.000 &	10.000 \\
\midrule
\multirow{8}{*}{Fourier} & \multirow{2}{*}{GCV} & $\widehat{K}$ & 6 & 4 & 4 & 6 & 4 & 8 \\
& & $\widehat{\lambda}$ & -1.9191 & -3.9595 & -5.0909 & -3.7979 & -2.7272 &	-2.7272 \\ \\
& \multirow{2}{*}{GIC} & $\widehat{K}$ & 10 & 8 & 10 & 10 & 10 & 10 \\
& & $\widehat{\lambda}$ & -3.3333 & -5.4141 & -6.0000 & -5.2525 & 3.9393 &	-4.7676 \\ \\
& \multirow{2}{*}{MAIC} & $\widehat{K}$ & 10 & 6 & 10 & 6 & 10 & 10 \\
& & $\widehat{\lambda}$ & -2.1212 & -3.9595 & -4.9292 & -4.1212 & -2.7272 &	-3.3333 \\ \\
& \multirow{2}{*}{GBIC} & $\widehat{K}$ & 6 & 7 & 6 & 7 & 7 & 6 \\
& & $\widehat{\lambda}$ & 10.000 & 8.000 & 8.000 & 8.000 & 10.000 &	10.000 \\
\bottomrule
\end{tabular}
\label{tab:param}
\end{table}

The results indicate that Gaussian and B-spline bases tend to use similar numbers of basis functions; in general, more than those estimated via Fourier basis functions. Another important finding is that while the GCV, GIC, and MAIC criteria produce reasonable values for the roughness penalty, the GBIC tends to have large $\log_{10}\widehat{\lambda}$ values for all bases. As an example, we present the discrete data and obtained smooth functions for \textit{Abashiri} station in Appendix~\ref{app:jmmd_fig}, which shows that the smooth functions produced by all basis functions with GBIC correspond to the standard regression line. For other information criteria, the smooth functions obtained by all basis functions provide clear pictures of the discrete data, but the Fourier basis fails to provide satisfactory smooth functions when GCV is used to smooth the data.

The functional regression model is constructed using the variables of 19 (about 70\%) randomly selected stations, as follows:
\begin{align*}
\Y^*_i(t) &= \int_{\Tau} \X^*_{i1}(s) \beta_1(s,t) + \int_{\Tau} \X^*_{i2}(s) \beta_2(s,t) + \int_{\Tau} \X^*_{i3}(s) \beta_3(s,t) \nonumber \\
&+ \int_{\Tau} \X^*_{i4}(s) \beta_4(s,t) + \int_{\Tau} \X^*_{i5}(s) \beta_5(s,t) +\epsilon_i(t),\qquad i = 1, \cdots, 19.\label{eq:real}
\end{align*}
The model is estimated by the LS and MPL methods, and GCV, GIC, MAIC, and GBIC are used to select the best model. The estimated models are then used to predict the monthly global solar radiation functions of the remaining eight (about 30\%) stations. In addition, the bootstrap procedure, introduced in Section~\ref{sec:num}, is used to construct pointwise confidence intervals for the response functions. The calculated AMSE values as well as the CP, width, and score values of the constructed bootstrap confidence intervals for the test stations are reported in Table~\ref{tab:resultsin}. 

\begin{table}[htbp]
\tabcolsep 0.27in
\centering
\caption{Estimated AMSE, CP, width, and score values for the Japanese meteorological data.}
\begin{tabular}{@{}l l l c c c c@{}}
\toprule
Basis & Method & Metric & GCV & GIC & MAIC & GBIC \\
\midrule
Gaussian & LS & AMSE &37.5687 & 30.7033 & 28.0015 & - \\
& & CP & 0.9791 & 1.0000 & 0.9895 & -  \\ 
& & width & 4.5491 & 5.0223 & 4.9257 & -\\ 
& & score & 4.5740 & 5.0223 & 4.9630 & - \\ \cmidrule{2-7}
& MPL & AMSE & 23.1697 & 22.242 & 21.8954 & - \\
& & CP & 1.0000 & 0.9475 & 1.0000 & - \\
& & width & 4.6718 & 3.9053 & 3.8362 & -\\
& & score & 4.6718 & 4.8564 & 3.8362 & - \\
\midrule
B-spline & LS & AMSE & 29.8616 & 29.0738 & 28.7892 & - \\
& & CP & 0.9166 & 1.0000 & 1.0000 & -  \\
& & width & 3.9848 & 4.6923 & 4.6456 & - \\ 
& & score & 5.0633 & 4.6923 & 4.6456 & - \\ \cmidrule{2-7}
& MPL & AMSE & 25.3246 & 19.9871 & 21.6628 & - \\
& & CP & 0.9791 & 1.0000 & 1.0000 & - \\
& & width & 3.8774 & 3.5463 & 3.5348 & -\\
& & score & 3.9280 & 3.5463 & 3.5348 & - \\
\midrule
Fourier & LS & AMSE & 21.4320 & 26.9130 & 27.0450 & - \\
& & CP & 0.9270 & 0.8645 & 0.8958 & -  \\
& & width & 4.5684 & 4.4220 & 3.9565 & -\\ 
& & score & 6.2928 & 6.9387 & 5.9446 & - \\ \cmidrule{2-7}
& MPL & AMSE & 20.5456 & 25.9475 & 25.5723 & - \\
& & CP & 1.0000 & 1.0000 & 1.0000 & - \\
& & width & 3.5859 & 2.9766 & 3.6238 & - \\
& & score & 3.5859 & 2.9766 & 3.6238 & - \\
\bottomrule
\end{tabular}
\label{tab:resultsin}
\end{table}

The results indicate that the MPL method performs better than the LS method in terms of AMSE values. Further, both the LS and MPL methods tend to have smaller AMSE values when Gaussian and B-spline bases are used to smooth the data and MAIC is used to control the roughness parameter. On the other hand, the Fourier basis function has minimum AMSE values when GCV is used as the information criterion. For all basis functions and information criteria, in general, the MPL method produces better coverage probabilities with smaller confidence interval widths and score values than those obtained by the LS method. As stated in Section~\ref{sec:num}, the GBIC tends to produce positive-larger $\lambda$ values compared with other information criteria. Thus, for the Japanese meteorological data, all the methods fail to fit the response functions for all stations when GBIC is used as the information criterion. Therefore, these results are not reported in Table~\ref{tab:resultsin}. As an example, we provide the observed and predicted smooth functions, as well as their bootstrap confidence intervals for a test station (\textit{Abashiri}) in Appendix~\ref{app:jmmd_fig}.

\subsection{North Dakota weekly weather data}

All three variables in the weekly weather data (wind power, mean temperature, and global solar radiation) are averaged over the whole period to construct a functional linear regression model, and the time series plots of the averaged variables are given in Appendix~\ref{app:ndwwd_fig}. 

As for the monthly Japanese meteorological data, we consider predicting global solar radiation but use only the wind and temperature variables. The data are smoothed using the smoothing techniques and the estimated $\widehat{K}$ and $\log_{10}\widehat{\lambda}$ values are reported in Table~\ref{tab:paramN}.

\begin{table}[!htbp]
\tabcolsep 0.26in
\centering
\caption{Estimated number of basis functions and penalty parameters for the weekly North Dakota weather data.}
\begin{tabular}{@{}llcccc@{}}
\toprule
& & Parameter & \multicolumn{3}{c}{Variables} \\
Basis & & & Wind & Temperature & Solar radiation \\
\midrule
\multirow{8}{*}{Gaussian} & \multirow{2}{*}{GCV} & $\widehat{K}$ & 11 & 15 & 17  \\
& & $\widehat{\lambda}$ & -1.8367 & -4.2244 & -2.6530  \\ \\
& \multirow{2}{*}{GIC} & $\widehat{K}$ & 11 & 14 & 11  \\
& & $\widehat{\lambda}$ & -1.4489 & -4.0612 & -4.2244  \\ \\
& \multirow{2}{*}{MAIC} & $\widehat{K}$ & 12 & 17 & 17 \\
& & $\widehat{\lambda}$ & -1.4489 & -4.2857 & -2.6530  \\ \\
& \multirow{2}{*}{GBIC} & $\widehat{K}$ & 8 & 8 & 8  \\
& & $\widehat{\lambda}$ & -2.4897 & -4.2857 & -3.4693   \\
\midrule
\multirow{8}{*}{B-spline} & \multirow{2}{*}{GCV} & $\widehat{K}$ & 11 & 18 & 13   \\
& & $\widehat{\lambda}$ & -1.8367 & -3.0612 & -3.4693  \\ \\
& \multirow{2}{*}{GIC} & $\widehat{K}$ & 11 & 14 & 11  \\
& & $\widehat{\lambda}$ & -1.1020 & -3.4081 & -3.5306  \\ \\
& \multirow{2}{*}{MAIC} & $\widehat{K}$ & 14 & 18 & 13  \\
& & $\widehat{\lambda}$ & -1.4285 & -3.4693 & -3.4693  \\ \\
& \multirow{2}{*}{GBIC} & $\widehat{K}$ & 8 & 8 & 8  \\
& & $\widehat{\lambda}$ & -2.6530 & -3.8775 & -3.8775  \\
\midrule
\multirow{8}{*}{Fourier} & \multirow{2}{*}{GCV} & $\widehat{K}$ & 6 & 10 & 10  \\
& & $\widehat{\lambda}$ & -3.8775 & -5.3673 & -5.0408  \\ \\
& \multirow{2}{*}{GIC} & $\widehat{K}$ & 6 & 12 & 10  \\
& & $\widehat{\lambda}$ & -3.8775 & -5.3673 & -5.3673  \\ \\
& \multirow{2}{*}{MAIC} & $\widehat{K}$ & 6 & 12 & 10  \\
& & $\widehat{\lambda}$ & -3.8775 & -5.2244 & -5.0408  \\ \\
& \multirow{2}{*}{GBIC} & $\widehat{K}$ & 5 & 5 & 5  \\
& & $\widehat{\lambda}$ & -3.8775 & -5.5714 & -5.3673  \\
\bottomrule
\end{tabular}
\label{tab:paramN}
\end{table}

Our results indicate that all variables are well smoothed by all smoothing techniques. The difference between the three smoothing methods is that only the Fourier basis produces different smooth functions for the wind variable, since it uses a smaller number of basis functions compared with the other two bases; for an example, see Appendix~\ref{app:ndwwd_fig}. 

For these data, the functional regression model is constructed using 32 (about 70\%) randomly selected stations, and the estimated model is used to predict the global solar radiation functions of the remaining 13 (30\%) stations. The regression model is:
\begin{equation*}
\Y^*_i(t) = \int_{\Tau} \X^*_{i1}(s) \beta_1(s,t) + \int_{\Tau} \X^*_{i2}(s) \beta_2(s,t) + \epsilon_i(t), \qquad i=1,\dots,32. \label{eq:realN}
\end{equation*}
The calculated AMSE, Cp, width, and score values for the test stations are presented in Table \ref{tab:resultsinN}. The results demonstrate that in general, the MPL method gives significantly smaller AMSE values compared with LS. For both methods, the minimum AMSE values are obtained when Gaussian and B-spline bases are used with GIC and GBIC. Additionally, they achieve the minimum AMSE values for the Fourier basis when GBIC is used to control the roughness parameter. For the proposed bootstrap method, the results show that MPL performs better compared with the LS method, because of the smaller score values. An example of the observed and smooth functions with bootstrap confidence intervals for a test station (\textit{Perley}) is presented in Appendix~\ref{app:ndwwd_fig}.

\begin{table}[htbp]
\tabcolsep 0.27in
\centering
\caption{Estimated AMSE, CP, width, and score values for the weekly weather data.}
\begin{tabular}{@{}lllcccc@{}}
\toprule
Basis & Method & Metric & GCV & GIC & MAIC & GBIC \\
\midrule
Gaussian & LS & AMSE &55.8492 & 11.5813 & 52.4646 & 6.9929 \\
& & CP & 0.5236 & 1.0000 & 0.8210 & 1.0000  \\
& & width & 1.6406 & 4.4631 & 2.4585 & 1.8937 \\ 
& & score & 12.8549 & 4.4631 & 6.6600 & 1.8937 \\ \cmidrule{2-7}
& MPL & AMSE & 20.4993 & 6.9023 & 20.3555 & 6.9925 \\
& & CP & 0.8757 & 0.9940 & 0.9792 & 0.9142 \\
& & width & 1.6705 & 3.3114 & 2.2852 & 0.9704 \\
& & score & 4.1828 & 3.3371 & 2.5640 & 1.4067 \\
\midrule
B-spline & LS & AMSE & 53.6537 & 13.2212 & 60.3397 & 7.3205 \\
& & CP & 0.9526 & 1.0000 & 0.4940 & 1.0000  \\
& & width & 3.1772 & 4.7176 & 1.5884 & 2.2696 \\ 
& & score & 4.1575 & 4.7176 & 16.4699 & 2.2696 \\ \cmidrule{2-7}
& MPL & AMSE & 11.6453 & 10.2249 & 18.7197 & 7.3183 \\
& & CP & 1.0000 & 0.9778 & 1.0000 & 0.9230 \\
& & width & 3.3167 & 3.2021 & 2.9493 & 0.8677 \\
& & score & 3.3167 & 3.2729 & 2.9493 & 1.0461 \\
\midrule
Fourier & LS & AMSE & 13.9520 & 16.7243 & 15.6946 & 5.2677 \\
& & CP & 1.0000 & 1.0000 & 1.0000 & 1.0000  \\
& & width & 4.5172 & 3.4071 & 6.6162 & 3.3116 \\ 
& & score & 4.5172 & 3.4071 & 6.6162 & 3.3116 \\ \cmidrule{2-7}
& MPL & AMSE & 16.1774 & 16.7240 & 7.2028 & 5.2677 \\
& & CP & 0.9215 & 1.0000 & 0.9822 & 1.0000 \\
& & width & 2.8325 & 3.4175 & 3.1534 & 3.3299 \\
& & score & 3.6245 & 3.4175 & 3.2197 & 3.3299 \\
\bottomrule
\end{tabular}
\label{tab:resultsinN}
\end{table}

\newpage

Overall, the results of our real-data examples indicate that the performances of the LS and MPL methods differ according to the choice of basis function and the pre-determined number of basis functions/smoothing parameter $\lambda$. There is no unique basis function/information criterion combination that achieves the most accurate model estimate. However, the results show that, in general, the MPL method outperforms LS, in genral, the estimation methods tend to produce their best performances when GIC is used to control the roughness parameter and evaluate the model for the first real-data example, while, for the second real-data example, they have their best performances when GBIC is used as the information criterion. Moreover, for the second dataset, the estimation methods generally have their second-best performances when GIC is used as the information criterion. Compared with GBIC, the LS and MPL produce more consistent results when GIC is used as the information criterion. Therefore, the results of our real-data examples suggest MPL as the estimation method and GIC as the information criterion. However, no clear result emerges for the choice of basis function.

\section{Conclusion\label{sec:conc}}

Many applications involve observing functional data in a graphical representation of curves that are sampled over a continuum measure. This motivates the development of FDA smoothing techniques for visualizing and analyzing such data. In particular, the function-on-function linear model is a frequently used technique in many applied scientific areas to explore the relationships between the functional response and predictors. However, the accuracy of these models is based on several factors, such as an optimal choice of basis function and corresponding roughness parameter, estimation method, and model evaluation criterion. 

In this study, we compare several smoothing techniques to demonstrate which basis function, estimation method and model evaluation criteria are the best suited to the function-on-function regression model. The comparisons are performed through Monte Carlo simulations and use two real-data examples. Our results show that the performances of the estimation techniques differ according to basis function/information criterion selection. The simulation results indicate that, when both the response and predictor variables have a non-periodic structure, both LS and MPL methods produce their best performances (in terms of AMSE values) when a B-spline-GIC basis function-information criterion combination is used to smooth the data and control the roughness parameter/evaluate the estimated model. However, when the response and/or predictor functions have a periodic structure, the estimation methods have their best performances when a Gaussian-GCV (or MAIC) basis function-information criterion combination is used. Generally, LS and MPL methods tend to have a similar performance in the Monte Carlo experiments, although MPL outperforms LS in empirical data examples. In addition, we proposed a case-resampling-based bootstrap method to construct a pointwise confidence interval for the response function. All the numerical analyses considered in this study indicate that the proposed bootstrap method is capable of producing a reliable confidence interval for the response function. 

There are several ways in which the methodology presented can be further extended; we briefly list two:
\begin{inparaenum}
\item[1)] we consider function-on-function regression, but the performances of the smoothing techniques can also be extended to scalar-on-function or function-on-scalar regressions;
\item[2)] we consider three commonly used basis functions, but other basis functions, such as the Bernstein polynomial basis and wavelet basis functions, may also be explored.
\end{inparaenum}

\clearpage

\section*{Appendices}\label{appendix}
\appendix
\section{Figures for the simulation studies}\label{app:sim}
\subsection{Case-I}\label{app:sim_Case_I}

\begin{figure}[!htbp]
  \centering
  \includegraphics[width=8cm]{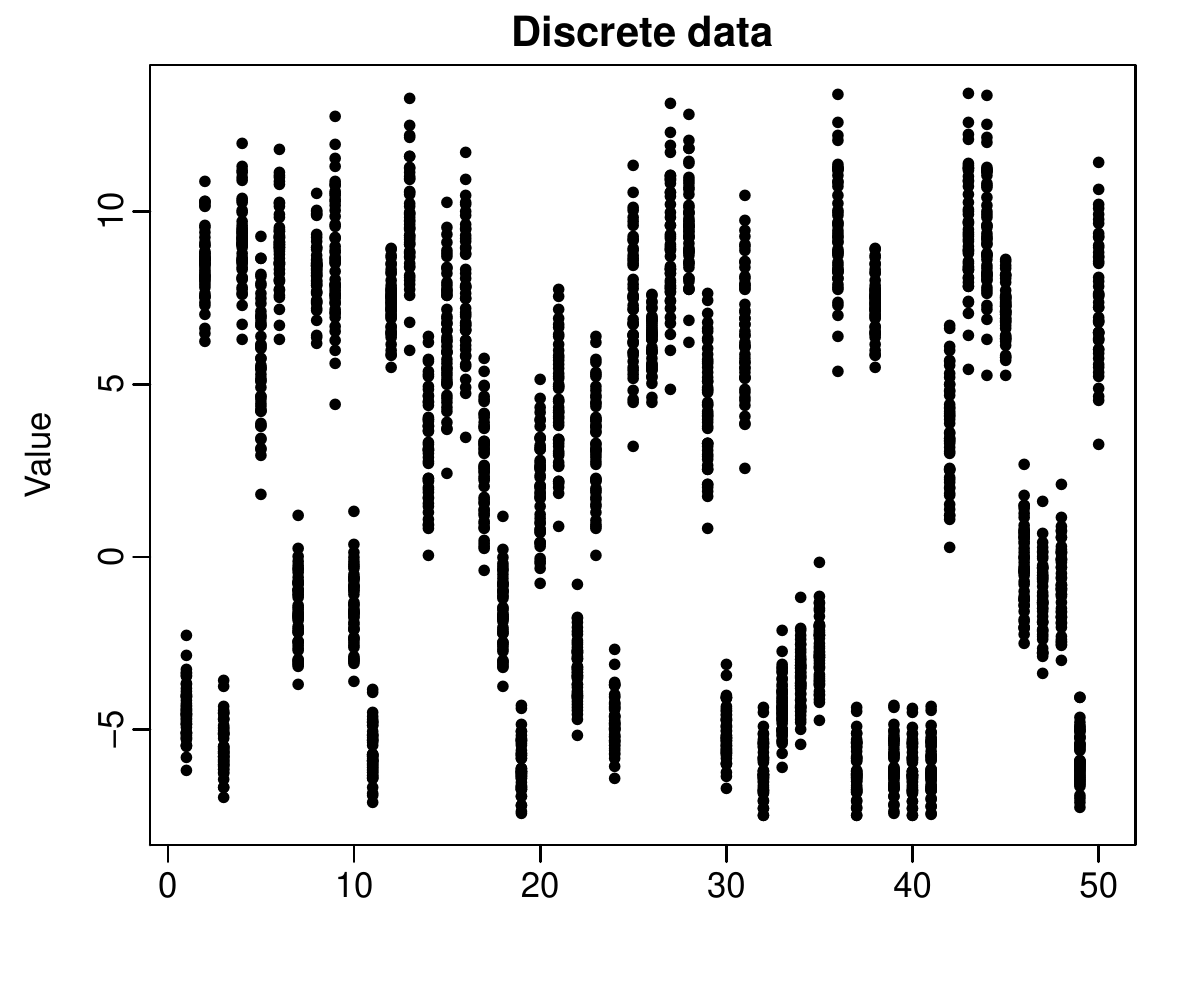}
  \includegraphics[width=8cm]{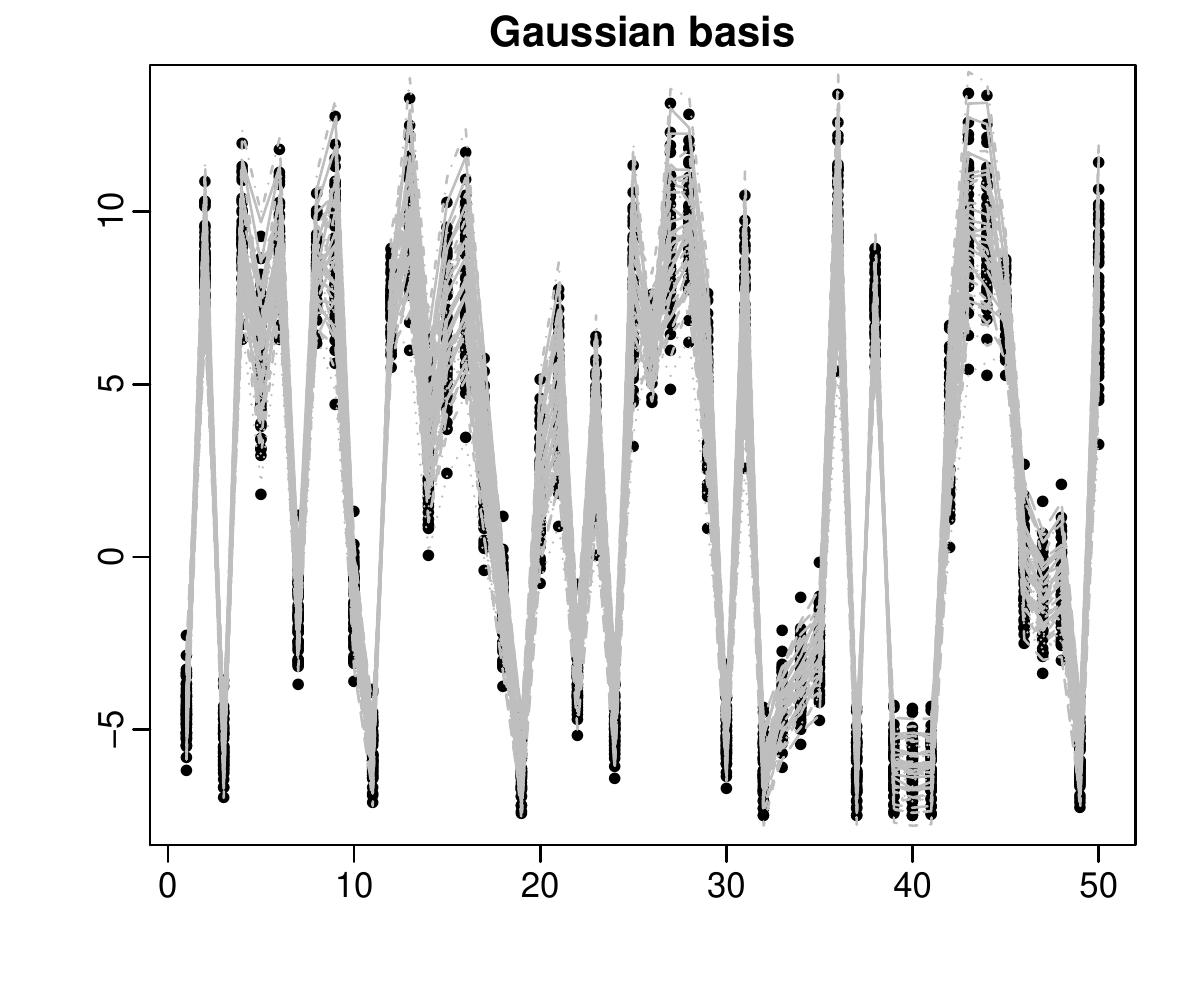}
\\
  \includegraphics[width=8cm]{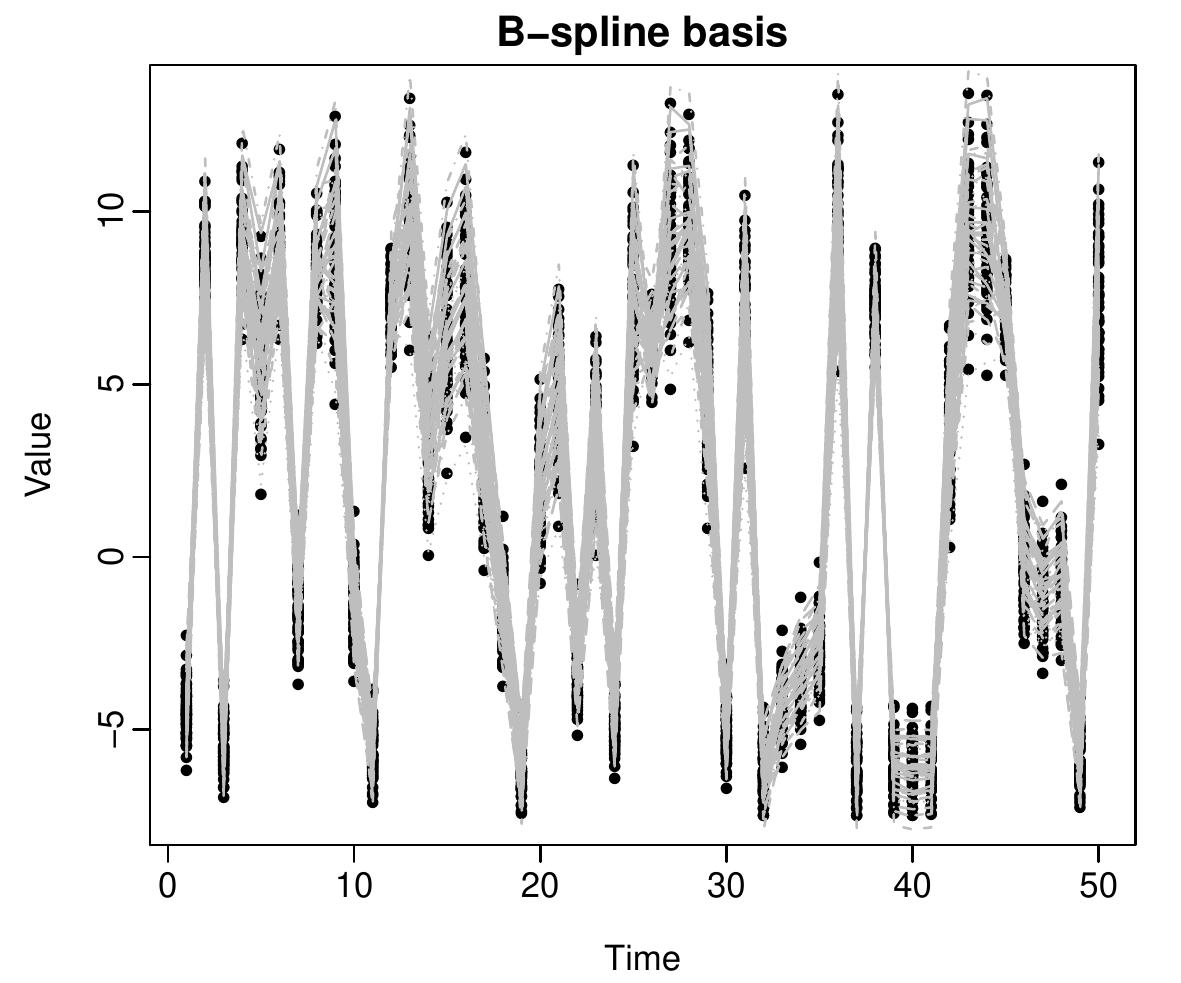}
  \includegraphics[width=8cm]{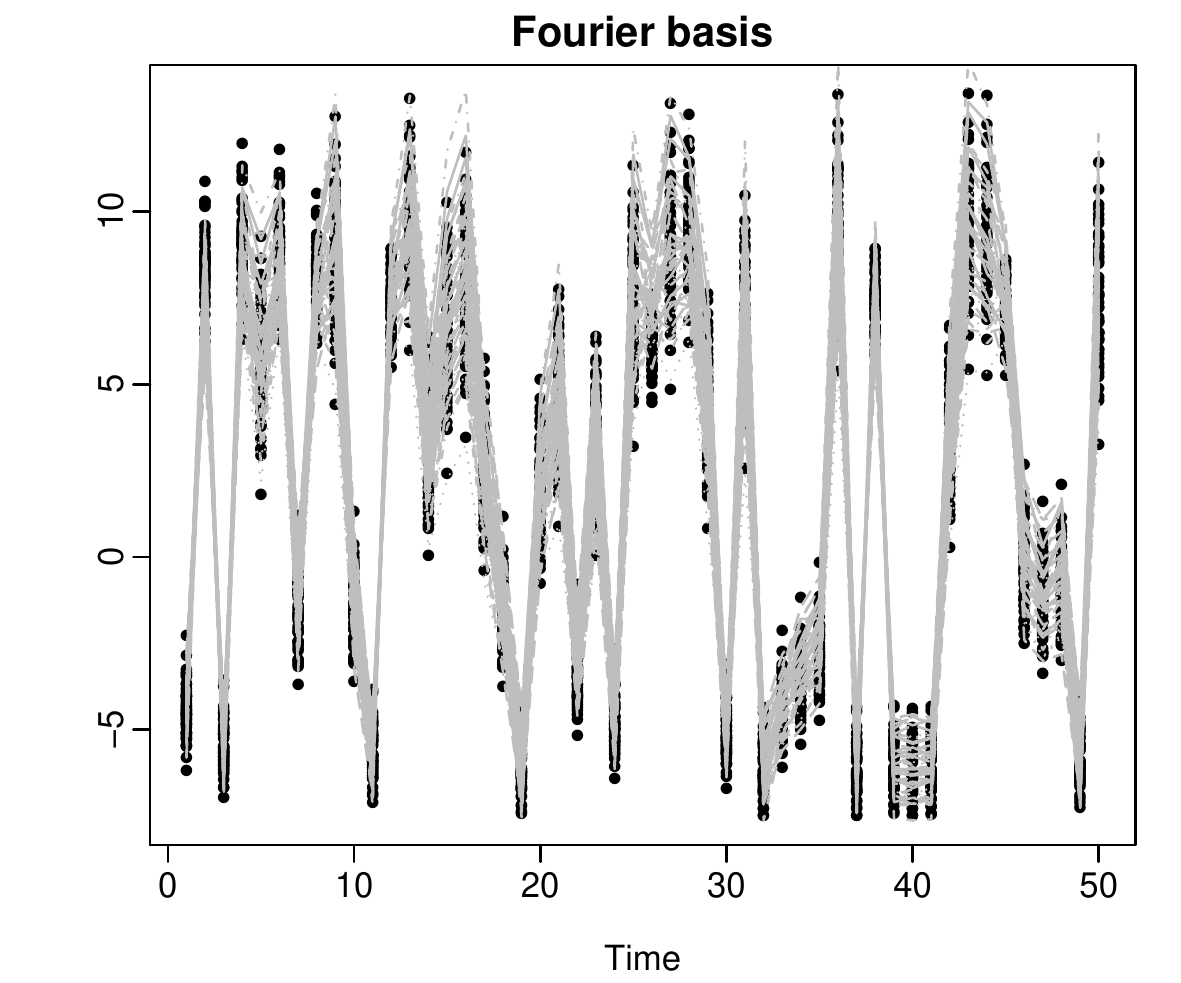}  
  \caption{Plots of the generated $N$ sets of discrete data and computed smooth functions when GCV is used to select the roughness parameter $\lambda$.}
  \label{fig:simdat}
\end{figure}

\begin{figure}[!htbp]
  \centering
  \includegraphics[width=8cm]{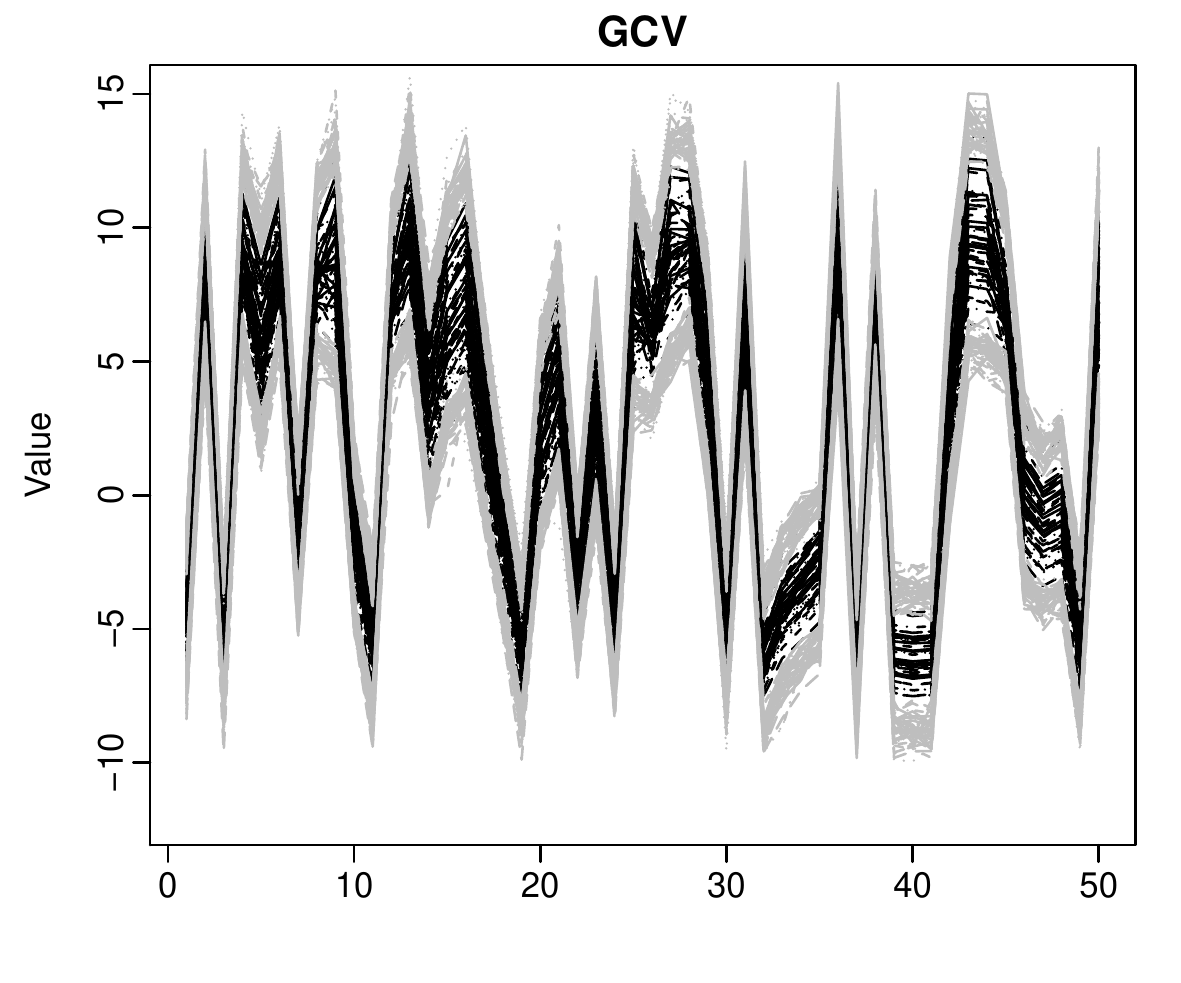}
  \includegraphics[width=8cm]{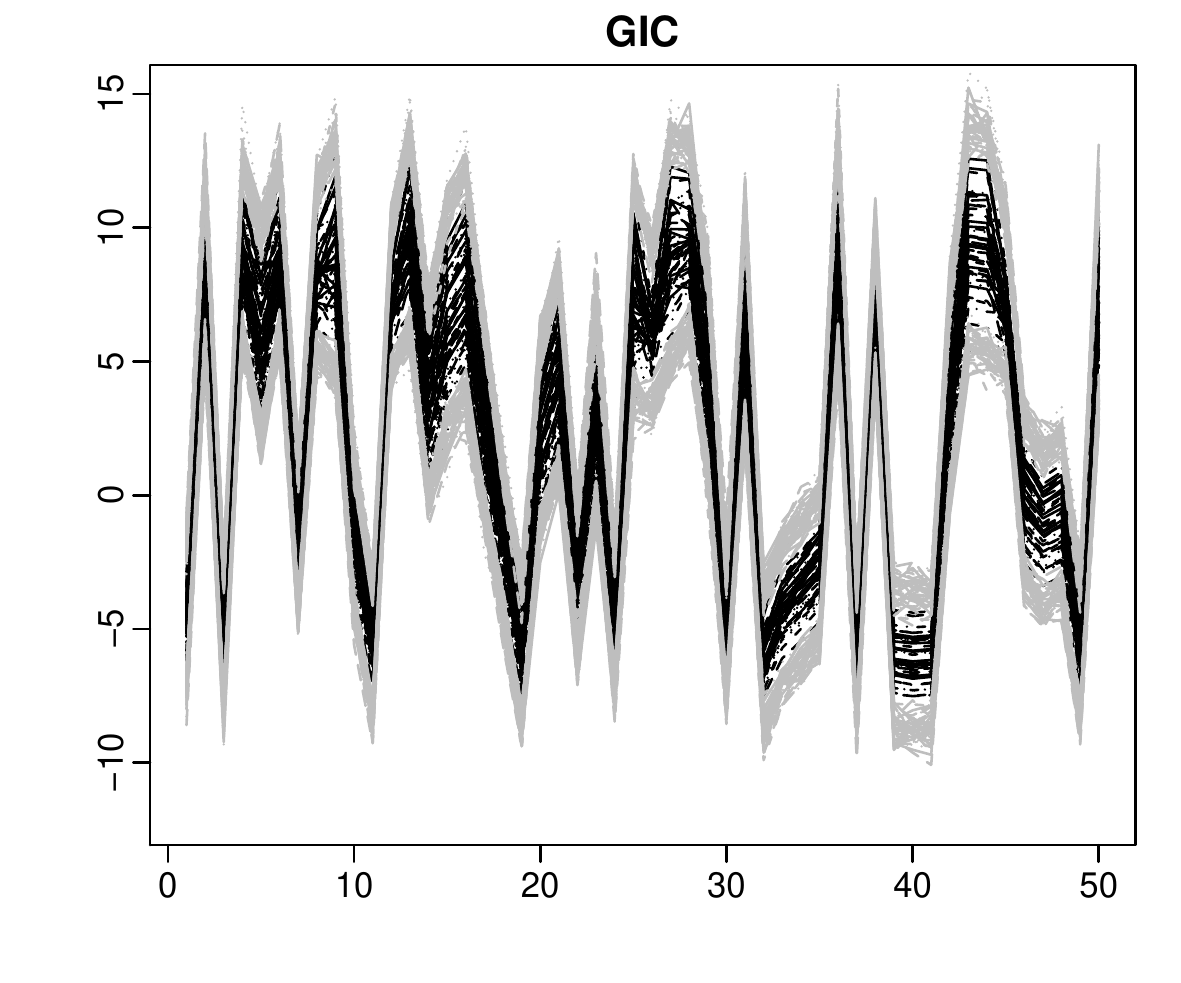}
\\
  \includegraphics[width=8cm]{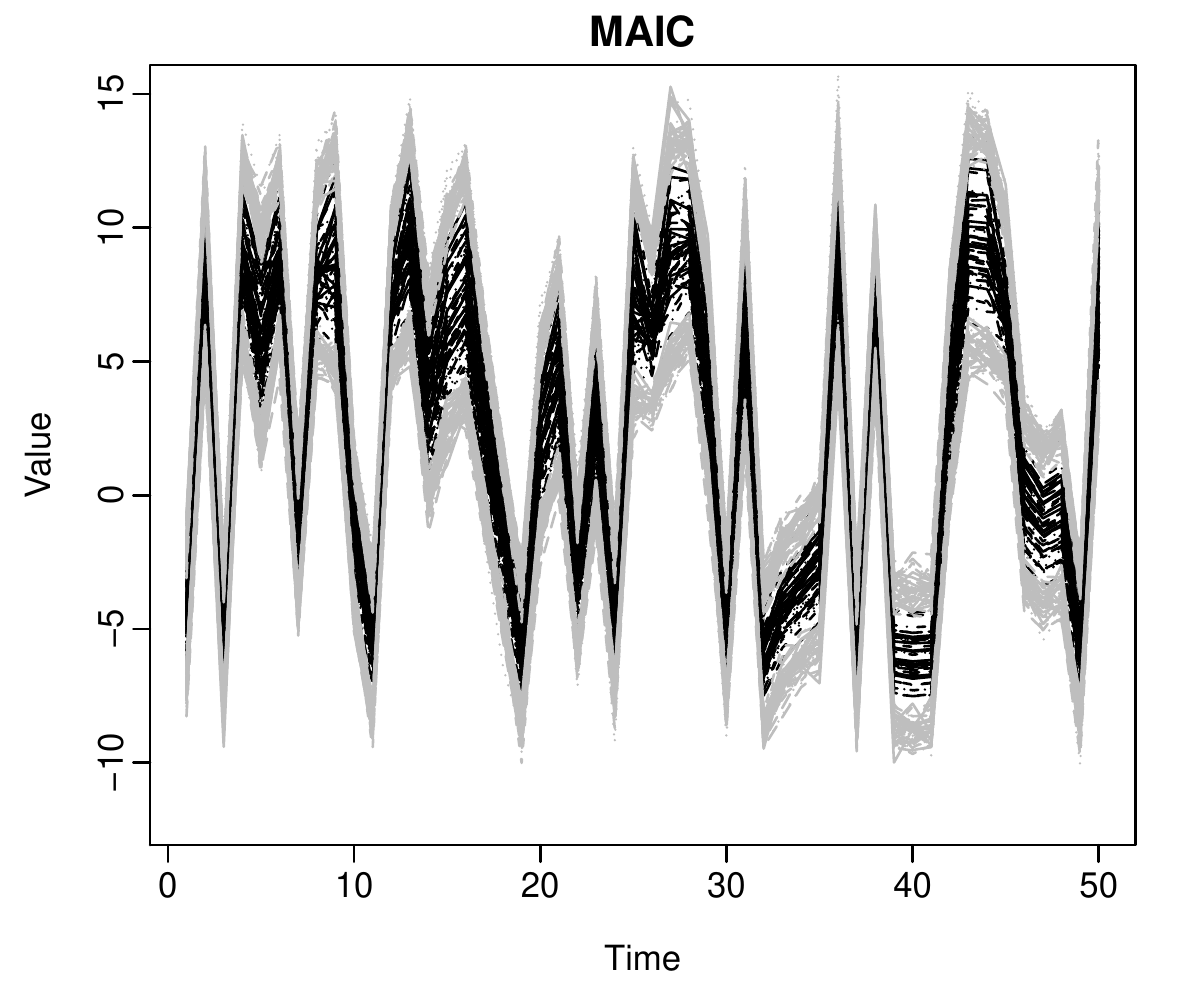}
  \includegraphics[width=8cm]{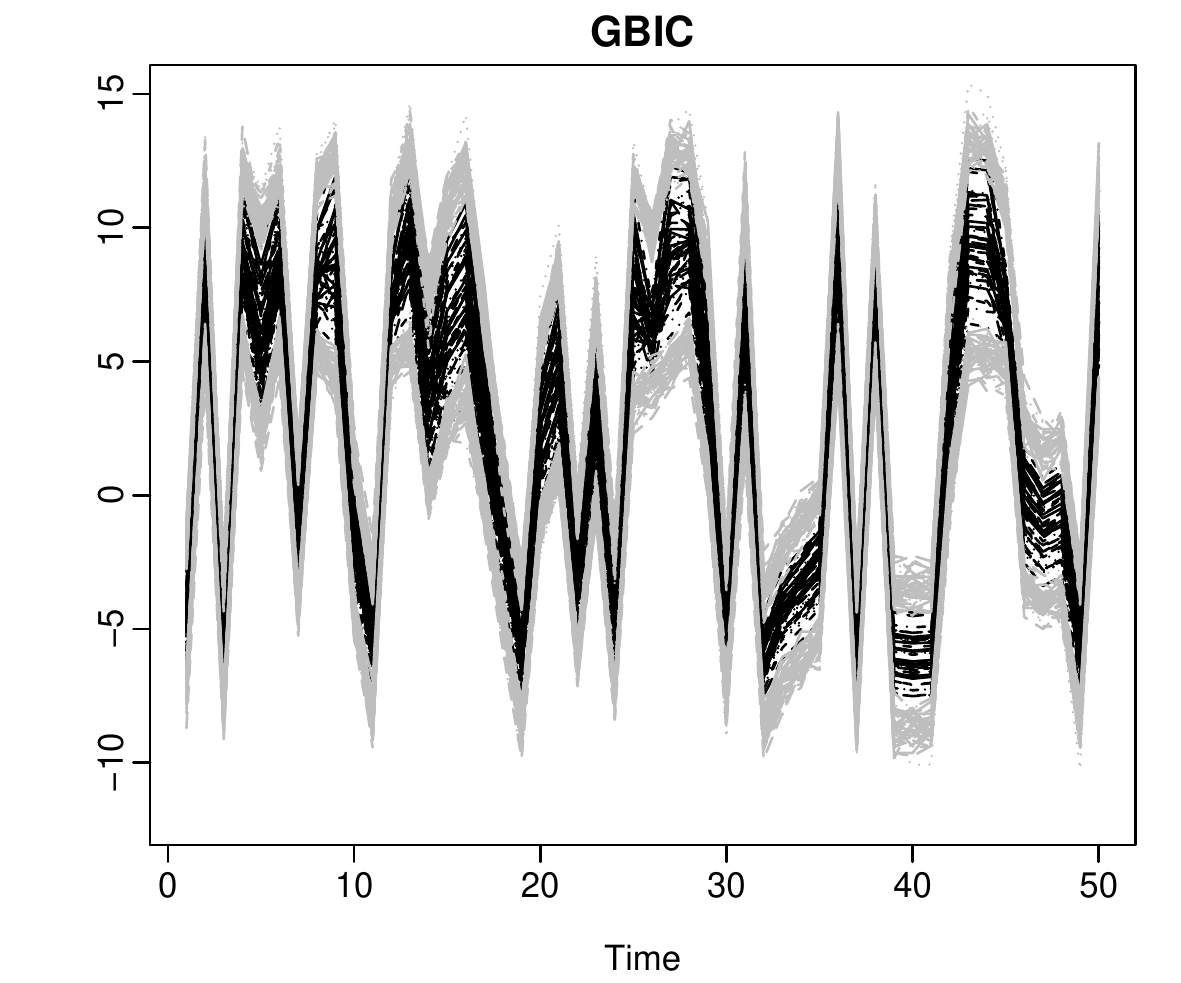}  
  \caption{Plots of the generated N sets of actual functions (black lines) and their corresponding bootstrap confidence intervals (gray lines) when $\rho = 2$. Note, the Gaussian basis is used to smooth the data, and the LS method is used to estimate the function-on-function regression model. The GCV, GIC, MAIC, and GBIC criteria are used to control the roughness parameter and evaluate the estimated model.}
  \label{fig:boot}
\end{figure}

\clearpage
\subsection{Case-II and Case-III}\label{app:sim_Case_II_III}

\begin{figure}[!htbp]
  \centering
  \includegraphics[width=6.1cm]{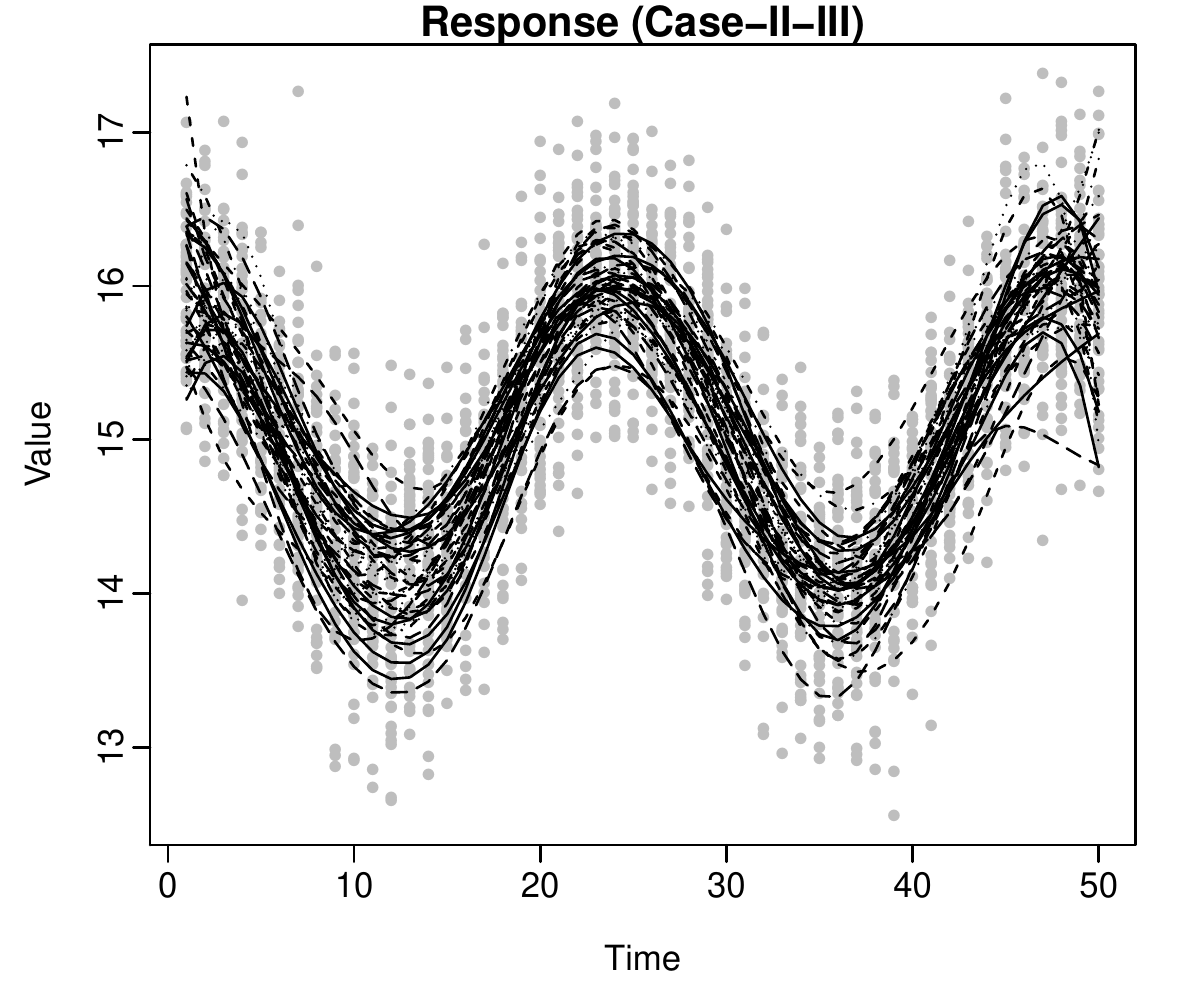}
  \includegraphics[width=6.1cm]{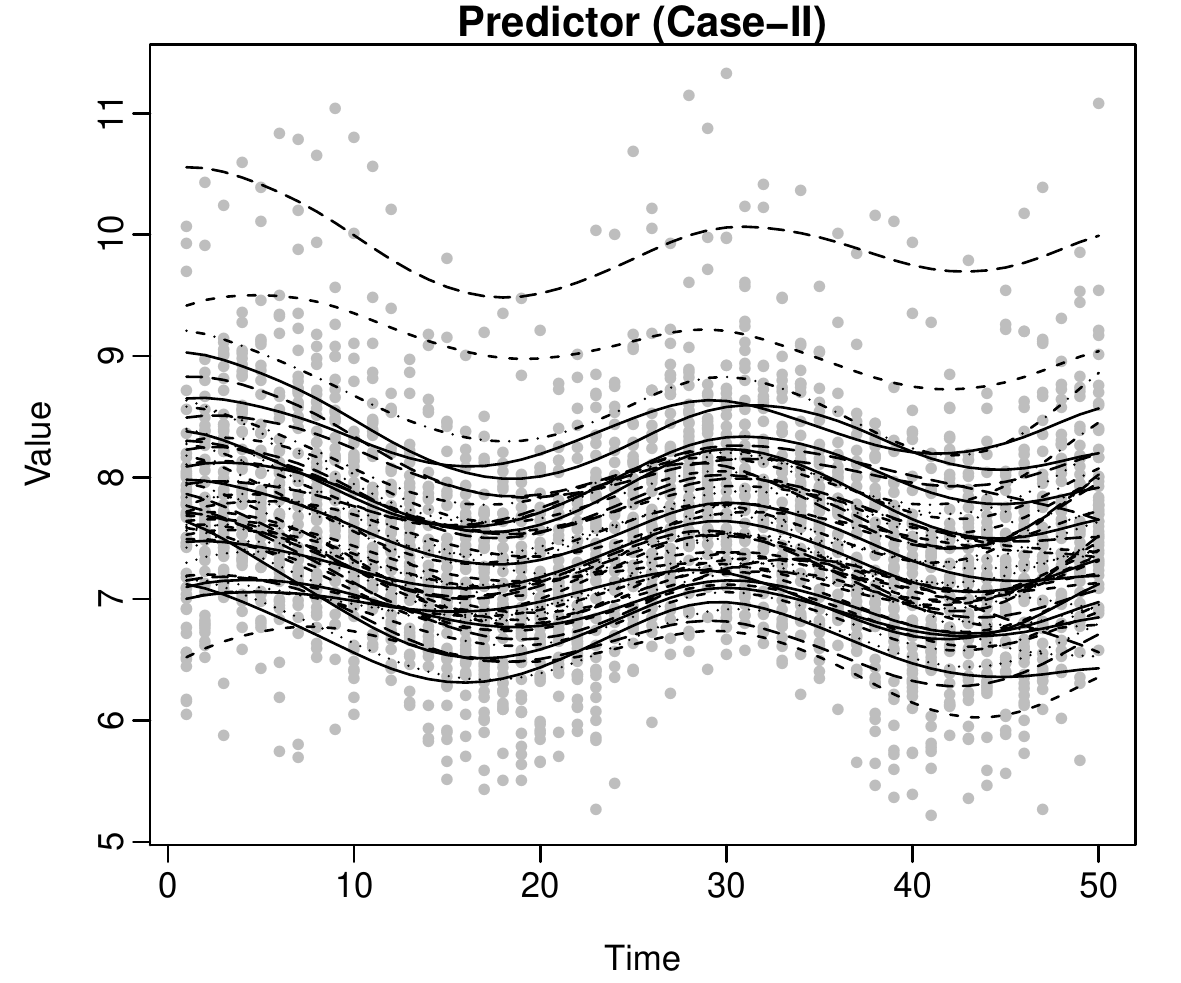}
  \includegraphics[width=6.1cm]{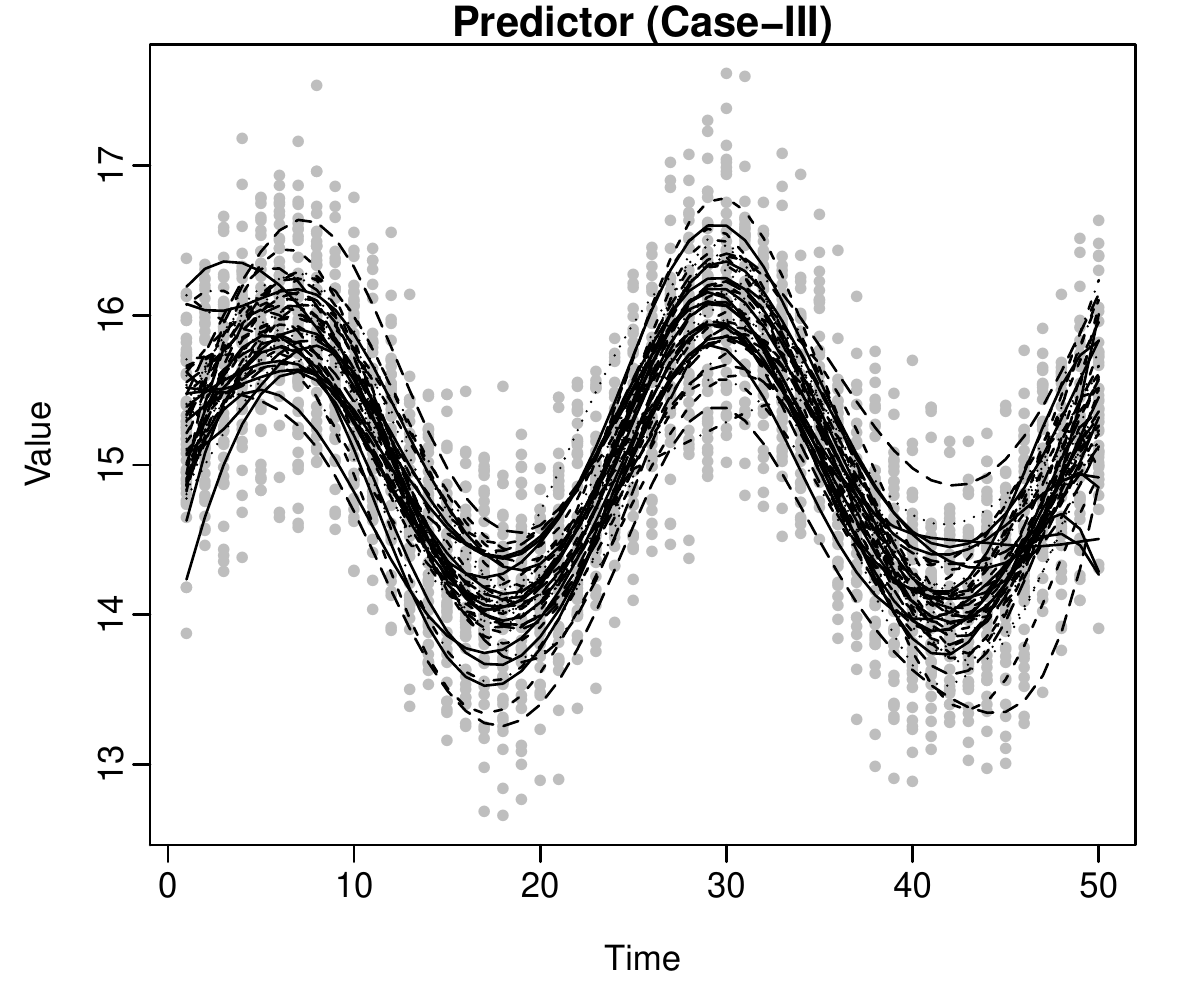}  
  \caption{Plots of the generated N sets of discrete data (gray points) and computed smooth functions (black lines) for Case-II and Case-III when GCV is used to select the roughness parameter $\lambda$.}
  \label{fig:Fig_na_data}
\end{figure}

\clearpage
\section{Additional figures and tables for the empirical data analyses}
\subsection{Japanese monthly meteorological data}
\subsubsection{Figures}\label{app:jmmd_fig}

\begin{figure}[htbp]
  \centering
  \includegraphics[width=17.5cm]{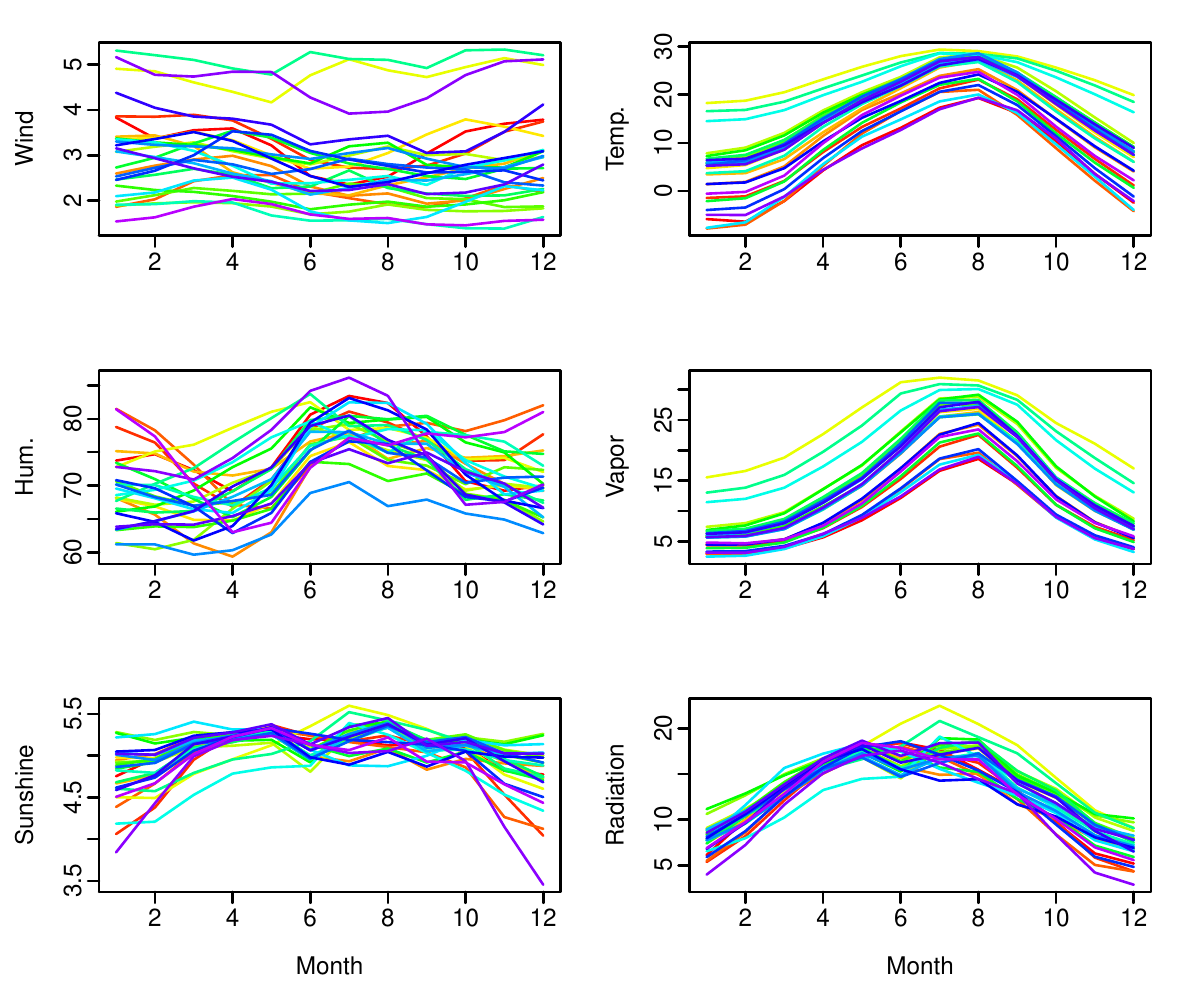}
  \caption{Time series plots of the averaged Japanese meteorological variables.}
  \label{fig:ts}
\end{figure}

\begin{figure}[htbp]
\centering
\subfloat[GCV]
{\includegraphics[width=8.8cm]{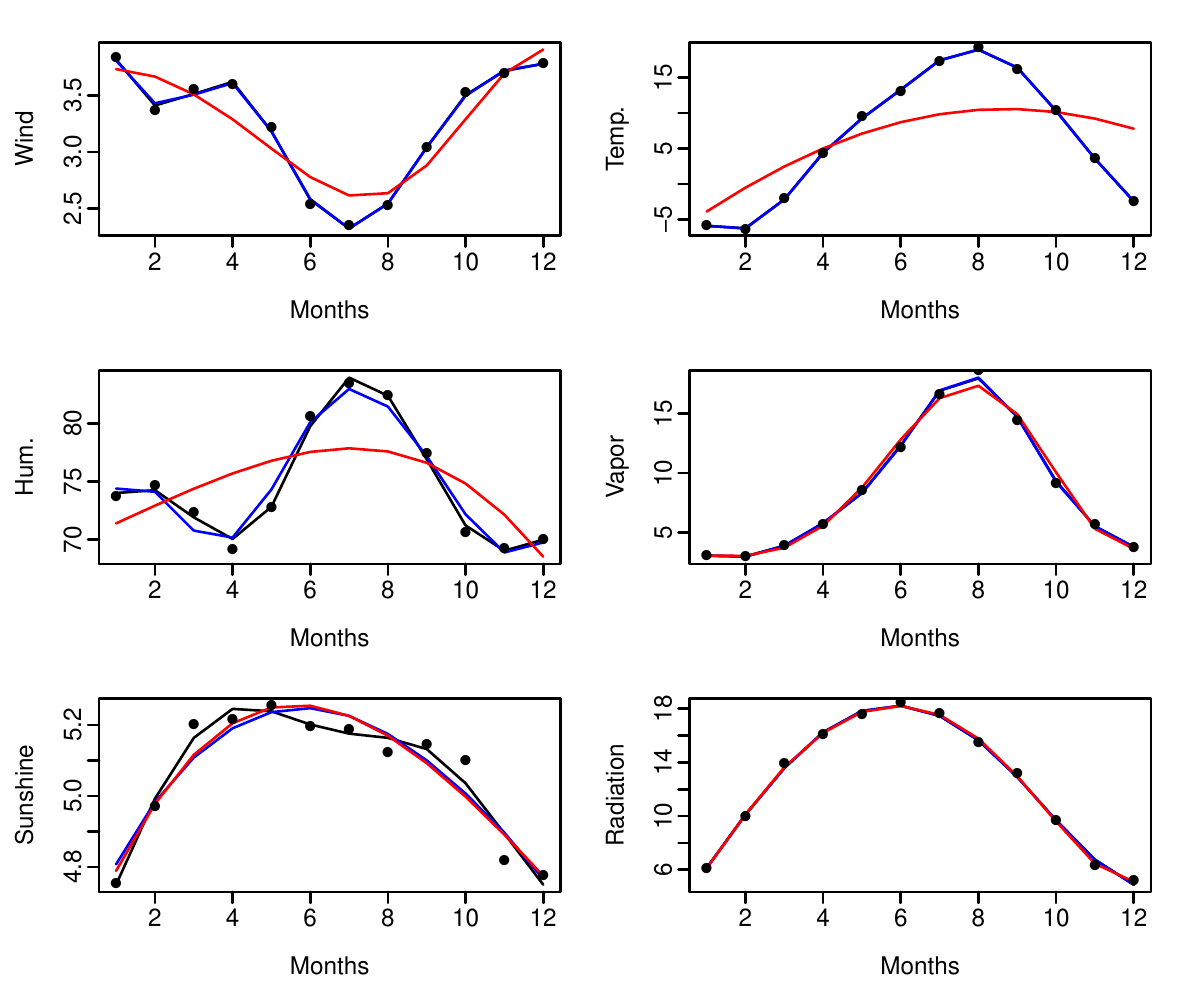}}
\quad
\subfloat[GIC]
{\includegraphics[width=8.8cm]{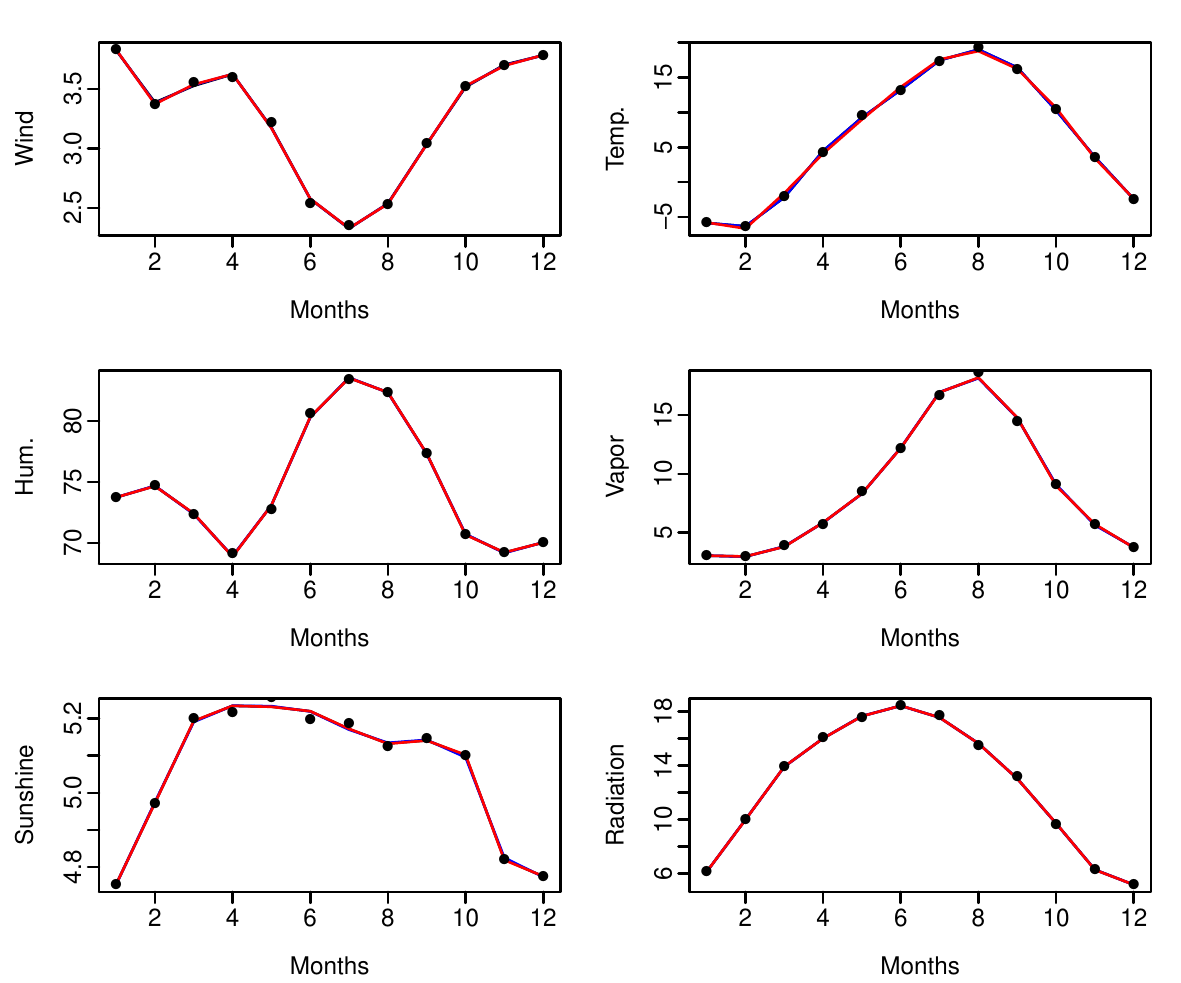}}
\\
\subfloat[MAIC]
{\includegraphics[width=8.8cm]{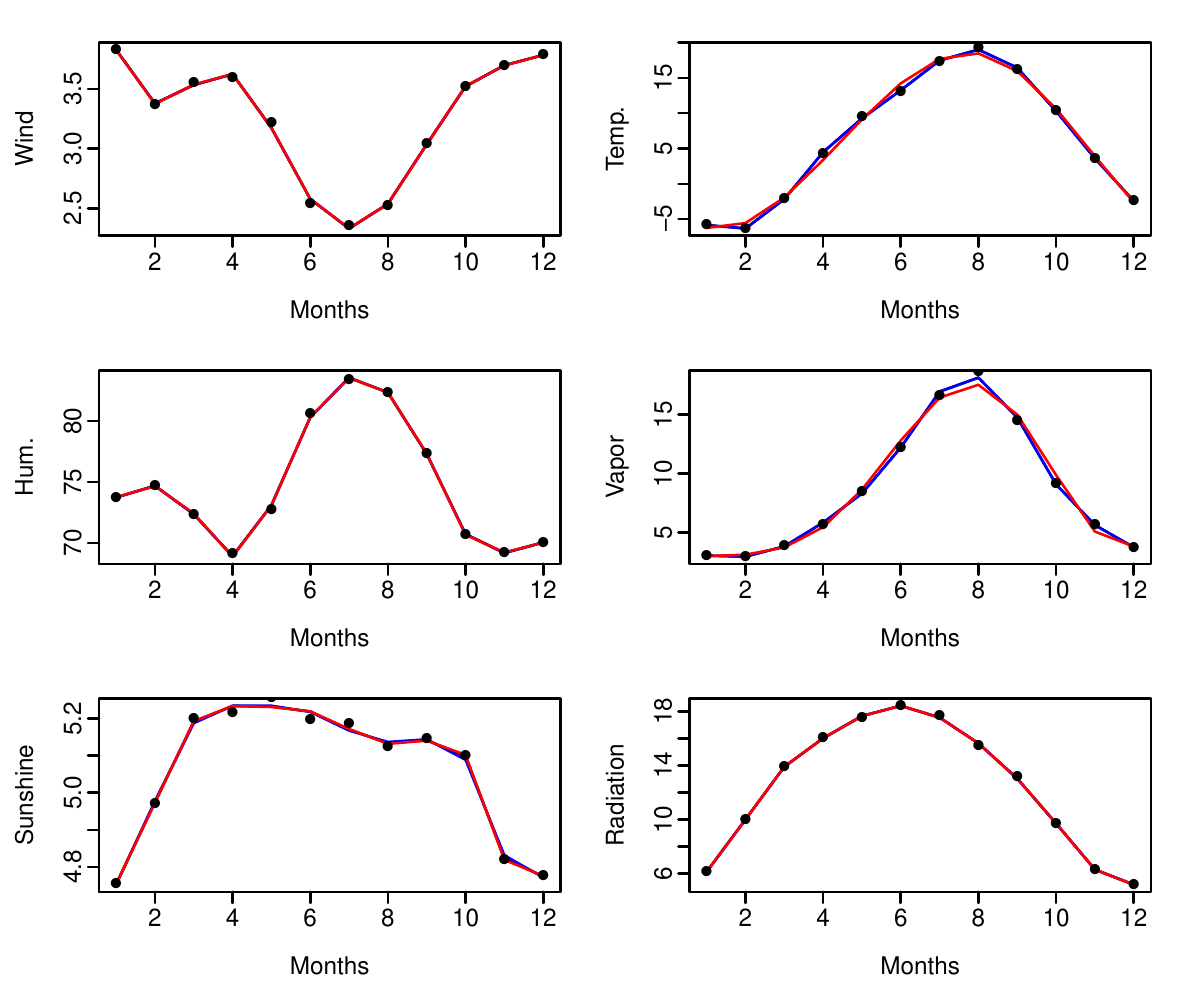}}
\quad
\subfloat[GBIC]
{\includegraphics[width=8.8cm]{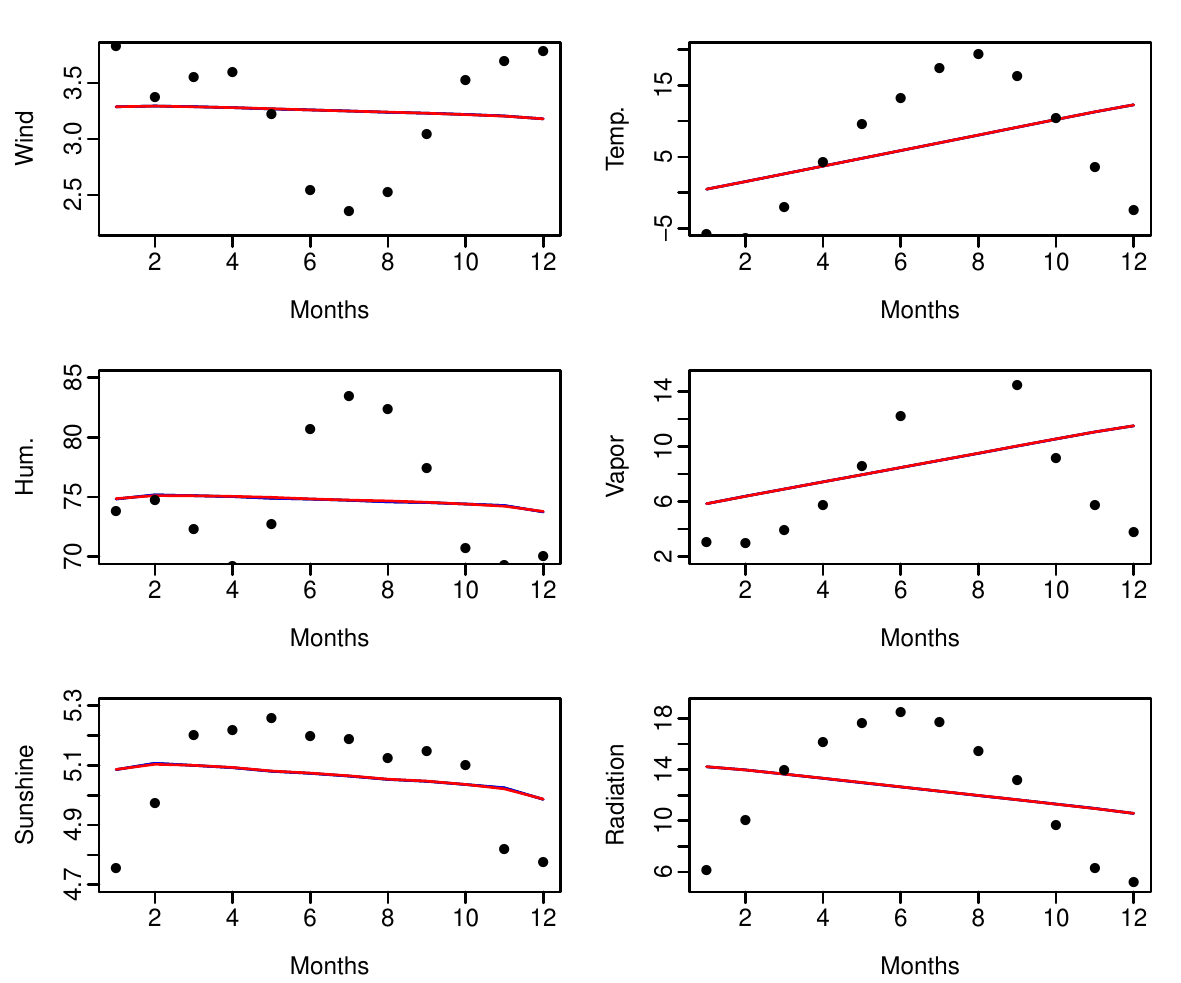}}
\caption{Plots of discrete Japanese meteorological variables (black points) and their smooth functions: Gaussian basis (black lines), B-spline basis (blue lines), and Fourier basis (red lines). The GCV, GIC, MAIC, and GBIC criteria are used to control the roughness parameter.}
\label{fig:smooth}
\end{figure}

\begin{figure}[htbp]
\centering
{{\includegraphics[width=5cm]{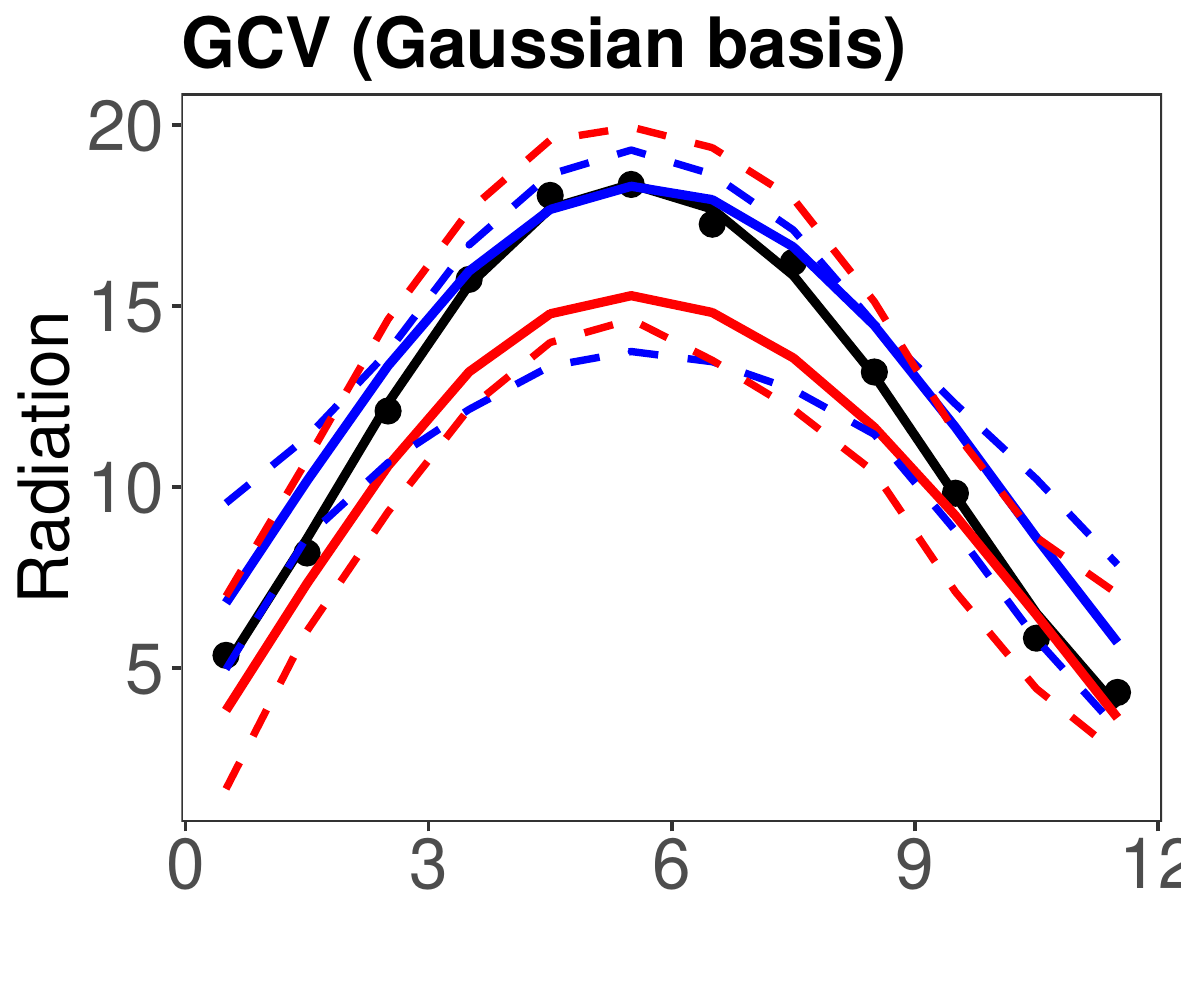}}}
\qquad
{{\includegraphics[width=5cm]{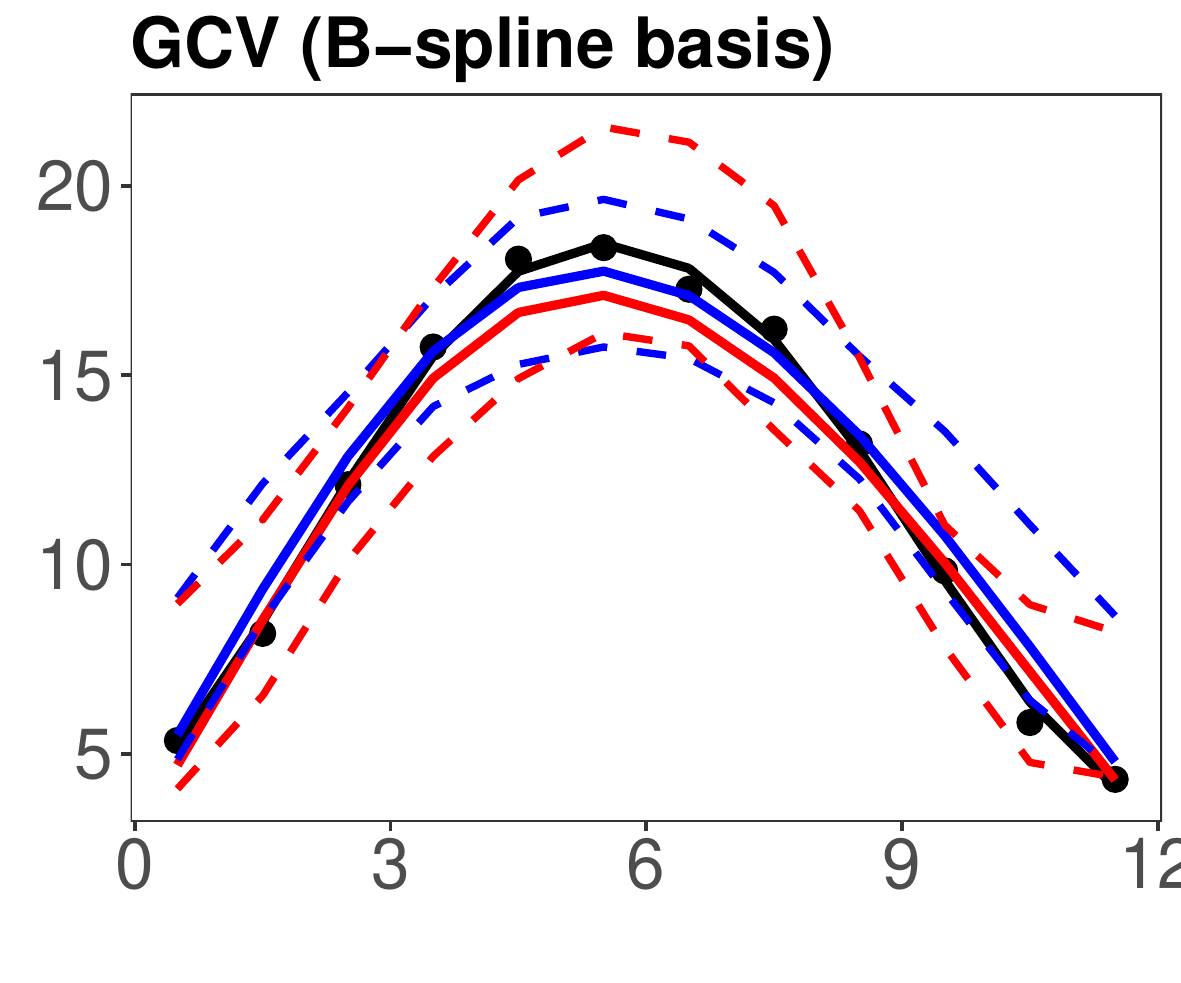}}}
\qquad
{{\includegraphics[width=5cm]{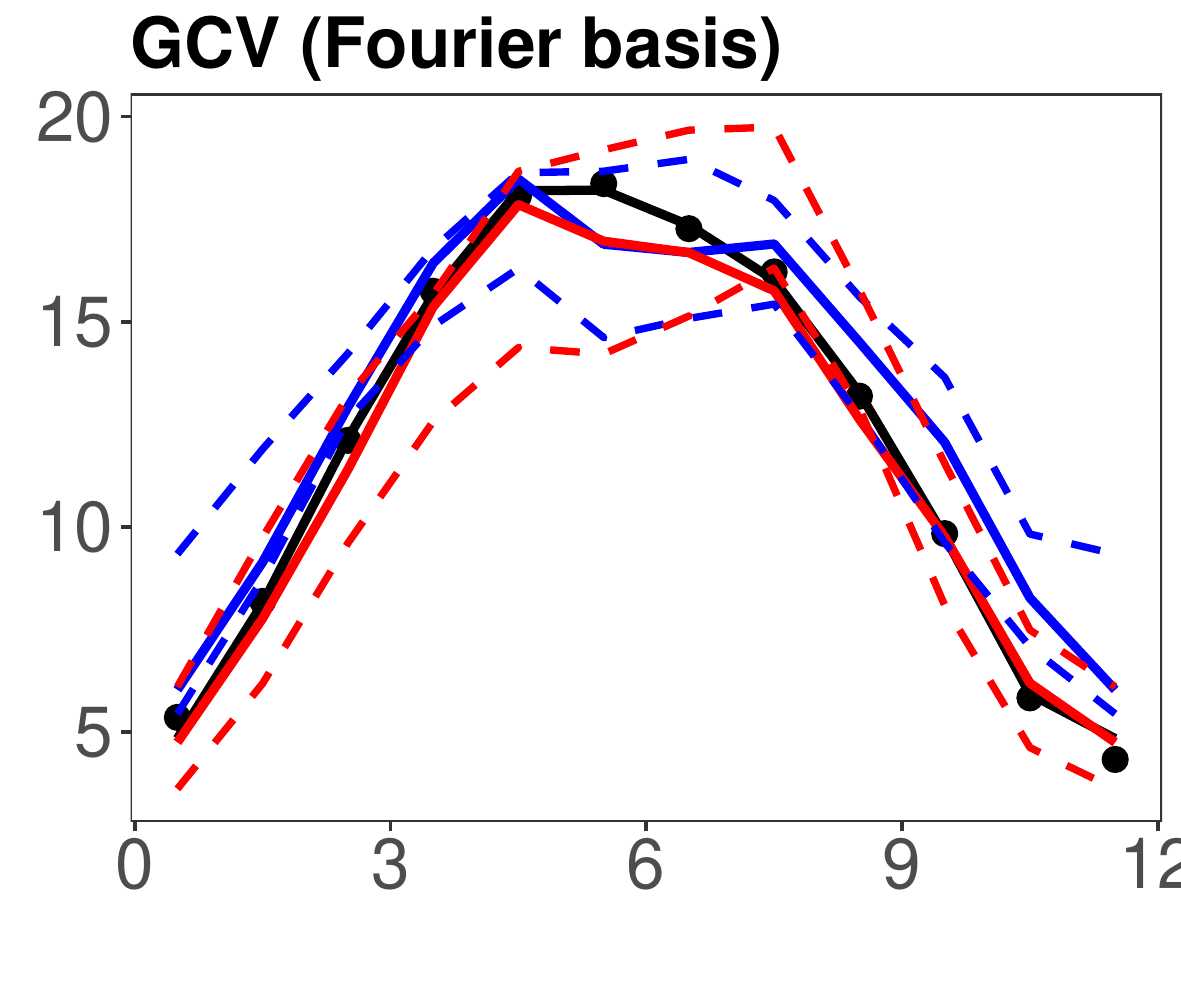}}}
\\
{\includegraphics[width=5cm]{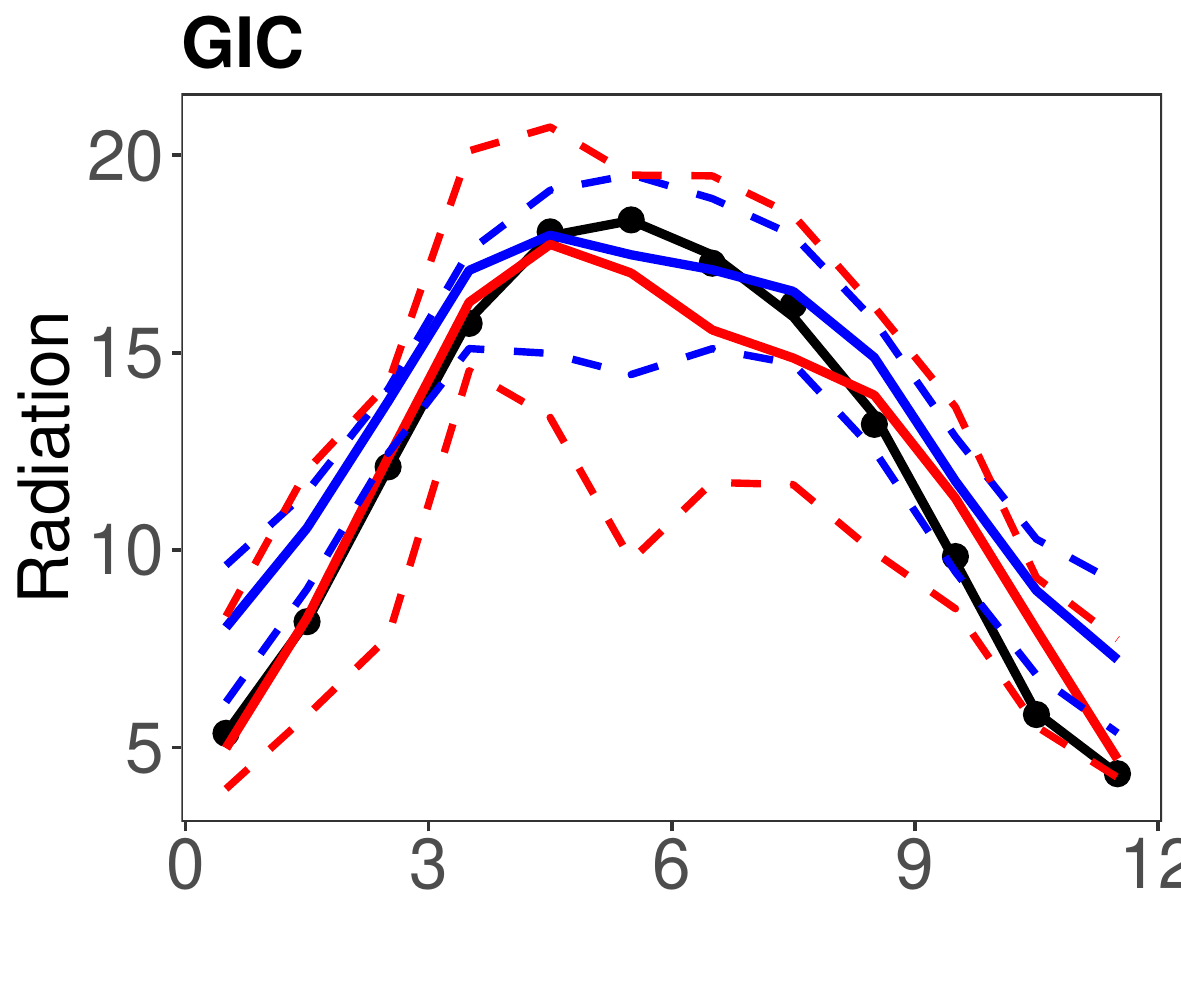}}
\qquad
{\includegraphics[width=5cm]{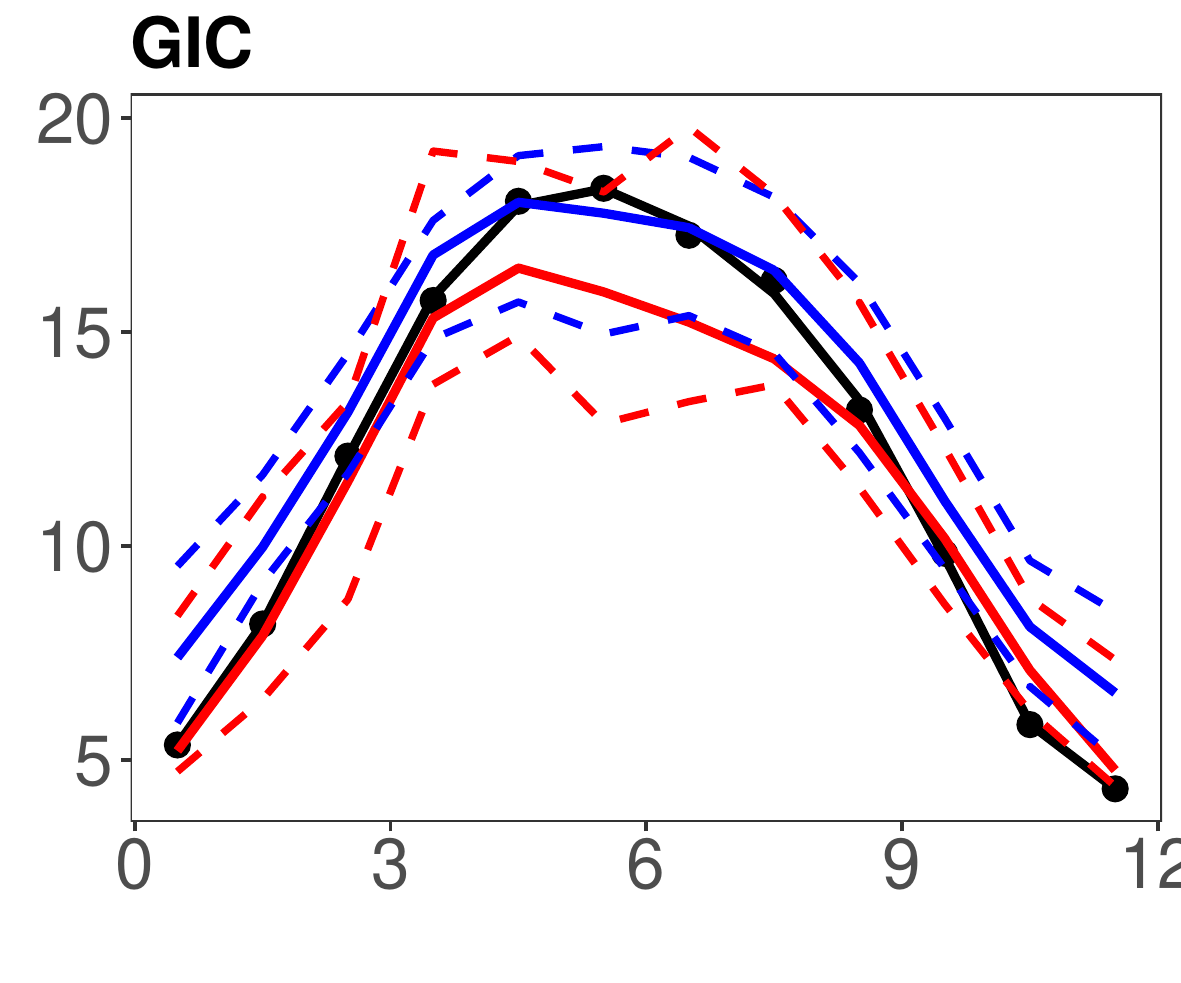}}
\qquad
{\includegraphics[width=5cm]{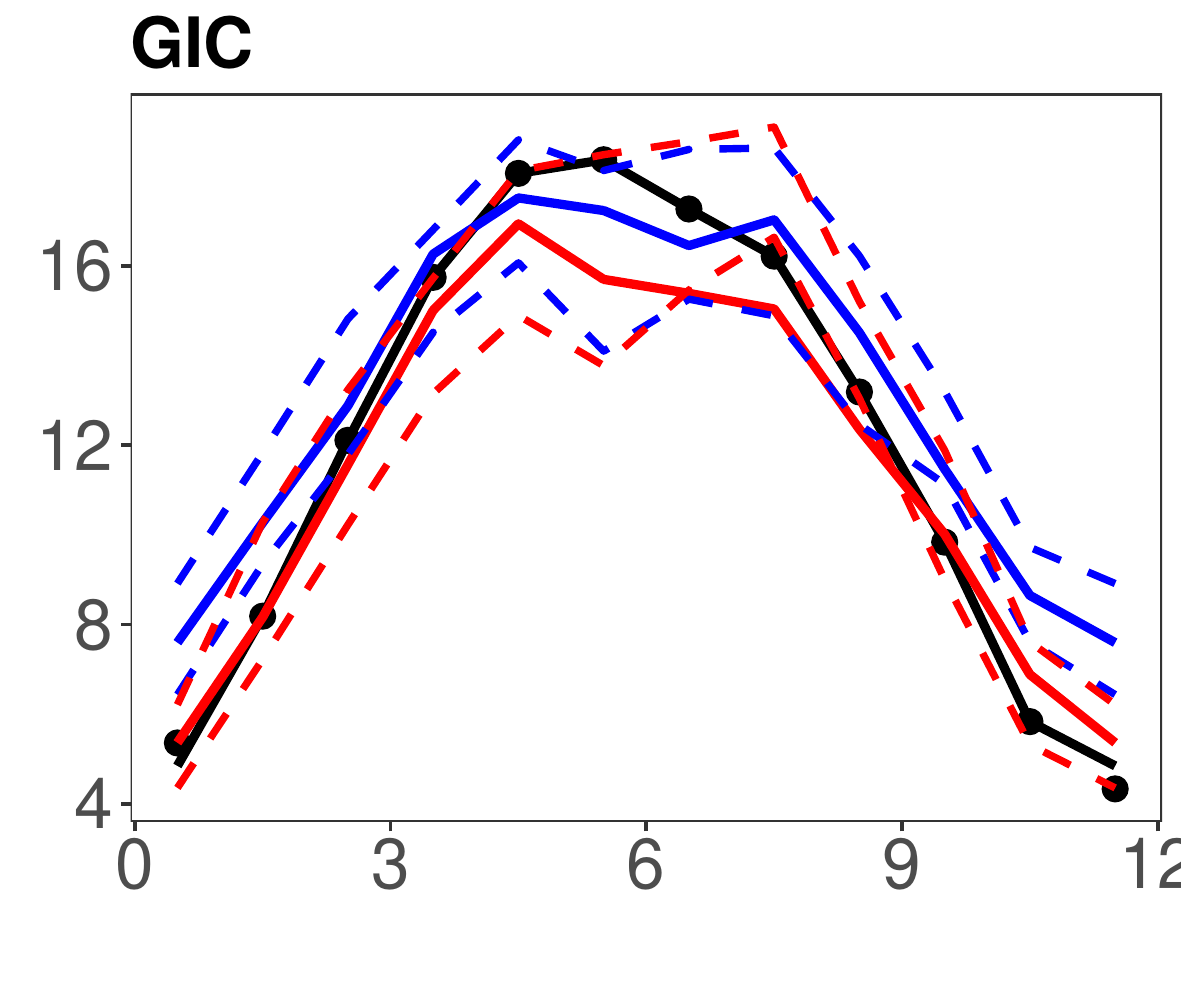}}
\\
{\includegraphics[width=5cm]{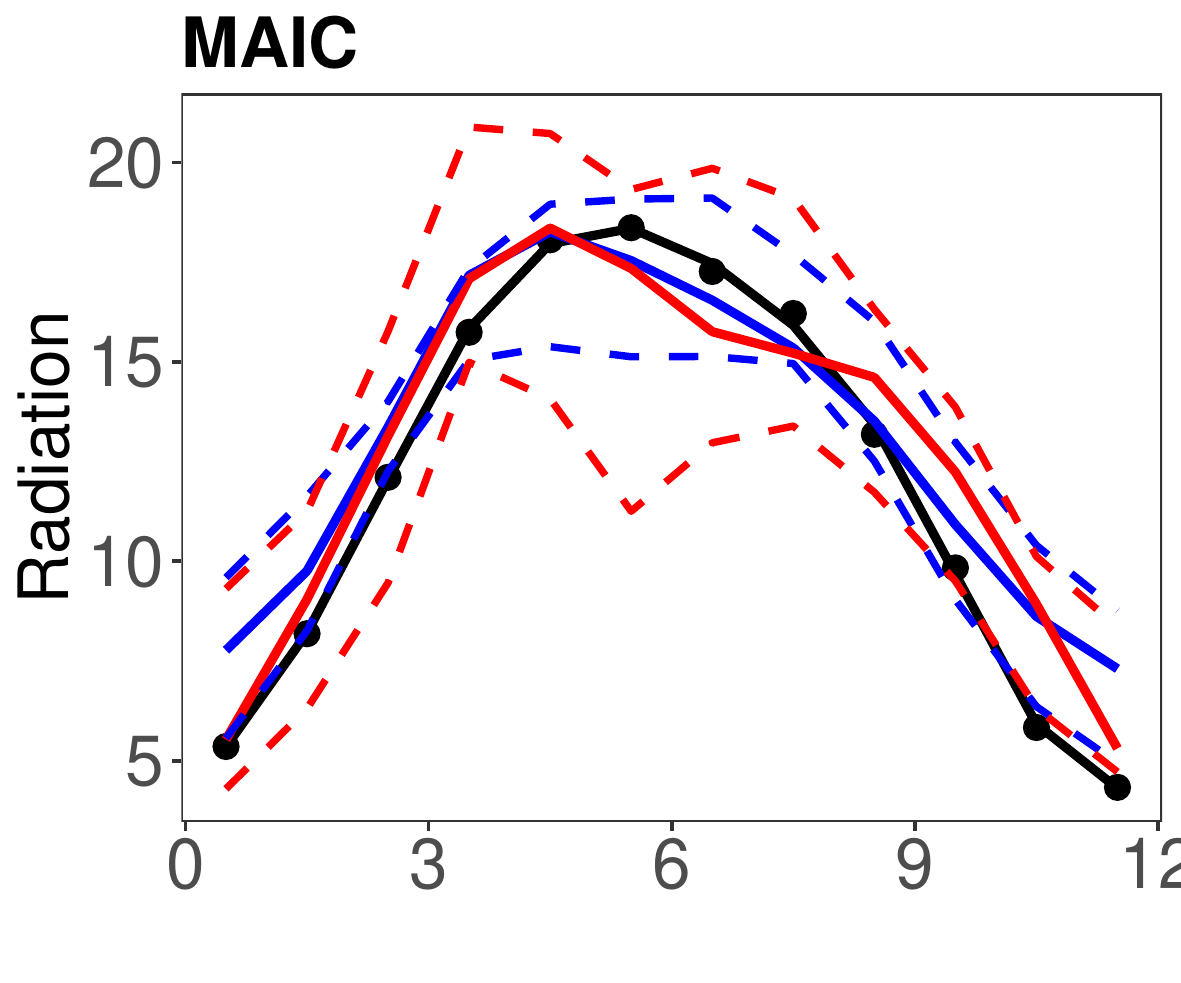}}
\qquad
{\includegraphics[width=5cm]{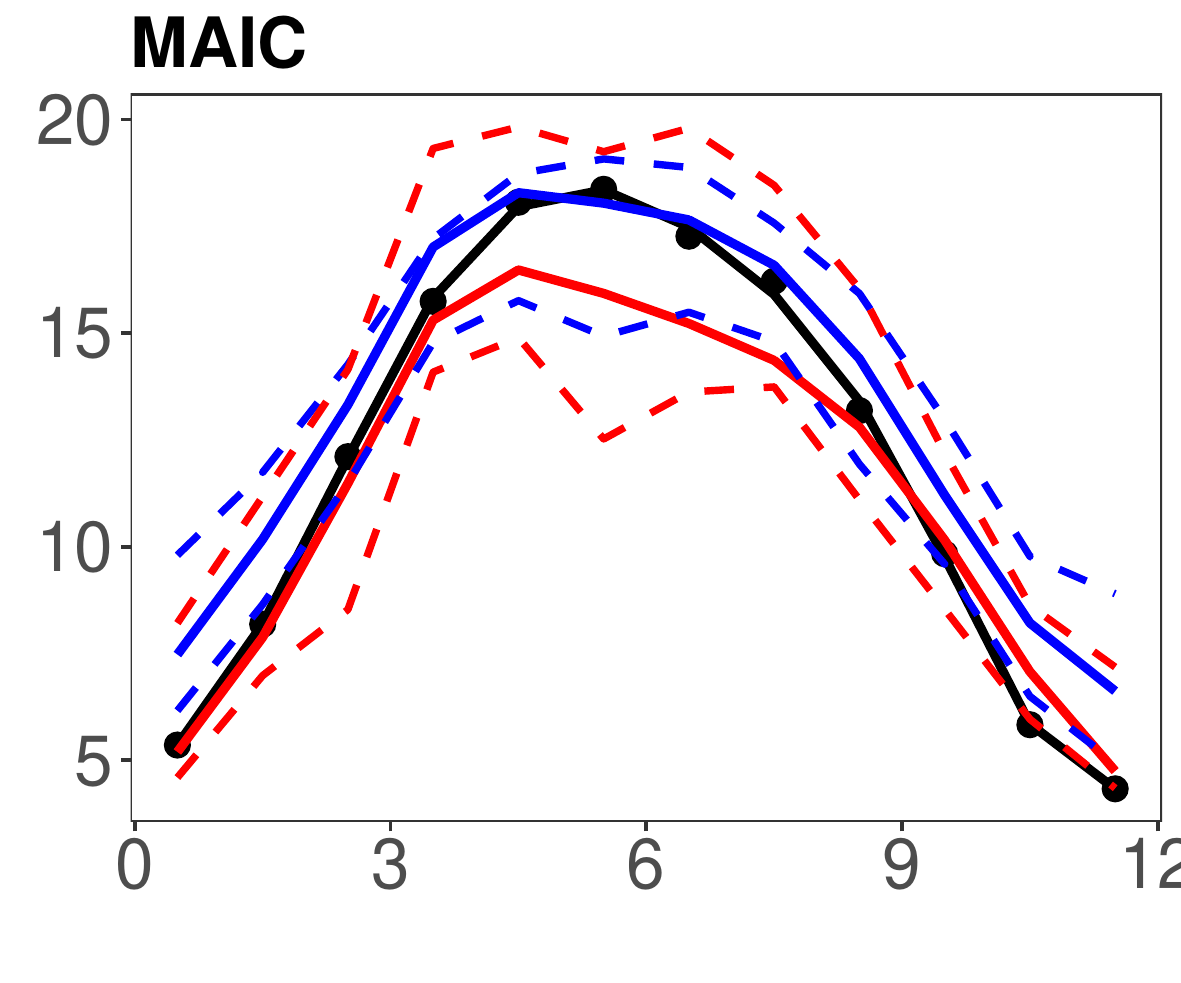}}
\qquad
{\includegraphics[width=5cm]{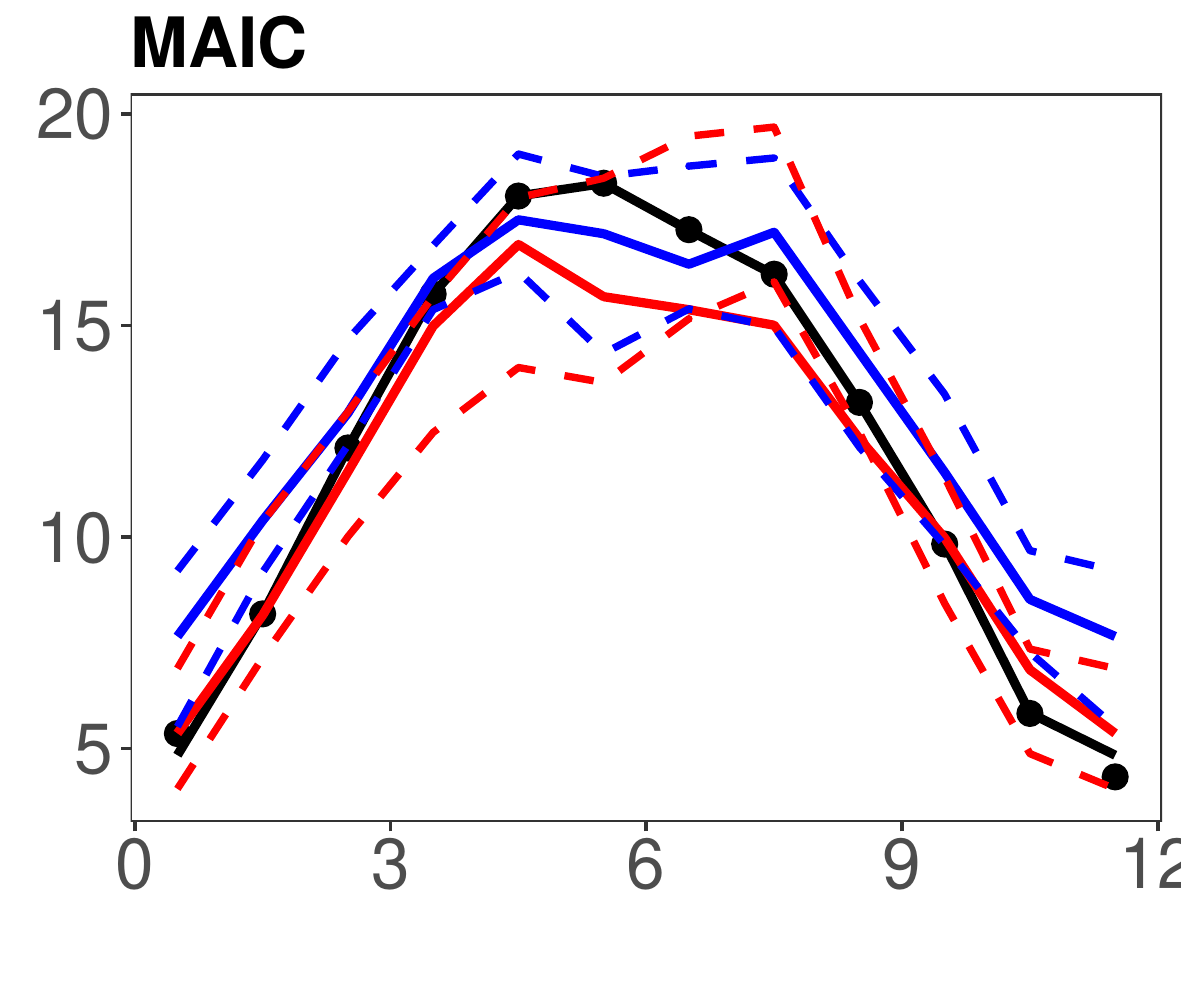}}
\\
{\includegraphics[width=5cm]{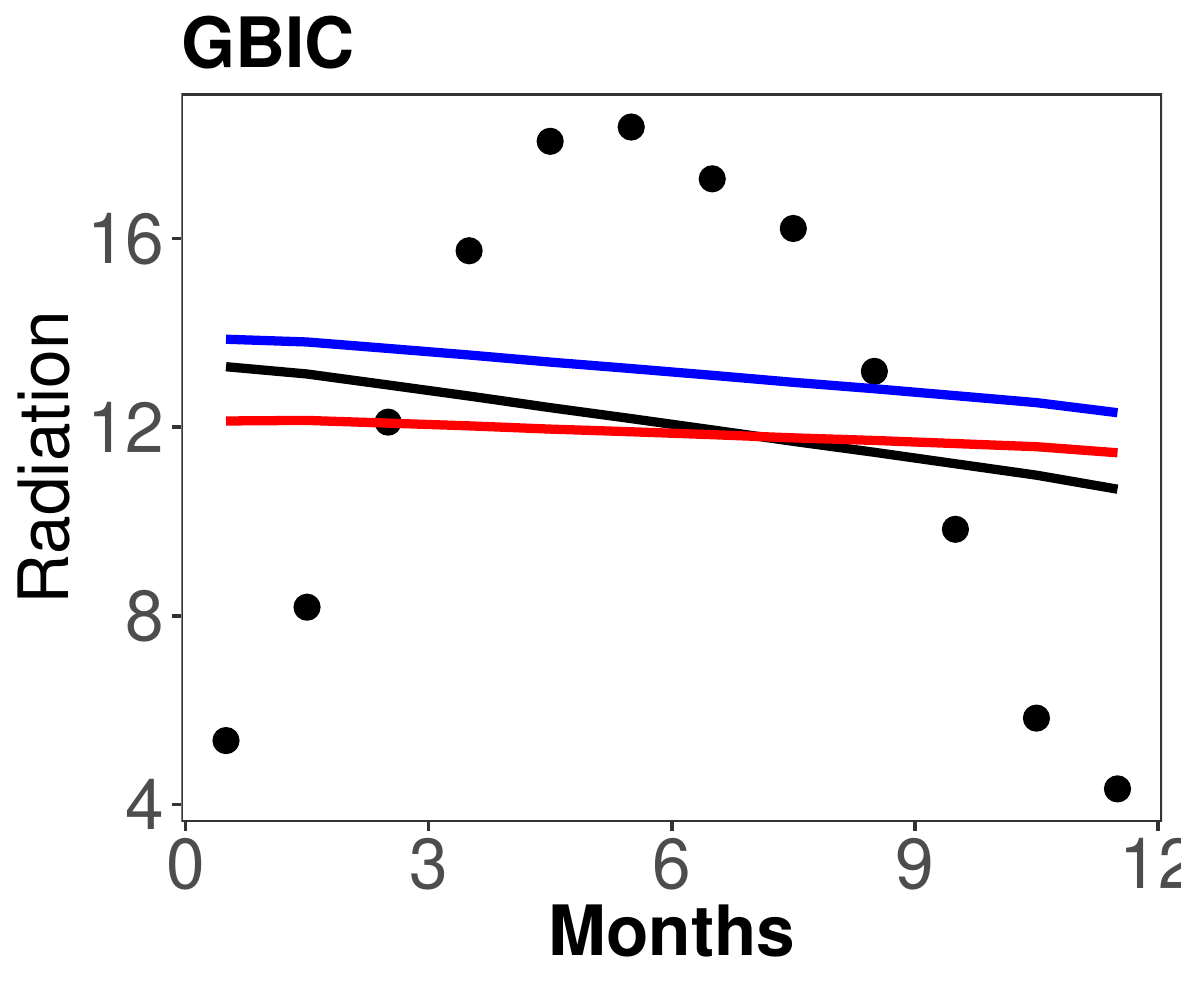}}
\qquad
{\includegraphics[width=5cm]{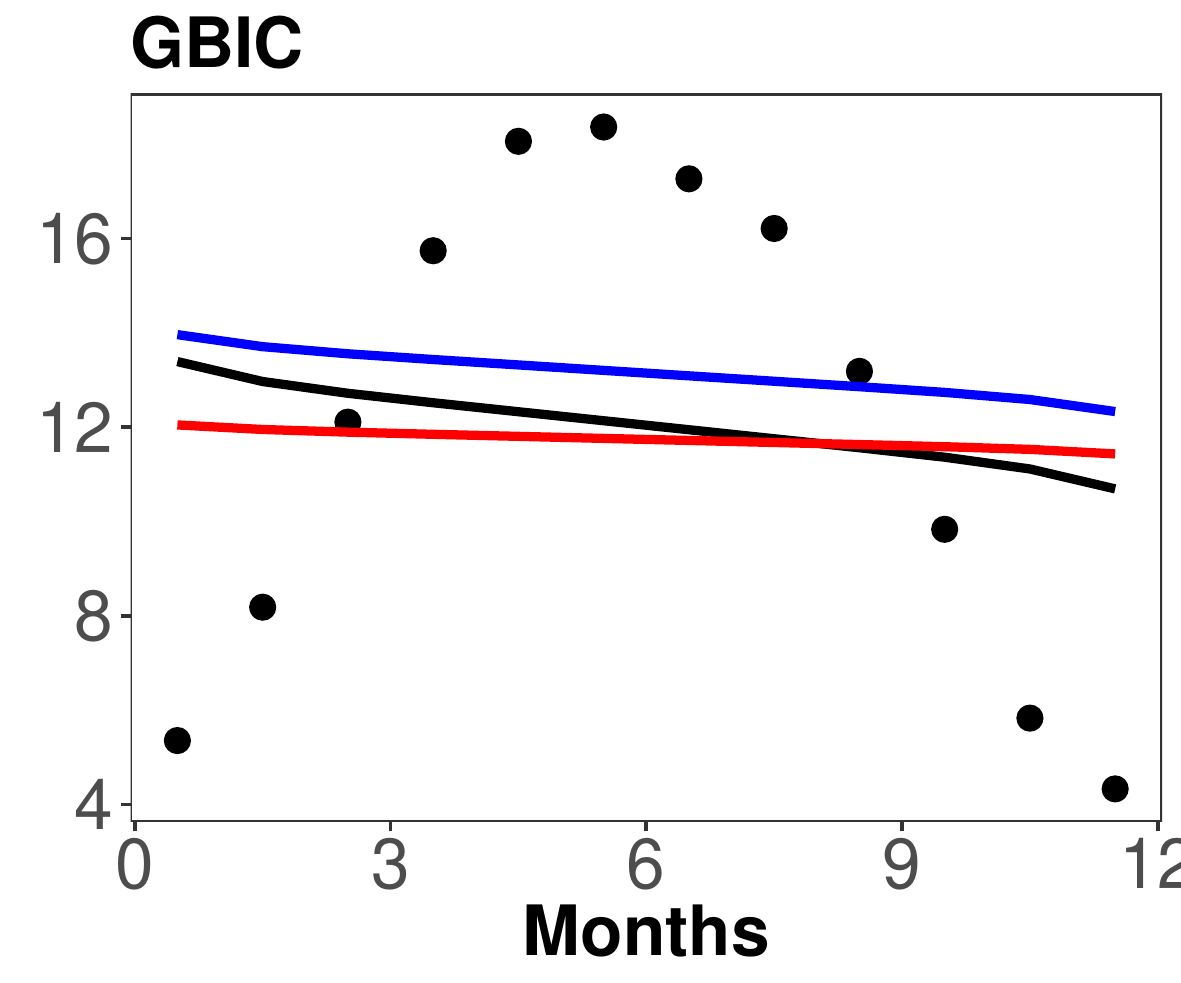}}
\qquad
{\includegraphics[width=5cm]{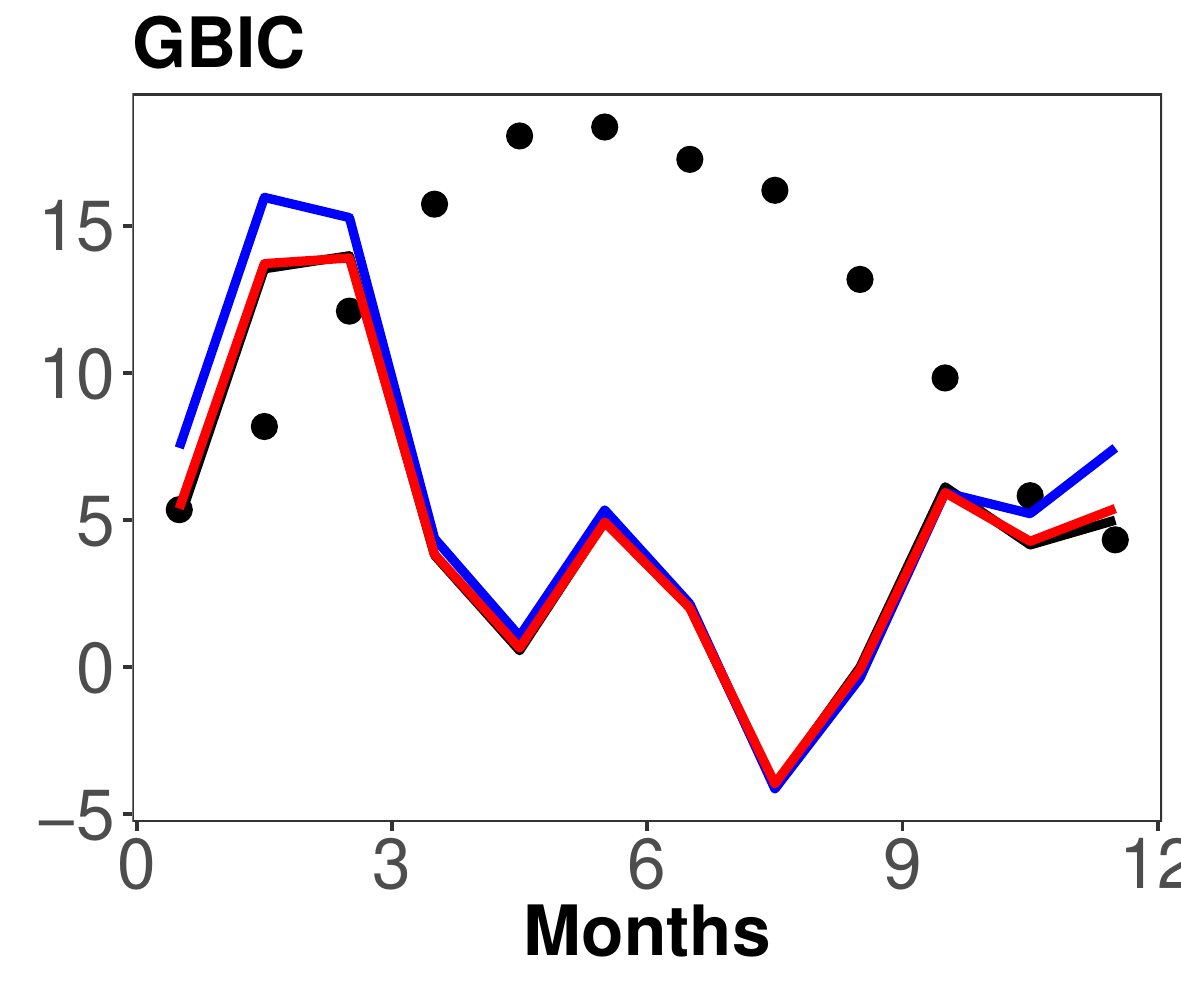}}
\caption{Plots of discrete data (black points), actual smooth functions (black solid lines) and predicted smooth functions for the Japanese meteorological data; MPL (blue solid lines) and LS (red solid lines); Gaussian basis (first column), B-spline basis (second column), and Fourier basis (third column). The dashed lines are the corresponding bootstrap confidence intervals. The GCV, GIC, MAIC, and GBIC criteria are used to control the roughness parameter and evaluate the estimated model.}
\label{fig:ci}
\end{figure}

\clearpage
\subsubsection{Tables}\label{app:jmmd_tab}

\begin{table}[!htbp]
\tabcolsep 0.36in
\centering
\caption{Station names for the Japanese monthly meteorological data.}
\begin{tabular}{@{}l l l c@{}}
\toprule
Station & Station & Station \\
\midrule
Abashiri		&  		Kumamoto	  & Oita			  \\
Aomori			&   	Matsuyama	  & Osaka			  \\
Asahikawa		&   	Miyazaki	  & Saga			  \\
Fukushima		&   	Morioka		  & Sapporo		  	\\
Hikone			&   	Nagasaki	  & Sendai			  \\
Hiroshima		&   	Naha		  & Shimonoseki	  		\\
Ishigakijima 	&   	Nara		  & Takamatsu		  \\
Kagoshima		&   	Naze		  & Wakkanai		  \\
Kochi			&   	Obihiro		  & Yamagata		  \\
\bottomrule
\end{tabular}
\label{tab:stations}
\end{table}

\clearpage
\subsection{North Dakota weekly weather data}
\subsubsection{Figures}\label{app:ndwwd_fig}

\begin{figure}[htbp]
  \centering
  \includegraphics[width=12.8cm]{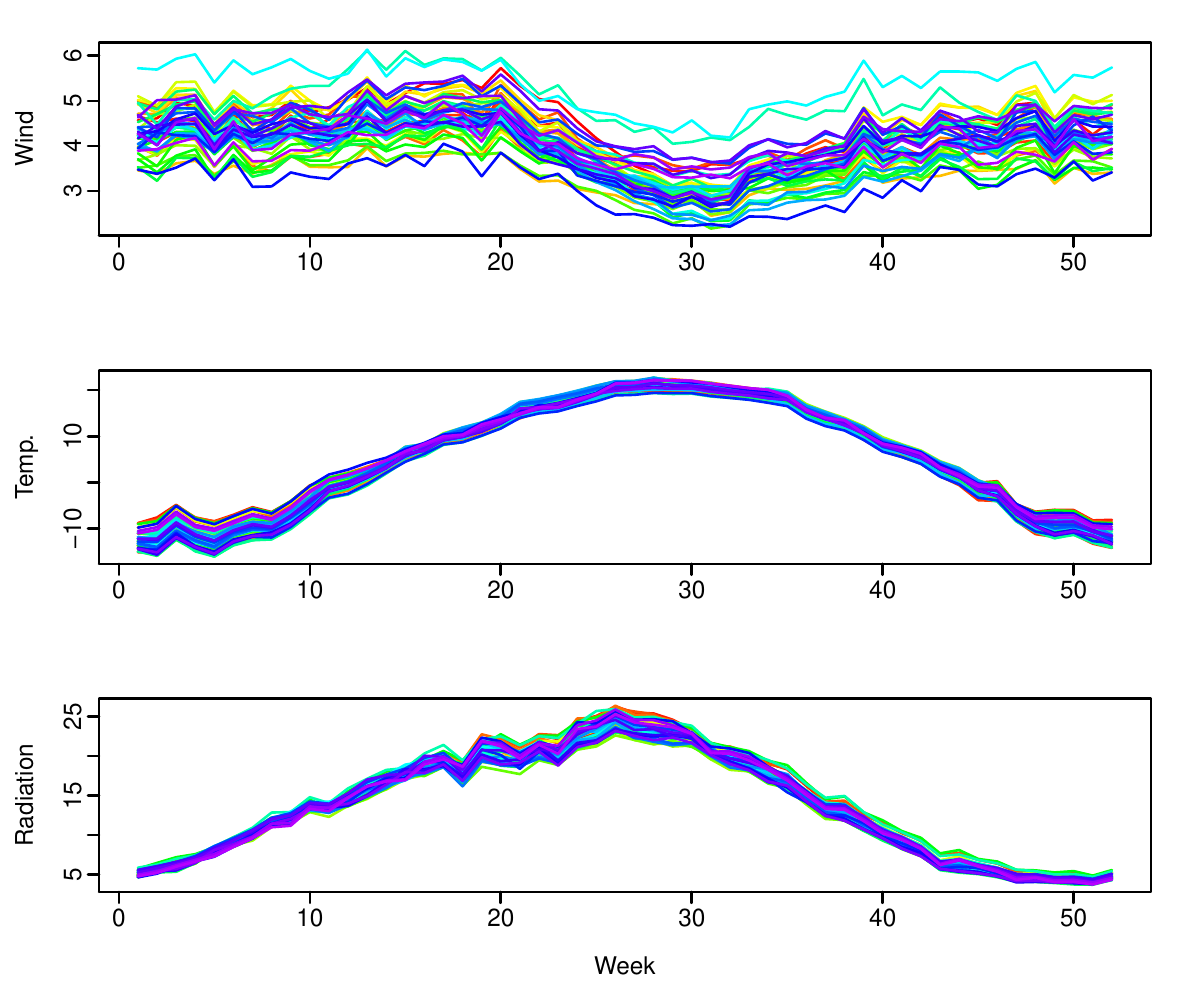}
  \caption{Time series plots of the averaged weekly weather variables.}
  \label{fig:tsN}
\end{figure}

\begin{figure}[htbp]
\centering
\subfloat[GCV]
{\includegraphics[width=88mm]{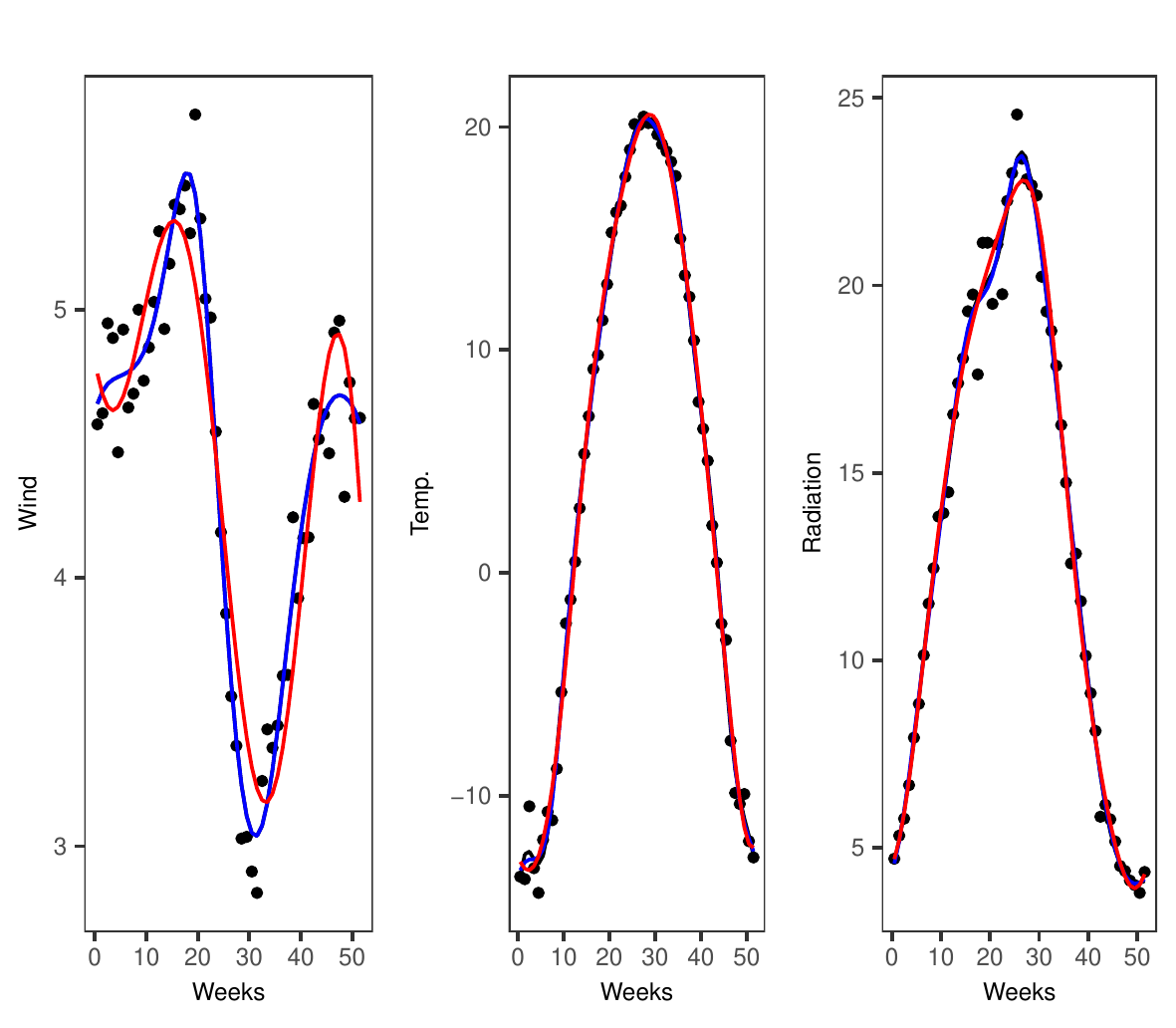}}
\quad
\subfloat[GIC]
{\includegraphics[width=88mm]{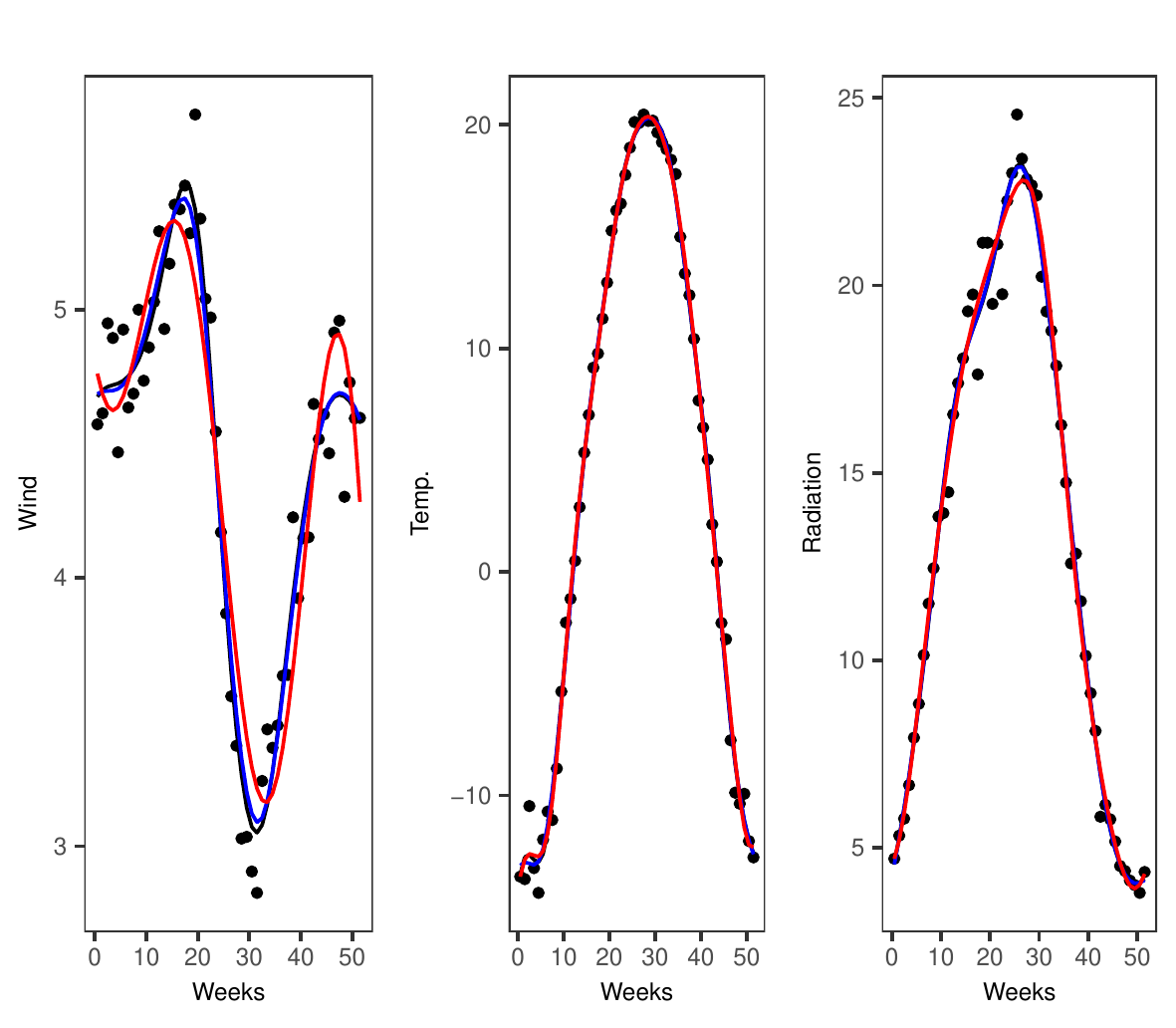}}
\\
\subfloat[MAIC]
{\includegraphics[width=88mm]{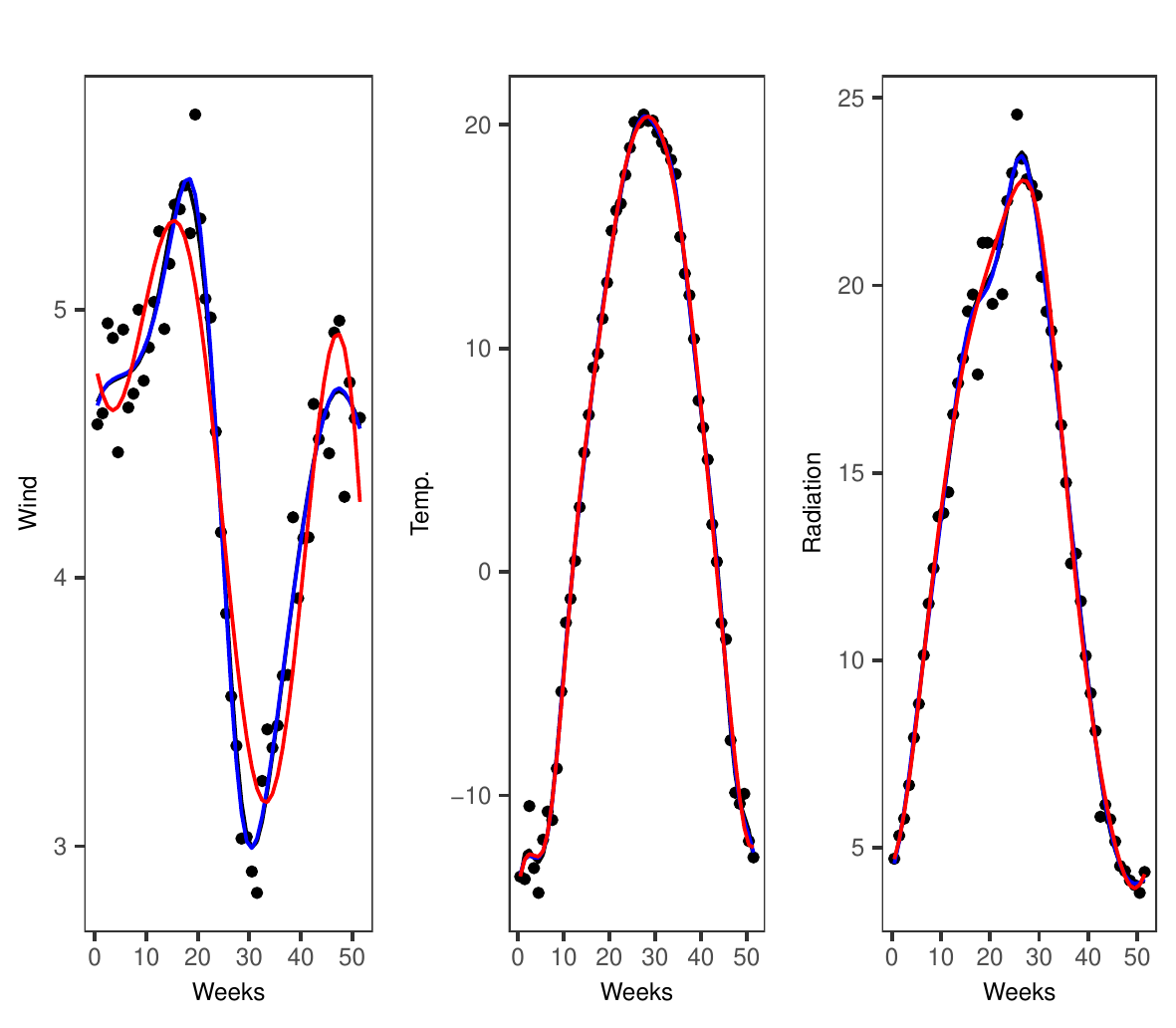}}
\quad
\subfloat[GBIC]
{\includegraphics[width=88mm]{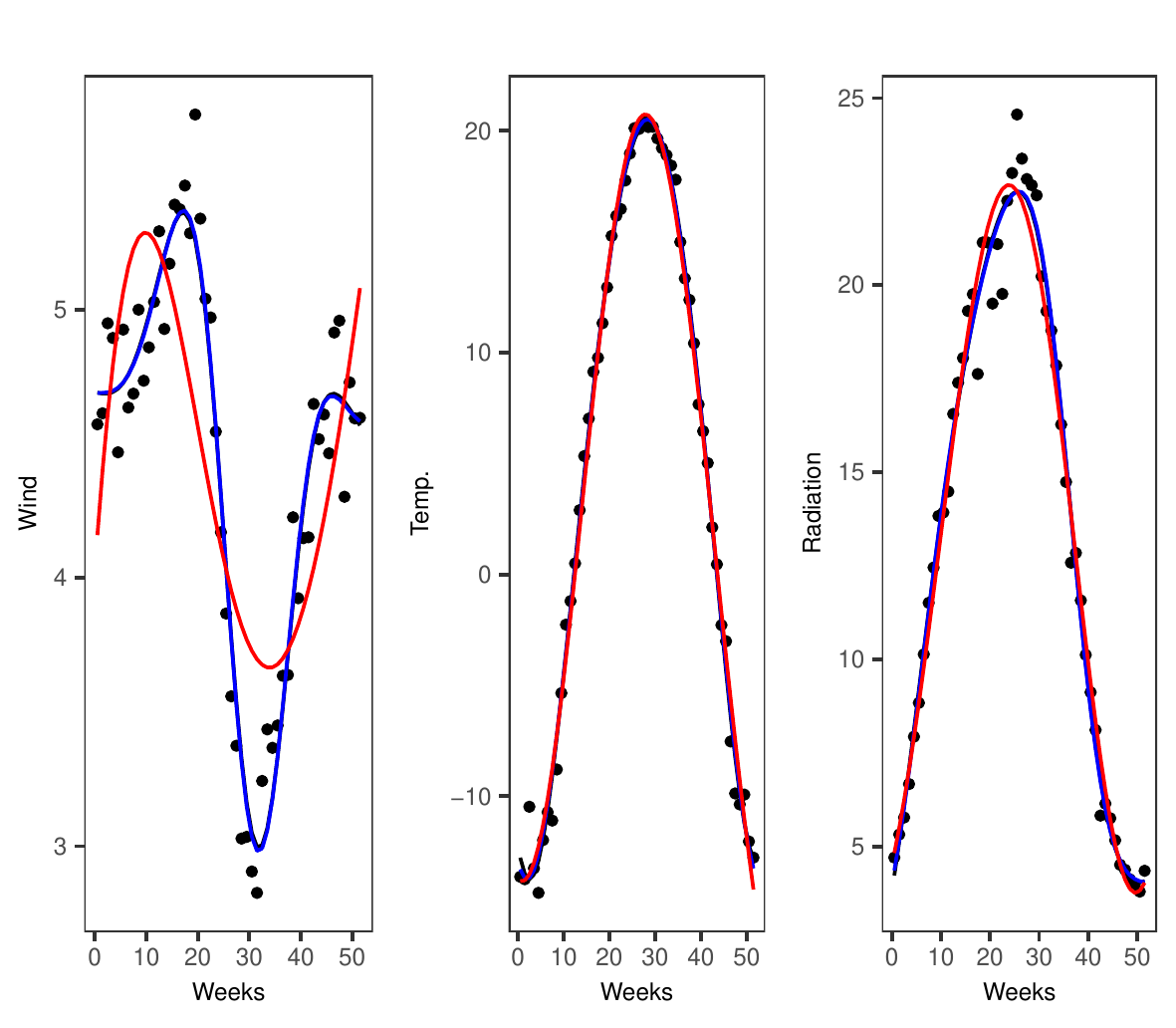}}
\caption{Plots of discrete weekly weather variables (black points) and their smooth functions: Gaussian basis (black lines), B-spline basis (blue lines), and Fourier basis (red lines). The GCV, GIC, MAIC, and GBIC criteria are used to control the roughness parameter and evaluate the estimated model.}
\label{fig:smoothN}
\end{figure}

\begin{figure}[htbp]
\centering
{{\includegraphics[width=5cm]{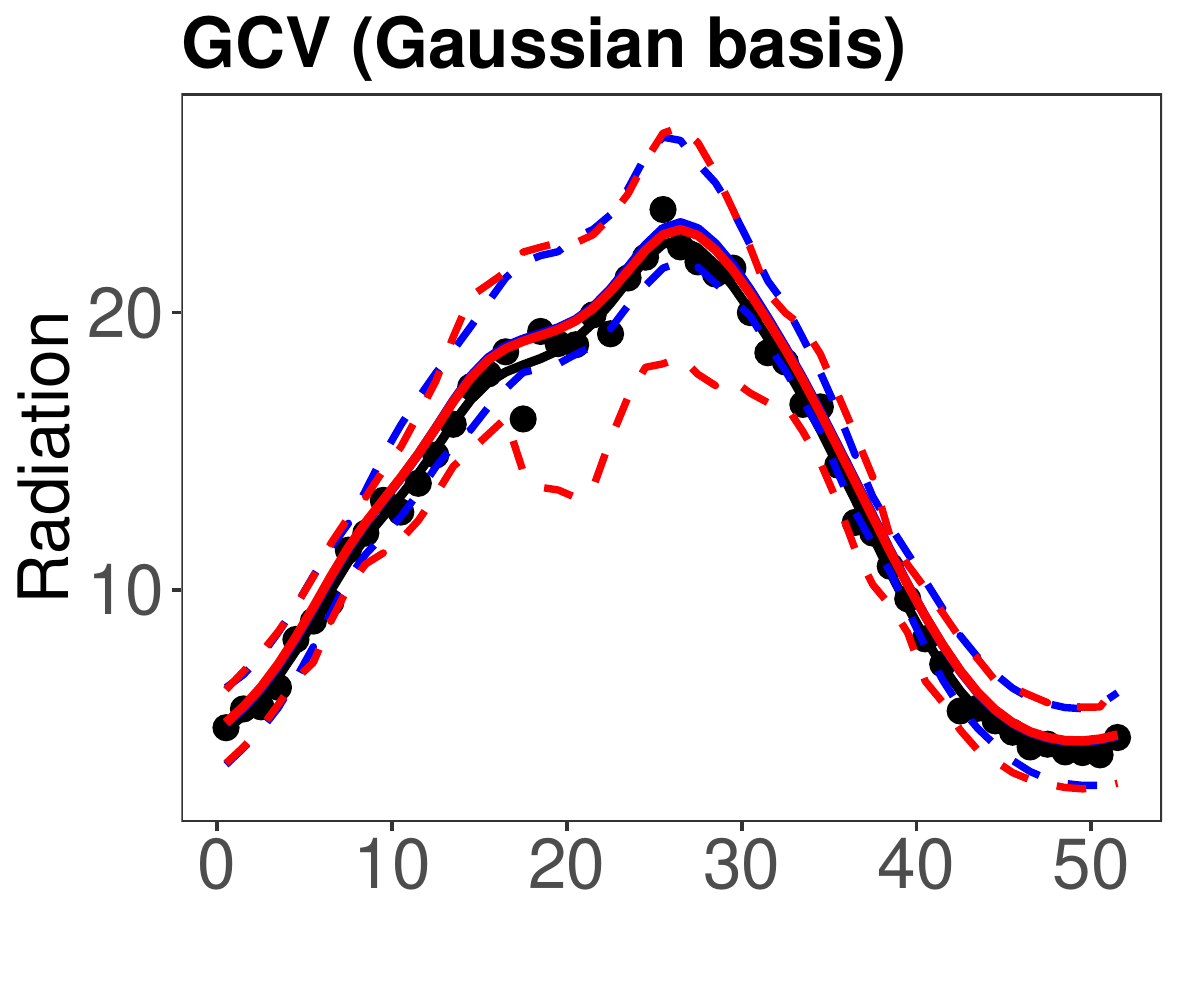}}}
\qquad
{{\includegraphics[width=5cm]{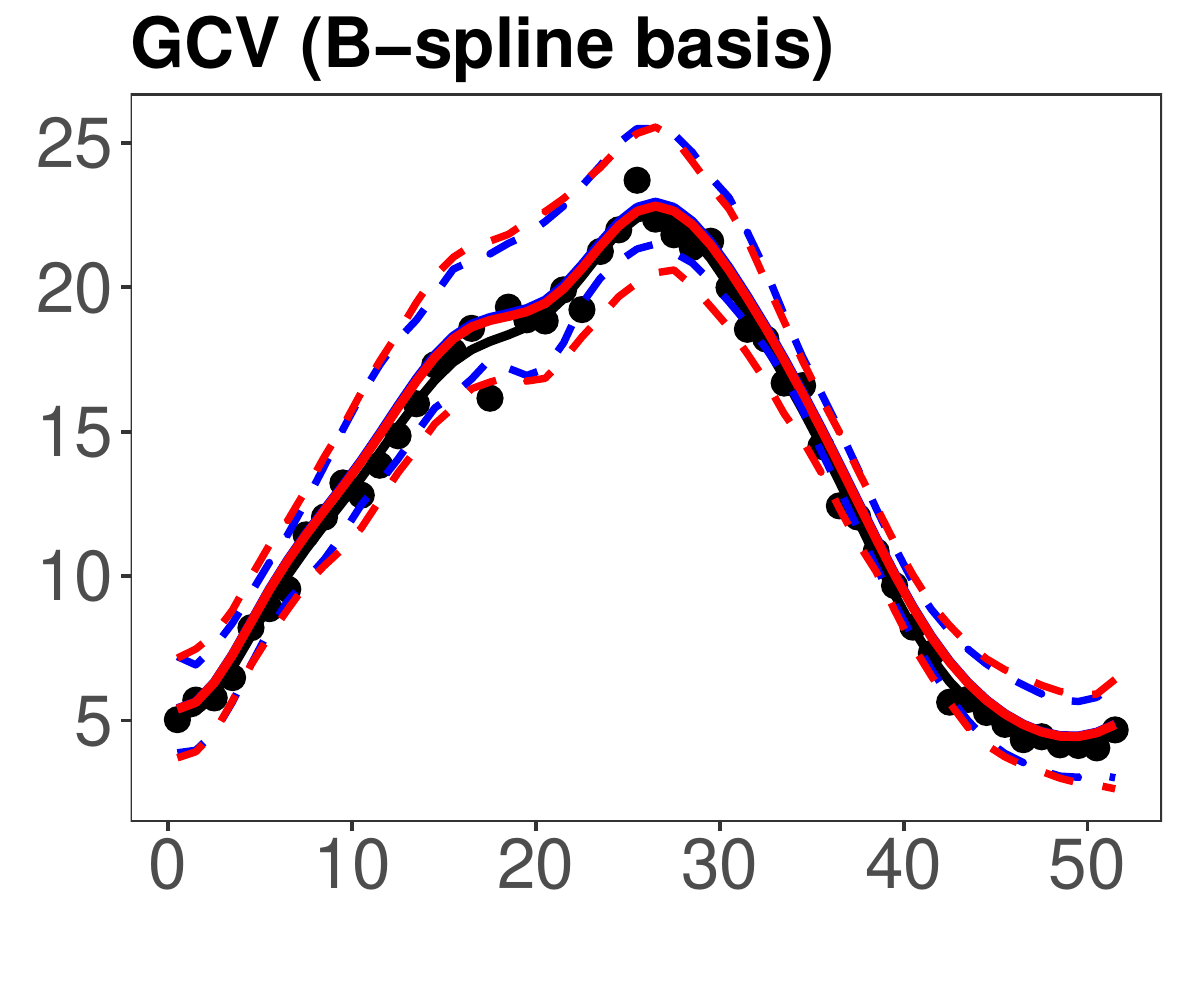}}}
\qquad
{{\includegraphics[width=5cm]{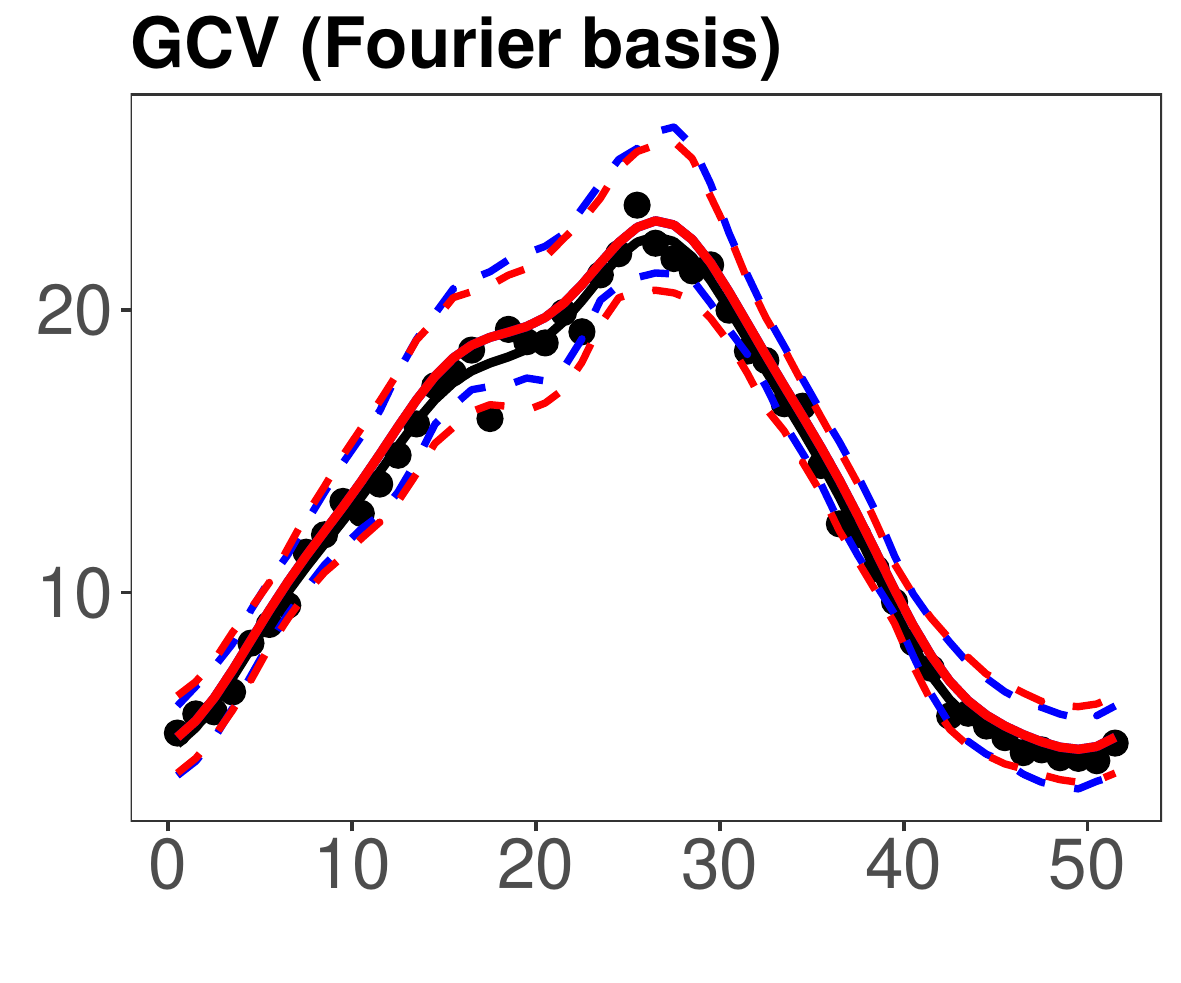}}}
\\
{\includegraphics[width=5cm]{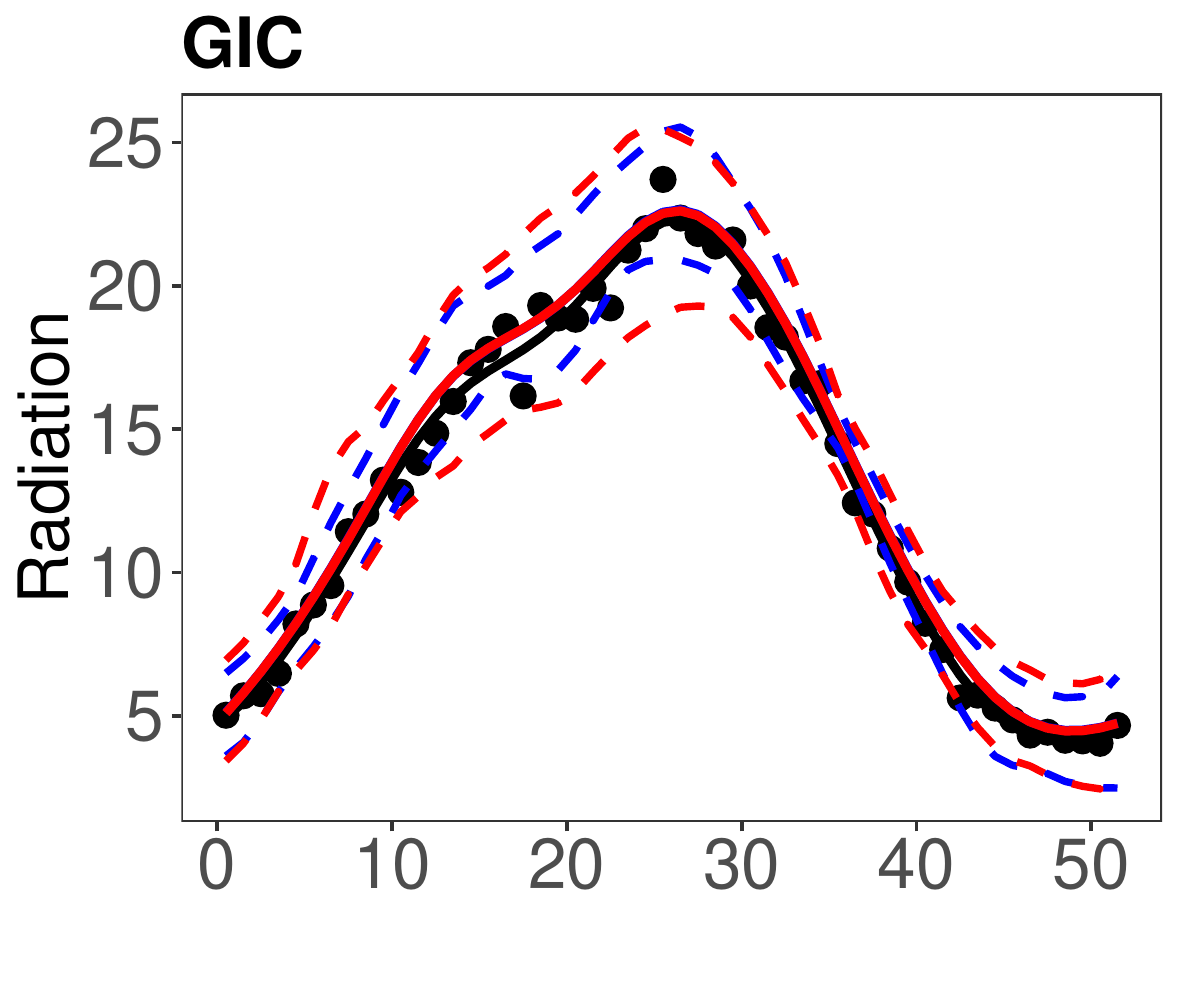}}
\qquad
{\includegraphics[width=5cm]{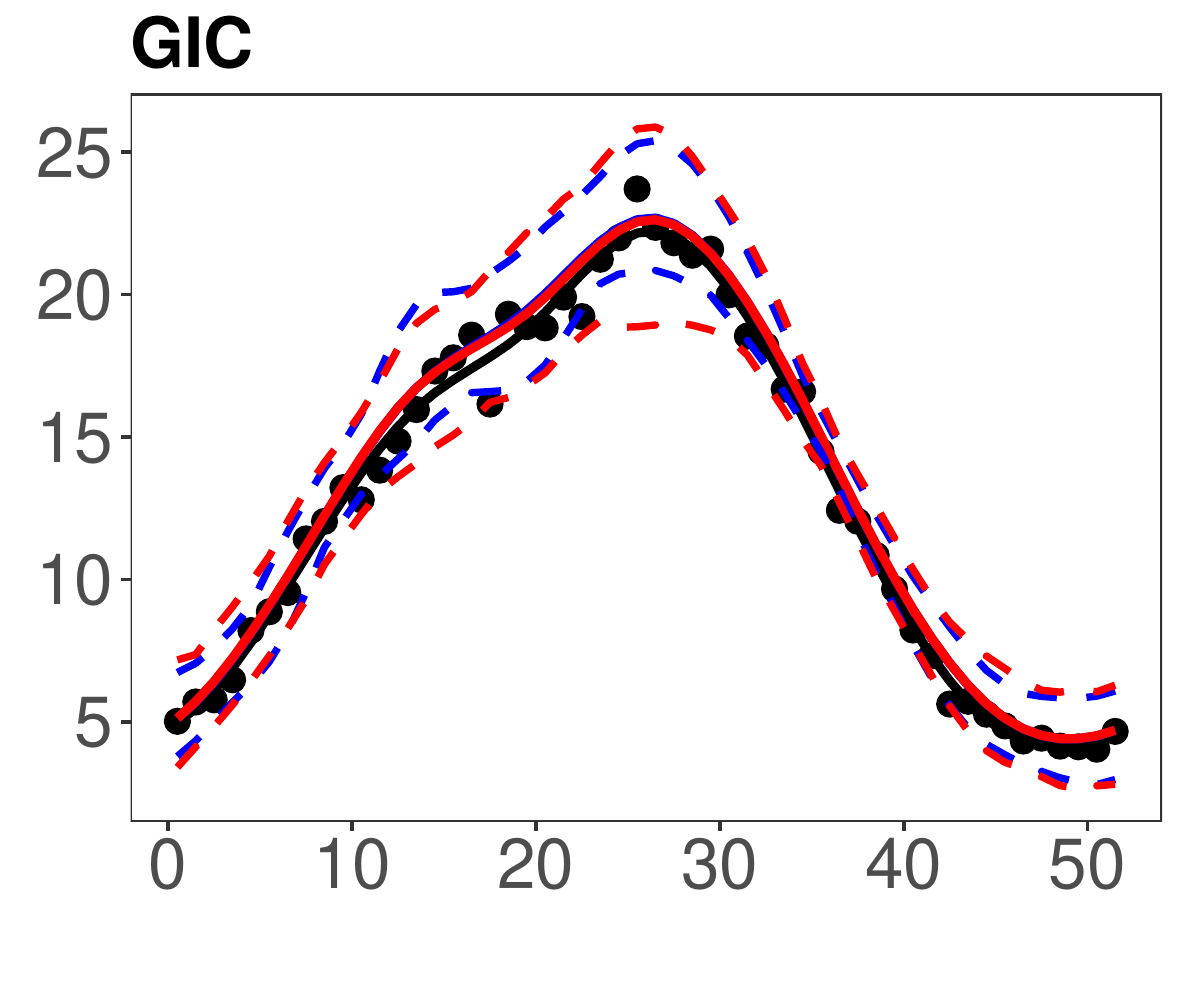}}
\qquad
{\includegraphics[width=5cm]{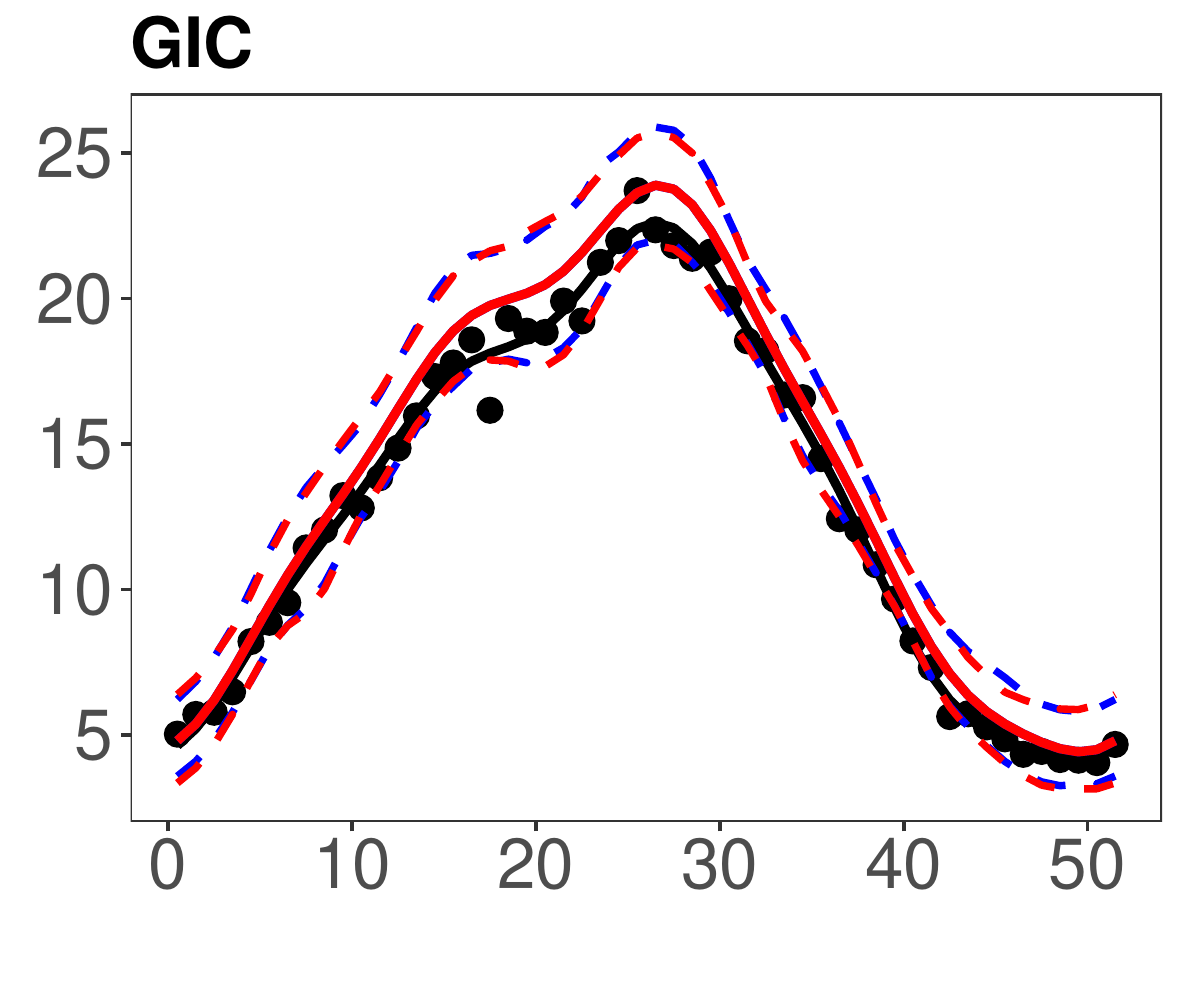}}
\\
{\includegraphics[width=5cm]{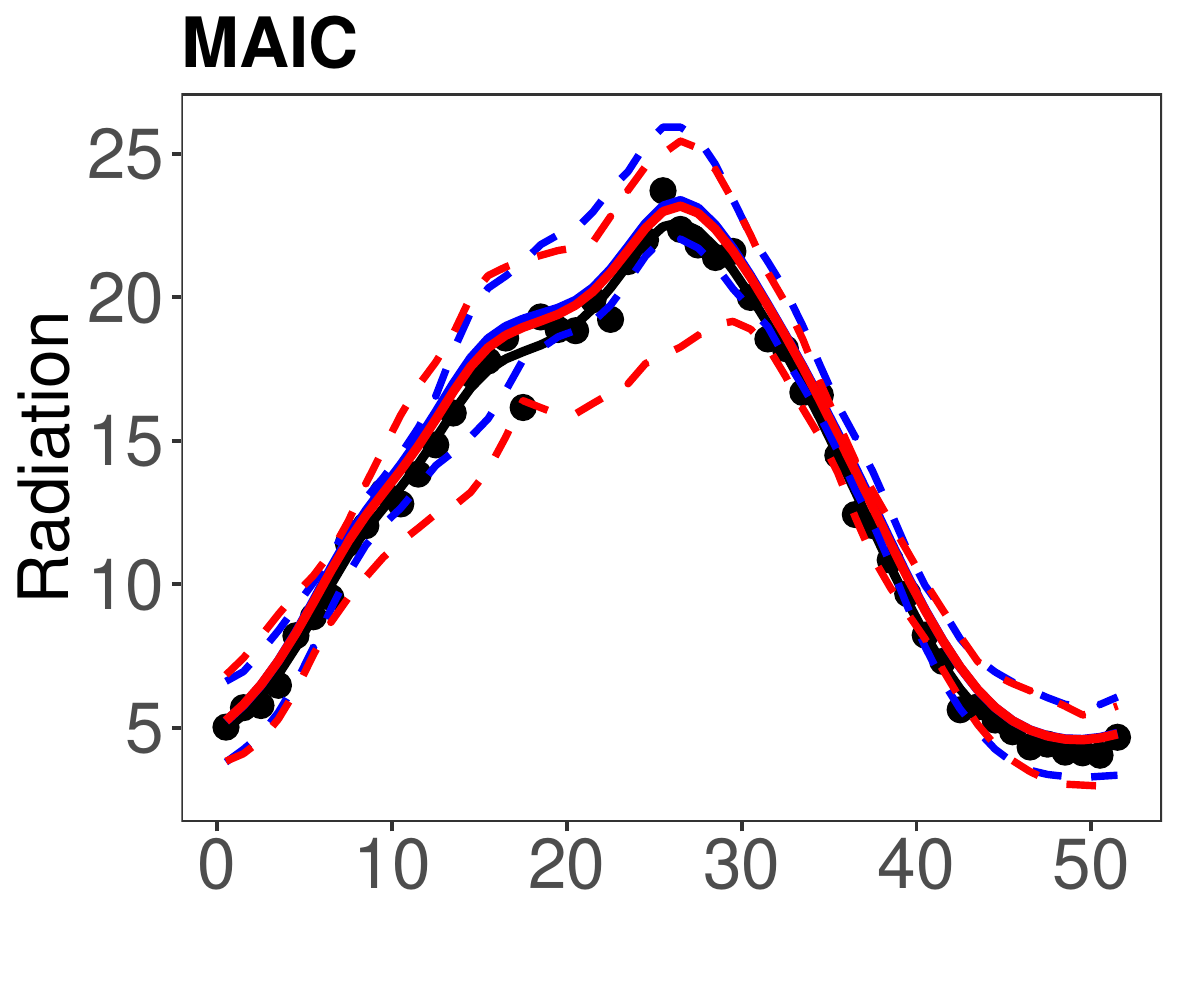}}
\qquad
{\includegraphics[width=5cm]{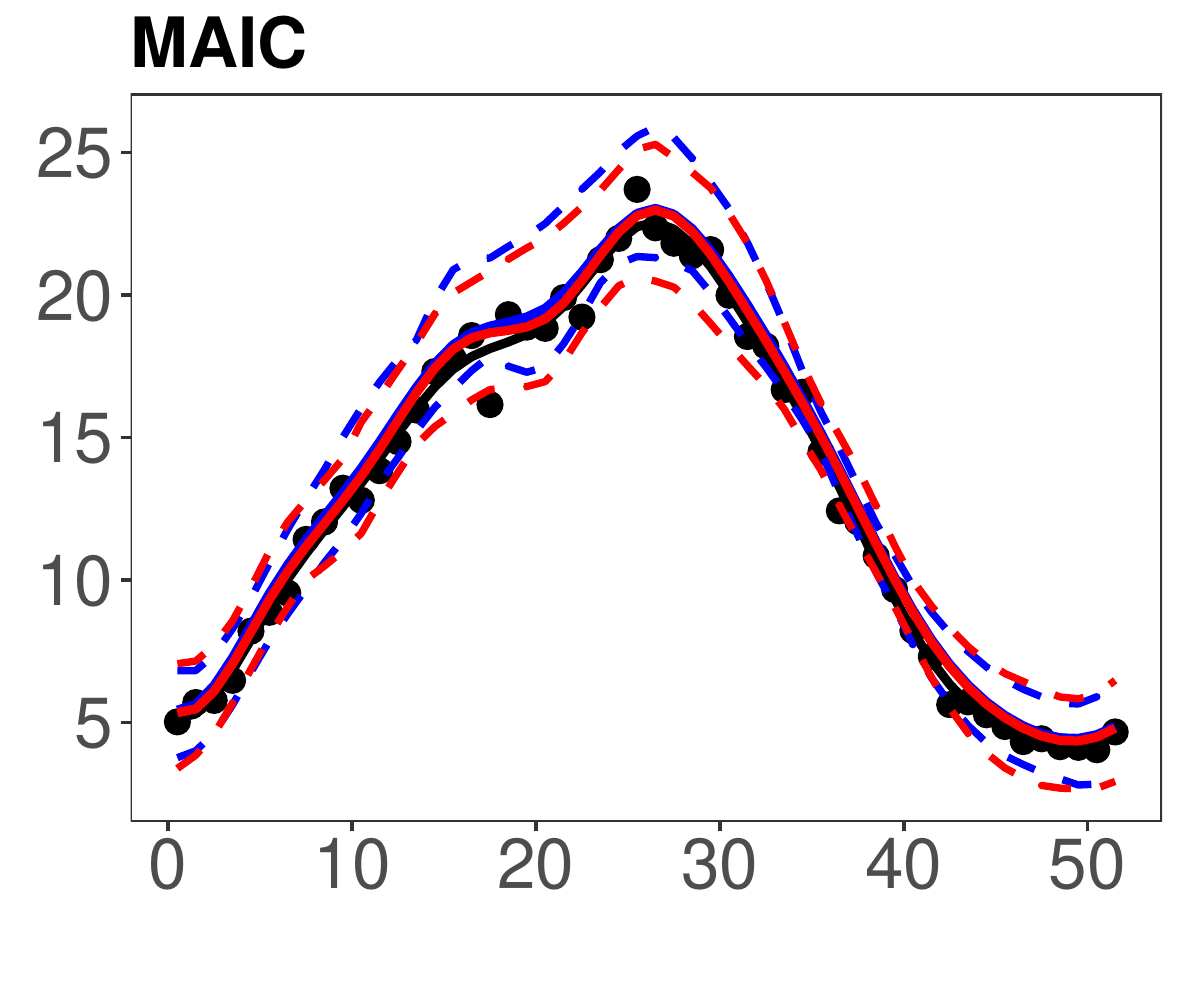}}
\qquad
{\includegraphics[width=5cm]{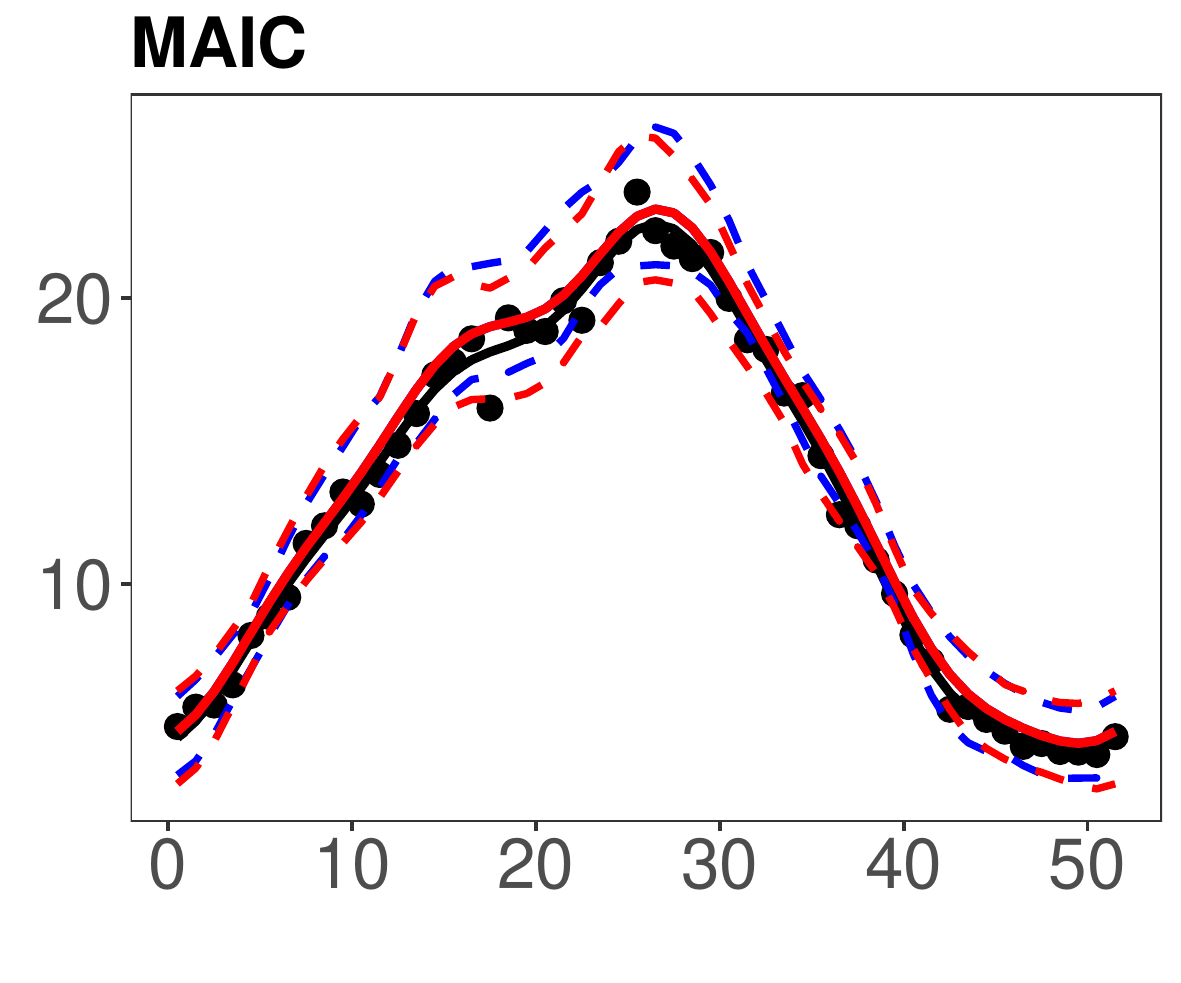}}
\\
{\includegraphics[width=5cm]{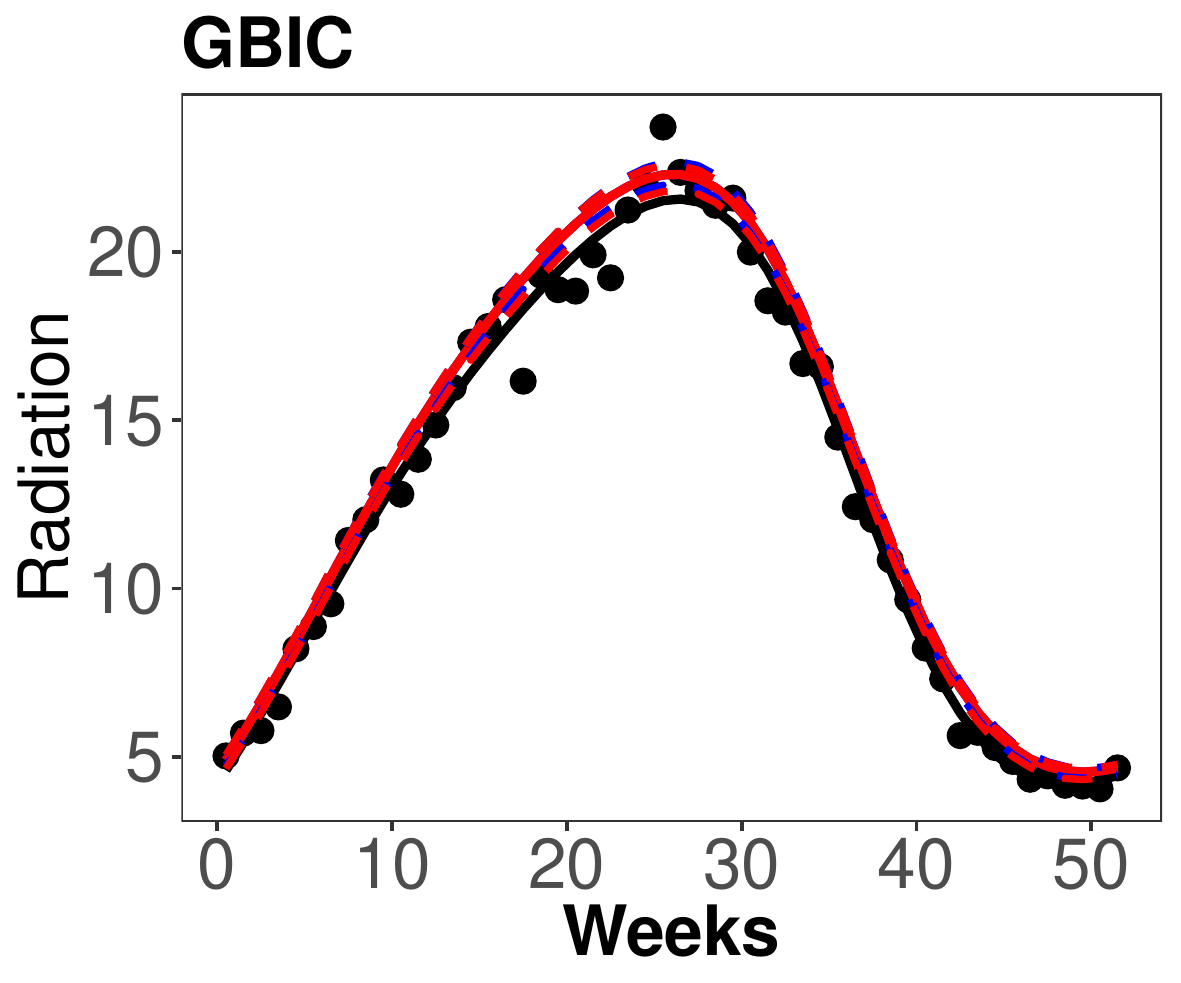}}
\qquad
{\includegraphics[width=5cm]{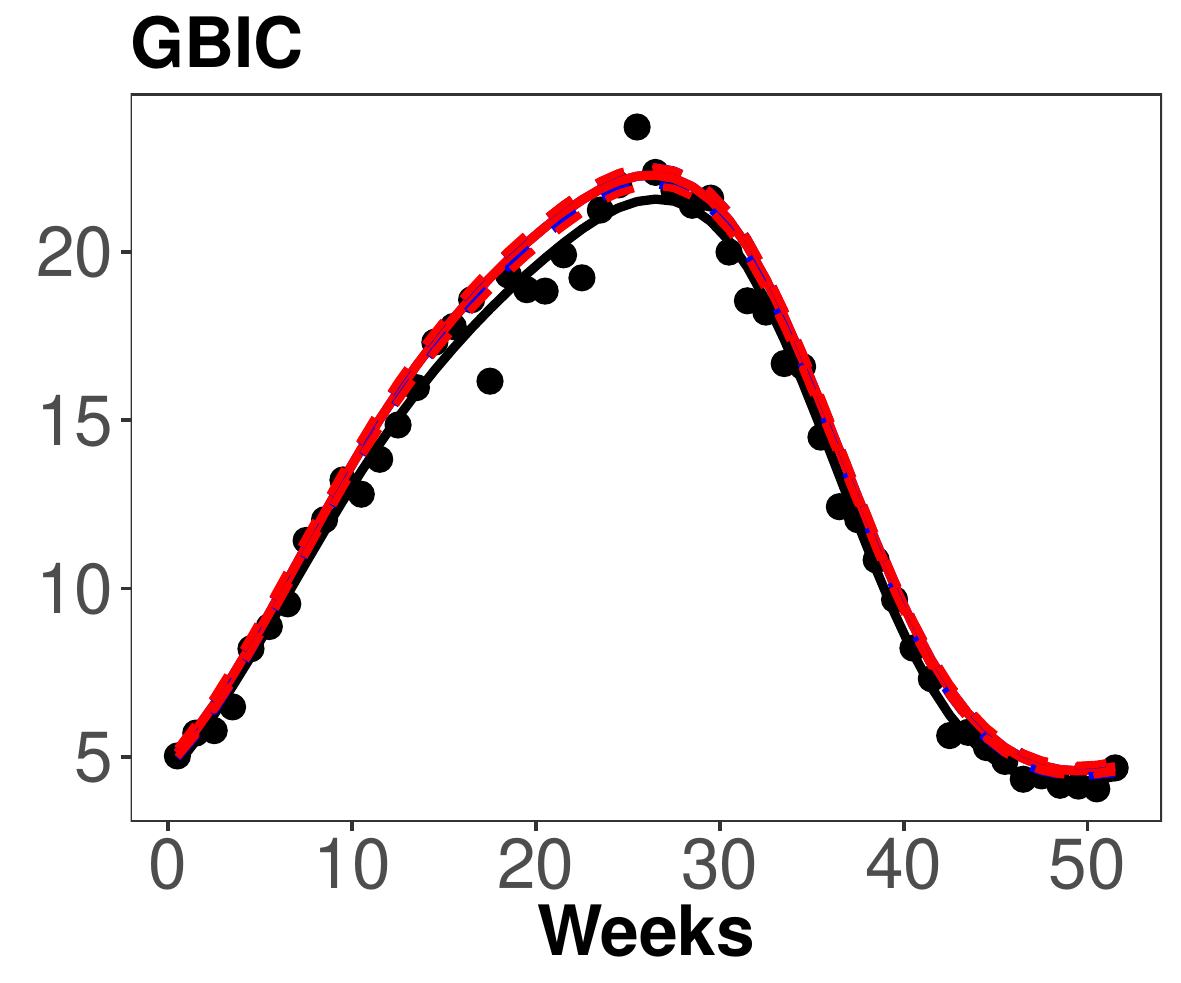}}
\qquad
{\includegraphics[width=5cm]{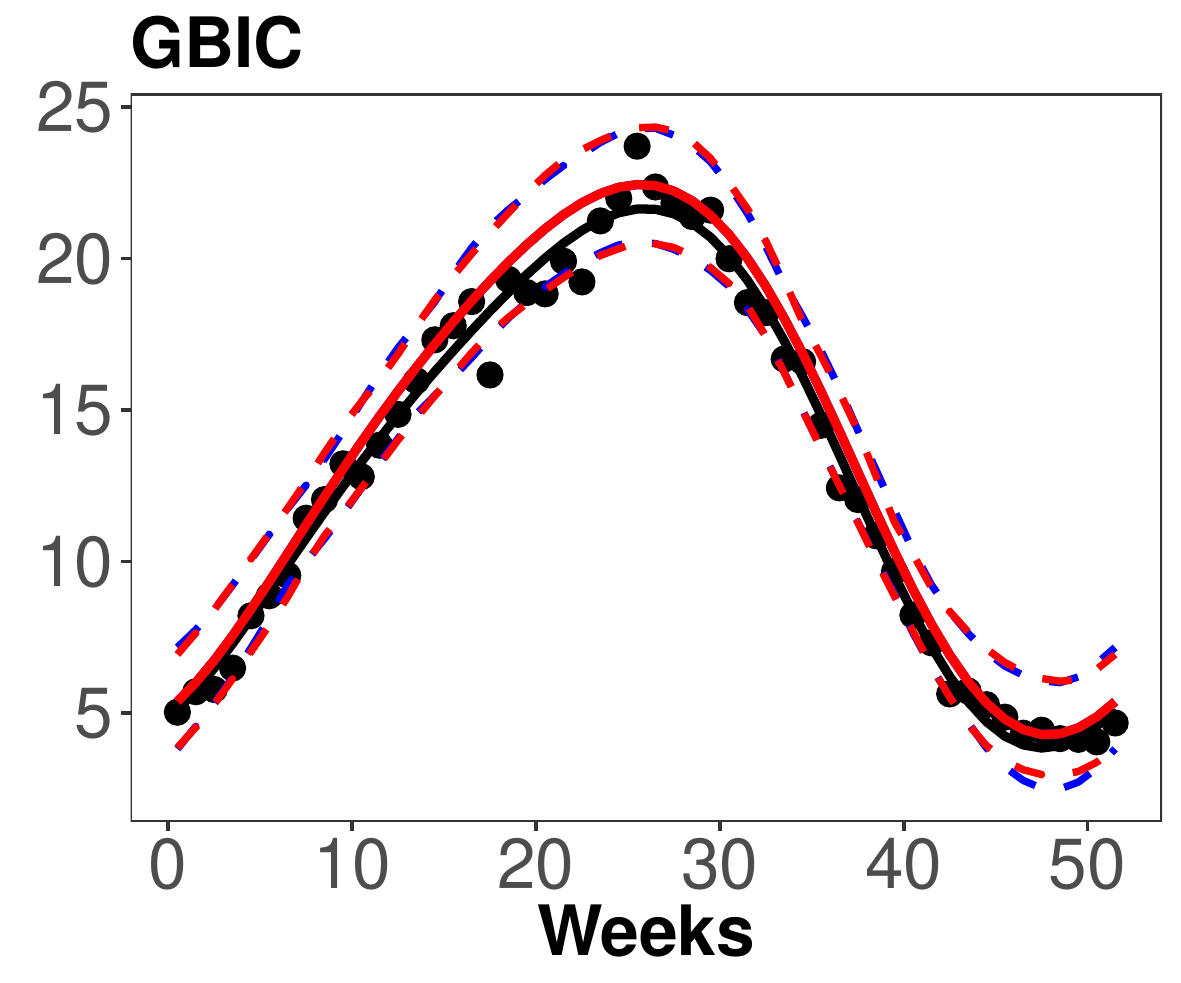}}
\caption{Plots of discrete data (black points), actual smooth functions (black solid lines), and predicted smooth functions for the weekly weather data; MPL (blue solid lines) and LS (red solid lines); Gaussian basis (first row), B-spline basis (second row), and Fourier basis (third row). The dashed lines are the corresponding bootstrap confidence intervals. The GCV, GIC, MAIC, and GBIC criteria are used to control the roughness parameter and evaluate the estimated model.}
\label{fig:ciN}
\end{figure}

\clearpage
\subsubsection{Tables}\label{app:ndwwd_tab}

\begin{table}[htbp]
\centering
\tabcolsep 0.2in
\caption{Station names for the North Dakota weekly weather data.}
\begin{tabular}{@{}llllll@{}}
\toprule
Station & Station & Station & Station & Station  \\
\midrule
Baker			&Dickinson		& Hillsboro		& Mohall  		& Streeter \\
Beach			&Edgeley	 	& Hofflund		& Mooreton 		& Turtle Lake\\
Bottineau		&Eldred	 		& Humboldt		& Oakes 		& Warren\\
Bowman			&Fargo	 		& Jamestown		& Perley 		& Watford City\\
Brorson			&Forest River	& Langdon		& Prosper		& Williston\\
Cando			&Galesburg		& Linton		& Robinson  \\
Carrington 		&Grand Forks	& Mandan		& Rolla  \\
Cavalier		&Harvey			& Mayville		& Sidney  \\
Crary			&Hazen	 		& McHenry		& St Thomas  \\
Dazey			&Hettinger	 	& Minot			& Stephen  \\
\bottomrule
\end{tabular}
\label{tab:stationsN}
\end{table}

\end{document}